%% file: Main.tex
\documentclass[twocolumn]{aastex62}
\usepackage{lineno}
\usepackage{hyperref}
\usepackage{amsmath}
\usepackage{amssymb}
\usepackage{graphicx}
\usepackage[utf8]{inputenc}

\DeclareRobustCommand{\ion}[2]{\textup{#1\,\textsc{\lowercase{#2}}}}

\newcommand\fei{\ion{Fe}{i}}
\newcommand\feii{\ion{Fe}{ii}}

\newcommand\euii{\ion{Eu}{ii}}

\newcommand\tiii{\ion{Ti}{ii}}
\newcommand\oi{\ion{O}{i}}

\newcommand\mgi{\ion{Mg}{i}}
\newcommand\ali{\ion{Al}{i}}
\newcommand\cai{\ion{Ca}{i}}
\newcommand\scii{\ion{Sc}{ii}}

\newcommand\cri{\ion{Cr}{i}}

\newcommand\mni{\ion{Mn}{i}}
\newcommand\coi{\ion{Co}{i}}
\newcommand\nii{\ion{Ni}{i}}
\newcommand\zni{\ion{Zn}{i}}
\newcommand\ki{\ion{K}{i}}

\newcommand\mystar{J0524$-$0336}
\newcommand\sii{\ion{Si}{i}}

\newcommand{\kms}{km\,s$^{-1}$}
\newcommand{\teff}{$T_{\rm eff}$}
\newcommand{\logg}{$\log g$}
\newcommand{\vt}{$\xi_{t}$}

\newcommand{\rproc}{$r$-process}
\newcommand{\sproc}{$s$-process}

\newcommand{\AB}[2]{$\mbox{[#1/#2]}$}
\newcommand{\feh}{\AB{Fe}{H}}

\usepackage[utf8]{inputenc}

\newcommand{\CC}[1]{{ \textcolor{black}{#1}}}

\shorttitle{The Most Li-rich Giant Star}
\shortauthors{Kowkabany et al.}
\begin{document}

\title{\bf Discovery of a Metal-Poor Red Giant Star with the Highest Ultra-Lithium Enhancement}                         

\newcommand{\alex}[1]{\textcolor{orange}{(APJ: #1)}}
\newcommand{\alexs}[2]{\textcolor{orange}{(APJ: \sout{#1} #2)}}

\correspondingauthor{Rana Ezzeddine}
\email{rezzeddine@ufl.edu}

\author{Jeremy Kowkabany}

\affiliation{Department of Astronomy, University of Florida, Gainesville, FL 32601, USA} 
\affiliation{Department of Physics, Florida State University, Tallahassee, FL 32306, USA}

\author{Rana Ezzeddine}

\affiliation{Department of Astronomy, University of Florida, Gainesville, FL 32601, USA}
\affiliation{Joint Institute for Nuclear Astrophysics - Center for Evolution of the Elements, USA}

\author{Corinne Charbonnel}

\affiliation{Department of Astronomy, University of Geneva, Chemin de P\'{e}gase
51, 1290 Versoix, Switzerland} 
\affiliation{IRAP, CNRS UMR 5277 \& Université de Toulouse, 14 av. E.~Belin, 31400 Toulouse, France}

\author{Ian U. Roederer}
\affiliation{Department of Physics, North Carolina State University, Raleigh, NC 27695, USA}

\author{Ella Xi Wang}
\affiliation{Research School of Astronomy and Astrophysics, Australian National University, Canberra, ACT 2611, Australia}
\affiliation{ARC Centre of Excellence for All Sky Astrophysics in 3 Dimensions (ASTRO 3D), Canberra, ACT 2611, Australia}

\author{Yangyang Li}

\affiliation{Department of Astronomy, University of Florida, Gainesville, FL 32601, USA}

\author{Zoe Hackshaw}
\affiliation{Department of Astronomy, University of Texas at Austin, 2515 Speedway, Austin, Texas 78712-1205, USA}

\author{Timothy C. Beers}
\affiliation{Department of Physics and Astronomy, University of Notre Dame, Notre Dame, IN 46556, USA}
\affiliation{Joint Institute for Nuclear Astrophysics - Center for Evolution of the Elements, USA}

\author{Anna Frebel}
\affiliation{Department of Physics \& Kavli Institute for Astrophysics and Space Research, Massachusetts Institute of Technology, Cambridge, MA 02139, USA}
\affiliation{Joint Institute for Nuclear Astrophysics - Center for Evolution of the Elements, USA}

\author{Terese T. Hansen}
\affiliation{Department of Astronomy, Stockholm University, AlbaNova University Centre, SE-106 91 Stockholm, Sweden}

\author{Erika Holmbeck}
\affiliation{Lawrence Livermore National Laboratory, 7000 East Ave, Livermore, CA 94550, USA}

\author{Vinicius M. Placco}
\affiliation{NSF NOIRLab, Tucson, AZ 85719, USA}

\author{Charli M. Sakari}
\affiliation{San Francisco State University, 1600 Holloway Avenue, San Francisco, CA, 94132 USA}


\begin{abstract}

 We present the discovery of 2MASS\,J05241392$-$0336543 (hereafter \mystar), a very metal-poor ([Fe/H]$=-2.43 \pm 0.16$), highly \rproc-enhanced ([Eu/Fe]$=+1.34 \pm 0.10$) Milky Way halo field red giant star, with an ultra high Li abundance of {A(Li)(3D,NLTE)${=6.15 \pm 0.25}$ and [Li/Fe]${=+7.64 \pm 0.25}$}, respectively. This makes \mystar\ the most lithium-enhanced giant star discovered to date. We present a detailed analysis of the star's atmospheric stellar parameters and chemical abundance determinations. Additionally, {we detect indications of infrared excess}, as well as observe variable emission in the wings of the H$_\alpha$ absorption line across multiple epochs, indicative of a potential enhanced mass-loss event with possible outflows.
 Our analysis reveals that \mystar\ lies either between the bump and the tip of the Red Giant Branch (RGB), or on the early-Asymptotic Giant Branch (e-AGB). 
We investigate the possible sources of lithium enrichment in \mystar, including both internal and external sources.  
Based on current models and on the observational evidence we have collected, our study shows that \mystar\ may be undergoing the so-called {\it lithium flash} that is expected to occur in low-mass stars when they reach the RGB bump and/or the early-AGB. 

\end{abstract}


 \section{Introduction}\label{intro}


Observational evidence has accumulated on the decrease of the $lithium$ (Li) photospheric abundance in low-mass stars  as they age. 
Li depletion is observed to occur already on the pre-main sequence for the very low-mass stars, and along the main sequence with  mass and age-dependent efficiency that can not be explained by so-called classical stellar-evolution models \citep[e.g.,][]{1990ApJ...352..681C,1993AJ....106.1080S,2016Ap.....59..411L,2021FrASS...8...22T,2022MNRAS.513.5727B}. The observed Li patterns instead  reveal the occurrence of 
 internal transport processes of chemical elements other than convection in low mass stars. \citep[e.g.,][]{2000IAUS..198...61D,2010IAUS..268..365T}. 
 Models including atomic diffusion and turbulence are almost fully bridging 
 the gap between the primordial (BBN) Li abundance and the Li observed in metal-poor warm turnoff stars along the Spite plateau and in globular clusters \citep[e.g.,][]{Richard2005,Korn2006,Nordlander2012,Gruyters2016,Gao2020,Deal_Martins_2021}. 
Also, the Li depletion in more metal-rich stars like the Sun and F- and G-type dwarfs
in open clusters can be self-consistently reproduced when accounting for different hydrodynamical  processes that transport both matter and angular momentum in stellar interiors \citep[e.g.,][and references therein]{2005Sci...309.2189C,2021A&A...654A..46D,2021A&A...646A..48D}.    

The photospheric Li keeps decreasing during the so-called first dredge-up when low mass stars evolve towards the red giant branch \citep[RGB; e.g.,][]{Icko1967}, as evidenced both 
in metal-poor stars from the halo and globular clusters and in metal-rich stars in the field and in open clusters \citep[e.g.,][]{1999A&A...345..936L,2000astro.ph..6280C,Charbonnel_2020,Lind2009,2011A&A...527A..94C,2021A&A...651A..84M,2022A&A...657A..33A,2022A&A...661A.153M}.
Later, Li decreases again sharply as soon as the stars reach the luminosity of the bump on the upper RGB \citep[e.g.,][]{1998A&A...332..204C,Gratton2000,Lind2009,Charbonnel_2020}.  This last Li depletion episode has been suggested to be attributed to 
thermohaline mixing, which is a double diffusive process that has been suggested to connect the base of the convective envelope to the hydrogen-burning shell of low-mass red giants and expected to decrease their photospheric Li abundance and carbon isotopic ratio simultaneously  \citep{2007A&A...467L..15C,2010A&A...522A..10C}. \citet{angelou2015} showed that their models, however, predicted that the mixing would start at a higher luminosity than predicted by that data, and that it was not possible to simultaneously reproduce the evolution of carbon and lithium abundance on the RGB. 
Given the universality of the first dredge-up in low mass stellar evolution, and the sensitivity of lithium to mixing processes as evidenced for both metal-poor and metal-rich dwarf and giant stars, 
one could reasonably expect to observe low lithium abundance in all low-mass giant stars, independently of their initial metal-content. While this is generally the case, a small fraction of low-mass red giants exhibit exceptionally high Li abundances. 

The enhancement of lithium in red giant stars is a rare and continuously not understood phenomenon since its discovery by \citet{Wallerstein1982}. Lithium being easy to observe in cool stars, many spectroscopic surveys have looked for Li-rich stars  \citep{BROWN1989,Jasniewicz1999,Charbonnel2000,Monaco2011,2012MNRAS.427...11L,MARTELL2013,2014ApJ...785...94L,2015A&A...577A..10B,2018A&A...615A..74Z,2018A&A...617A...4S,Desilva2015,2019ApJS..245...33G,Charbonnel_2020,2020MNRAS.494.1348D,martell2020galah,2021A&A...655A..23M}. Depending on the adopted criteria for a star to be classified as Li-rich (see discussions in \citealt{Charbonnel_2020} and \citealt{2022ApJ...933...58C}),  
all surveys show that roughly ${1\%}$ of red giant stars have enhanced photospheric lithium abundances with respect to their Li-depleted counterparts. 
Of the  
lithium-rich giants discovered in the last three decades \citep[e.g.,][]{casey2019,gao2019,martell2020galah,cai2023}, 
only about {3\%} 
have been observed to be super Li-rich, with an abundance of A(Li)\footnote{A(Li)$=\log_{10} (N_{Li}/N_{H}) + 12$} $>3.3$\CC{~dex, i.e., higher than the proto-solar Li abundance}. Accounting for non-local
thermodynamic equilibrium (NLTE) for Li abundances, only a handful of stars have been found to have abundance of A(Li) $>4$ \citep{MARTELL2013,CASEY,Yan2018,BALACH,2015A&A...574A..31S,2019MNRAS.482.3822S,susmitha2024}. 

\CC{Two general ideas have been explored to account} 
for the excess amounts of lithium present in this minority of stars, 
\CC{namely, the production of fresh Li inside the stars themselves, or the accretion of Li-rich material from an external source.} 
\CC{Internal Li production in red giants requires a fast transport process 
in the radiative layers between the hydrogen-burning shell where the unstable $^7$Be can be produced through the pp-chains and the base of the convective envelope where Li can survive\footnote{This is the so-called \citet{Cameron1971} process which is expected to occur in the convective envelope of Li-rich stars on the Asymptotic Giant Branch (AGB; \citealt{1992ApJ...392L..71S,1997A&AS..123..241F}).} \citep[][]{1999ApJ...510..217S,Charbonnel2000,Denissenkov,Palacios2001,Denissenkov2004,2014ApJ...784L..16S,Yan2018,Casey_2019,mori2021}.} 
A number of plausible external mechanisms have also been suggested, including planetary engulfment or novae debris contaminating the outer layers of these giants \citep{1967Obs....87..238A,Siess1999,Andrievsky1999,2016ApJ...829..127A,CASEY,mallik2022}.
\CC{The occurrence and the efficiency of these different mechanisms are expected to depend on the evolution stage of the Li-rich stars.}  Recently, \citet{deepak2021,yan2021,2022ApJ...933...58C} showed using asteroseismology that the majority of the Li-rich giant stars are in their core helium burning phase (Red Clump), while a smaller fraction of Li rich stars have been found to be high on the Red Giant branch (Red Bump), including the extremely high Li-rich star TYC 429-2097-1 \citep{Yan2018}. It is thus yet to be established whether Li-enrichment is strictly attributed to one or more evolutionary status of stars, to be evidenced by the influx of more asteroseismic data for the newly discovered Li-rich star. 

It is within this context that we present our discovery of a unique ultra-enhanced lithium giant metal-poor Milky Way halo star, 2MASS~J05241392$-$0336543 (hereafter referred to as \mystar), discovered serendipitously as part of the $R$-Process Alliance (RPA) survey.  
Initial high-resolution spectroscopic data collected for this star suggested a lithium abundance far greater than the Li-rich standard of A(Li) $\sim 1.5$. We then further collected higher S/N high resolution data for the star to confirm this enhancement, and several epochs of observations to monitor its radial velocity. The fundamental properties of \mystar\ are listed in Table\,\ref{tab:basic}, with detailed description explained in the sections to follow.         

This paper is outlined as follows: In Section\,\ref{sec:obs}, we describe our observations, data reduction, and radial velocity determinations for the candidate star. In Section\,\ref{sec:stell_param}, we thoroughly investigate the fundamental atmospheric stellar parameters, using different methods including (1D, LTE) as well as (NLTE) radiative transfer analyses, and use photometric measurements to establish the stellar evolutionary status of \mystar. In Section\,\ref{sec:abund}, we present the detailed chemical abundance determinations in our star, including abundances of the light, $\alpha$, and $r$-process elements, as well as comparison with previously discovered lithium-rich giants. In Section\,5, we discuss the evolutionary status of \mystar\ based on tailor-made stellar evolution models. Finally, we present our discussion of the results and conclusions in Sections\,\ref{sec:disc} and \ref{sec:conc}, respectively.

\section{Observations and Data reduction}\label{sec:obs}

\input{Basics}



\mystar\ was initially identified and vetted as an \rproc\ enhanced candidate by the RPA collaboration \citep{rpa1,rpa2,rpa3,rpa4}, from the Large Sky Area Multi-Object Fibre Spectroscopic Telescope (LAMOST) survey \citep{LAMOST}. The star was observed with the Magellan Inamori Kyocera
Echelle (MIKE) spectrograph \citep{bernstein2003} on the Magellan-Clay Telescope at Las Campanas Observatory on three separate nights: for 1200\,s on 08 March, 2018, for 7600\,s, on 26 October, 2019 as well as for 120s on 03 March, 2022. All observations were taken with the 0.7'' slit with the $2\times2$ binning setup, yielding nominal resolving powers of $R\sim35,000$ in the red ($\lambda > 5000$\,{\AA}) and $R\sim 41,000$ in the blue, respectively.

The spectra were then reduced using the latest versions of the  \texttt{Carnegie Python {\bf distribution}}\footnote{https://code.obs.carnegiescience.edu/mike} \citep{kelson2003}. Each order of each spectrum was afterwards normalized and merged into a final spectrum, covering a wavelength range of $\sim 3320$\,{\AA}-9165\,{\AA}. For our final analysis of \mystar, we use the highest S/N spectrum from October 2019, with S/N$\sim 25$ (pixel$^{-1}$) at 3950\,{\AA}, $\sim 70$ at 4550\,{\AA}, $\sim 85$ at 5200\,{\AA}, and $\sim 240$ at 6750\,{\AA}. The final spectrum was radial-velocity, v$_{\mathrm{rad}}$, shifted by cross-correlation with the \ion{Mg}{i} lines near 5100\,{\AA} from the spectrum of the benchmark metal-poor star HD\,122563, using the spectroscopic analysis tool \texttt{Spectroscopy Made Hard} \citep[SMH;][]{casey2014}. 
A heliocentric velocity (v$_{\mathrm{rad}}^{\mathrm{helio}}$) correction was then determined with the \texttt{rvcorrect} package in \texttt{IRAF}\footnote{NOIRLab IRAF is distributed by the Community Science and Data Center at NSF NOIRLab, which is managed by the Association of Universities for Research in Astronomy (AURA) under a cooperative agreement with the U.S. National Science Foundation.} \citep{tody}, with  v$_{\mathrm{rad}}^{\mathrm{helio}} = 103.10 \pm 0.90$\,\kms. This value is in good agreement with the $Gaia$ DR3 heliocentric radial velocity v$_{\mathrm{rad}}^{Gaia} = 102.79 \pm 1.35$ \,\kms. We also determine v$_{\mathrm{rad}}^{\mathrm{helio}} = 102.10 \pm 1.00$\,\kms\ from the 2022 data and v$_{\mathrm{rad}}^{\mathrm{helio}} = 101.40 \pm 0.90$\,\kms\ from the 2018 data. Based on the present data there is no evidence to support that \mystar\ is in a binary system.

%
%


\section{Stellar Properties \& Fundamental Atmospheric Parameters}\label{sec:stell_param}

\begin{figure*}[ht!]
\begin{center}
\includegraphics[scale=0.55]{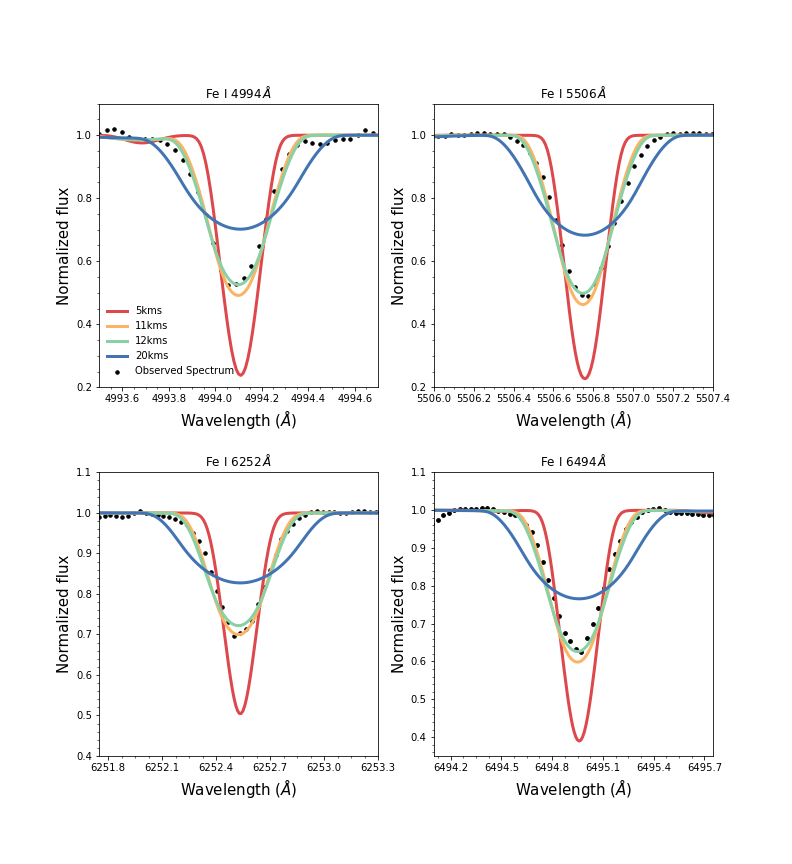}
\caption{\label{fig:rot} Synthetic spectral line profiles shown for four \fei\ lines in \mystar, computed with different stellar projected rotational velocities $v \sin i$, ranging from 5-20\,\kms\ (solid colored lines) shown against the observed spectra (dotted black points).}
\end{center}
\end{figure*}
Due to the significant Li enhancement of our star ({(Li)(3D,NLTE)}${=6.15 \pm 0.20}$; see Section\,\ref{sec:abund}), we conduct a thorough investigation of its stellar parameters, in an attempt to accurately identify its stellar evolutionary status.
We, thus, investigate the fundamental atmospheric stellar parameters (namely, the effective temperature \teff, surface gravity \logg, metallicity [Fe/H], and microturbulent velocity \vt) of \mystar\ using spectroscopic LTE and NLTE radiative transfer models, as well as with non-spectroscopic methods implementing photometry and fundamental equations. 
Additionally, we investigate its stellar properties relevant to the current study, including its luminosity, radius, projected rotational velocity $v_{\sin i}$, infrared excess, and H$_{\alpha}$ emission. Below, we describe the methods and report on each of our derived properties and parameters separately. 

\subsection{LTE Stellar Parameters}\label{sec:lte}
We determine the fundamental  atmospheric stellar parameters of \mystar\ under the assumption of LTE using abundances of 178 \fei\ and 17 \feii\ lines, and employing the 2017 version of the LTE radiative transfer code \texttt{MOOG} \citep{sneden1973}, which includes a Rayleigh scattering treatment following \citet{sobeck2011}\footnote{https://github.com/alexji/moog17scat}. The \fei\ and \feii\ linelist was adopted from \citet{Roederer_2018}, with $\log gf$ values compiled from several sources (see their Table 2 and references therein).
Abundances were computed using 1D, LTE, $\alpha$-enhanced stellar atmospheric models from \citet{castelli2004}, including standard $\alpha$-element enhancement of [$\alpha$/Fe] = +0.4. The abundance of \fei\ and \feii\ were determined using the equivalent width (EW) curve of growth method. 
The EW measurements were done by fitting Gaussian line profiles to the absorption lines using \texttt{SMH}. 
\teff\ was determined by establishing excitation equilibrium of the \fei\ abundance lines as a function of excitation potential, $\chi$. The \logg\ was determined by establishing an ionization equilibrium between abundances derived from the \fei\ and \feii\ lines. The \vt\ was estimated by requiring no trend between the abundances derived from the \fei\ lines and the reduced equivalent widths ($\log(EW/\lambda)$). \feh\ was determined from the average of the \fei\ and \feii\ line abundances. 
The derived LTE stellar parameters are: \teff\ = $4300 \pm 150$\,K, \logg\ $=0.02 \pm 0.3$\,, \vt\ = $3.14 \pm 0.2$\,\kms, \feh\ $= -2.57 \pm 0.14$\,. We estimate parameter uncertainties for \teff, \logg, and \vt\ assuming systematic uncertainties
following the analysis in \citet{ji2016b}. The [Fe/H] uncertainties were determined from the standard deviations of the \fei\ and \feii\ abundances.





\subsection{NLTE Stellar Parameters}\label{sec:nlte}
Atmospheric stellar-parameter determinations for metal-poor stars from LTE spectroscopic methods are affected by unaccounted-for departures from statistical equilibrium that can introduce significant systematic uncertainties, since line formation and populations of non-dominant species (in this case \fei) can potentially deviate from the Saha–Boltzmann equilibrium assumed in LTE \citep{lind2012,amarsi2016,ezzeddine2017}. To account for such departures, it is necessary to investigate the formation of iron lines (and thus stellar parameters) in NLTE.

Therefore, we also determine stellar parameters for \mystar\ using 1D, NLTE radiative transfer models.
The NLTE abundances were computed for \fei\ and \feii\ lines from their EWs using the radiative transfer code \texttt{MULTI} in its 2.3 version \citep{Carlsson1986,carlsson1992}, and employing 1D \texttt{MARCS} model atmospheres \citep{gustafsson1975,gustafsson2008} interpolated to the corresponding parameters. Blanketing from background opacities, excluding Fe lines, was employed from the \texttt{MARCS} opacity package (B. Plez, private communication).

The \fei/\feii\ atomic model used in the NLTE calculations is described in \citet{Ezzeddine2016b,ezzeddine2017}. This model was built by adopting up-to-date atomic data, taking into account inelastic collisions with neutral hydrogen rates for excitation and charge-exchange processes as implemented from \citet{barklem2018}. These collisions play an important dominant role (over electrons) in NLTE calculations of cool stars. 

The NLTE stellar parameters were derived using the 1D, NLTE atmospheric stellar parameters optimization tool \texttt{LOTUS}\footnote{https://github.com/Li-Yangyang/LOTUS} \citep{lotus2023}. The tool utilizes the same optimization conditions described in Section\,\ref{sec:lte} to derive the parameters (i.e., excitation and ionization equilibrium), and employs a global curve of growth method to take into account the inter-dependence of the EW of each \fei\ and \feii\ line on the corresponding atmospheric stellar parameters. Additionally, error bars were constrained using a Markov Chain Monte Carlo (MCMC) algorithm. The NLTE parameters are \teff=${4540 \pm 150}$\,K, \logg=${1.09 \pm 0.26}$, [Fe/H]=$-2.54 \pm 0.17$\, and \vt=$2.37 \pm 0.08$\,\kms. 

\subsection{Fundamental Stellar Parameters}\label{sec:gaia}

{We derive a luminosity of ${L=371 \pm 90}$\,${L_{\odot}}$, with ${\log(L/L_{\odot})=2.57 \pm 0.11}$.  The luminosity was calculated using the fundamental equations,

\begin{equation}
    {-2.5 \log\Big(\frac{L}{L_{\odot}}\Big) = M_{V} - M_{V,\odot}}
\end{equation}

\begin{equation}
    {M_V = m_v + 5 - 5\log d - A_V} 
\end{equation}


where $M_V$ and $m_V$ are the absolute and apparent magnitudes, respectively, $M_{V,\odot} = 4.83$ the Solar absolute magnitude, $d$ the distance from $Gaia$ DR3 and $A_v \approx R_V * E(B-V)$ the extinction. 
The apparent visual magnitude ($m_{V}=13.894$) is from $LAMOST$ (compares well to the $m_{V}= 13.904$ which was derived the from $Gaia$ EDR3 $m_{G}= 13.348$), and the distance, $d=9065.45 \pm 1228.1$\,pc was adopted from \citet{bailer-jones2021} derived from the $Gaia$ EDR3 parallax. The extinction value, $E(B-V) = 0.229$, is from \citet{Fink}.  {With the luminosity determined, the radius ${R = 32 \pm 5}$\,R$_{\odot}$ is then derived using the equation:}

\begin{equation}
    {R=\sqrt{\frac{L}{4\pi\sigma\ T^{4}}}}
\end{equation}

assuming a blackbody and utilizing the NLTE temperature. We adopt a mass $M = 0.8$\,M$_{\odot}$, typical of evolved giant stars at the metallicity of our star.}



Interestingly, \mystar\ was also recently flagged as a variable star in the $Gaia$ DR3, with a $G$ magnitude variability difference between \texttt{max\_mag\_g\_fov} and \texttt{min\_mag\_g\_fov} of 0.153, with a median $G$ magnitude of 13.343, and a period of 59.9 days. We note that \mystar\ is also listed as variable in the ASAS-SN database \citep{shappee2014}, with $V\textsubscript{max}-V\textsubscript{min}=0.28$.

\subsection{Adopted Atmospheric Fundamental Parameters}
The derived fundamental atmospheric parameters of \mystar\ using the different methods outlined in Sections \,\ref{sec:lte} - \ref{sec:gaia} are listed in Table \ref{tab:stell_param}.  The derived NLTE \teff\ are $\sim 240$\,K higher than the LTE \teff, and the \logg\ in NLTE is $\sim 1.0$\,dex higher than in LTE. Given that NLTE models are more realistic than LTE, we therefore adopt for our final parameters the NLTE values for \mystar, with \teff = $4540 \pm 150$ K, \logg = $1.09 \pm 0.26$, \vt = $2.37 \pm 0.08$ \kms, and \feh = $-2.54 \pm 0.17$ . We will use these parameters throughout the rest of the paper.  



\input{StellarParameters}

\begin{figure*}[ht!]
\begin{center}
\includegraphics[scale=0.55]{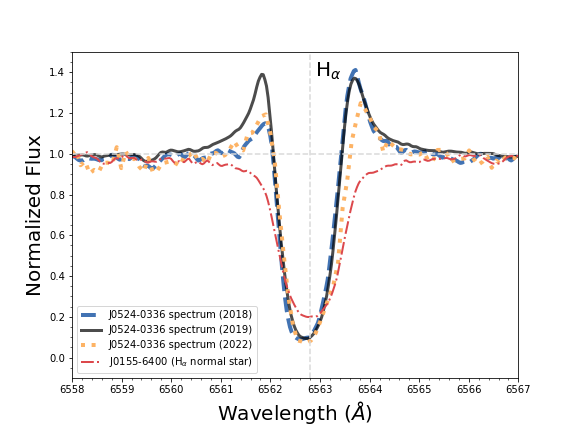}
\caption{{\label{fig:halpha} H$_{\alpha}$ line profile observed at different epochs in 2018, 2019 and 2022, respectively, in \mystar. Also shown, for comparison, is the H$_{\alpha}$ profile for the H$_{\alpha}$-normal metal-poor star J0155$-$6400 with similar stellar parameters as \mystar.}}
\end{center}
\end{figure*}

\subsection{Projected Rotational Velocity ($v_{\sin i}$)}\label{sec:rot}
Rotational velocity has been linked to both external and internal lithium enhancement \citep{Carlberg_2012,Charbonnel_2020}. A rapidly rotating star is defined as having a projected rotational velocity $v\sin\,i> 5$ \kms\,  \citep{tayar2015A}. We determine the projected rotational velocity, $v_{\sin i}$, of \mystar\ following two  methods: (i) using the Full Width Half Maximum (FWHM) of several \fei\ lines around 6400\,\AA\ region following the method outlined in \citet{bruntt}, as well as (ii) fitting the \fei\ lines using synthetic spectra computed with different $v_{\sin i}$ values. {We note that for both methods we fix the stellar parameters to those derived and adopted in Section\,\ref{sec:nlte}.}

We used six isolated iron lines ranging from 6400-6500\,\AA\ (S/N $\sim$ 240), and corrected their FWHM for instrumental broadening using the correction values from \citet{bruntt}. The lines were chosen as they were strong, unblended, had a high signal to noise ratio, following \citet{bruntt}. This yielded the intrinsic total broadening of the spectrum, which is contributed by both rotation and macroturbulence. In order to solve for the rotational velocity, we first estimated our star's macroturbulence ($v_{\mathrm{macro}} = 4.96 \pm 0.45$\,\kms) using the equation for luminosity class III stars from \citet{Hekker_2007}, and implementing the stellar parameters of our star. We determine $v_{\sin i} = 10.7 \pm 1.8$\,\kms. 
We also independently fit synthetic line spectra computed with different $v_{\sin i}$ values to several \fei\ lines in the 4000-6500\,{\AA} region of \mystar, as shown in Figure \ref{fig:rot}. To account for possible broadening sources, we take into account microturbulent velocity broadening, radiative (Doppler) and inelastic hydrogen collisional broadening (Van der Waals), as well as instrumental broadening based on the MIKE resolution (determined by convolving a Gaussian profile with each of the \fei\ lines). 
Based on the fits, we estimate a projected rotational velocity, $v_{\sin i} \sim 11 \pm 2$\,\kms, which is consistent with our FWHM calculations following \citet{bruntt}. This classifies \mystar\ as a rapidly rotating red giant metal-poor star with a projected rotational velocity $> 10$ \kms.

\subsection{H$_{\alpha}$ Emission \& IR Excess}\label{sec:ir-halpha}
Lithium enhancement in stars has been suggested to be linked to stellar properties such as infrared (IR) excess and H$_{\alpha}$ emission, as proposed in some studies such as \citet{1998AJ....116.2466F}, \citet{rebull2015}, and \citet{mallik2022}. Both these properties, if observed in stars, are indicators of mass-loss events, which have been connected to multiple channels of lithium enhancement \citep{de_la_Reza__1996,1997ApJ...482L..77D}. We therefore investigate both IR excess and H$_{\alpha}$ emission in \mystar. Figure\,\ref{fig:halpha} shows the H$_{\alpha}$ line profile in \mystar, observed at three different epochs, in 2018, 2019, and 2022, respectively. Also shown, for comparison, is the H$_{\alpha}$ profile for the metal-poor star, J0155$-$6400, with similar stellar parameters as \mystar\ and no Li detection. {We observe emission in the H$_{\alpha}$ wings in all observation epochs of \mystar.} Interestingly, however, each of these emission profiles are different for each epoch, signifying strong and variable activity, possibly due to mass-loss events or the presence of a circumstellar disk around the star. Additionally, we note the assymetric nature of the emission around the wings, signifying possible outflows. Consequently, we also looked into the infrared photometry for \mystar\ from the  WISE\footnote{https://wise2.ipac.caltech.edu/docs/release/allsky/} all-sky data colors \citep{WISE}. The IR- excess or enhancement in stars typically has WISE bands difference of $W1-W4>0.5$ \citep{Yan2018, martell2020galah}. {We note the WISE flags for our star: the contamination and confusion flag \texttt{cc\_flags}, and the photometric quality flag \texttt{ph\_qual} are set to \texttt{0000} and AAAC respectively, implying a non-contaminated WISE detection with S/N ratio of 2-3. 
WISE reports $W1 = 10.402 \pm 0.023$ and $W4=
8.773 \pm 0.362$ for \mystar, which yields $W1-W4=1.629$, indicating a possible strong IR excess, which further points to stellar activity and possible mass-loss events in \mystar. We warrant though that the W4 magnitude reported in WISE has been flagged as being less certain than the W1 color, with 4\% error bar reported for W4 versus 0.2\% for W1. We, therefore suggest that a detailed SED fitting and more precise IR colors are needed to determine whether IR excess is confirmed in \mystar.} Notably, IR excesses have been detected in 1\% of all giant stars, as compared to the 7\% lithium-enhanced giant stars discovered in the Milky Way \citep{rebull2015,martell2020galah}. No clear connection has yet been made between IR excess and Li-rich stars, however, non-traditional mixing mechanisms enhancing photospheric Li abundances (such as rotation-driven mixing) might need to be invoked to explain this phenomenon.

\begin{figure*}[ht!]
\begin{center}
\includegraphics[scale=0.5]{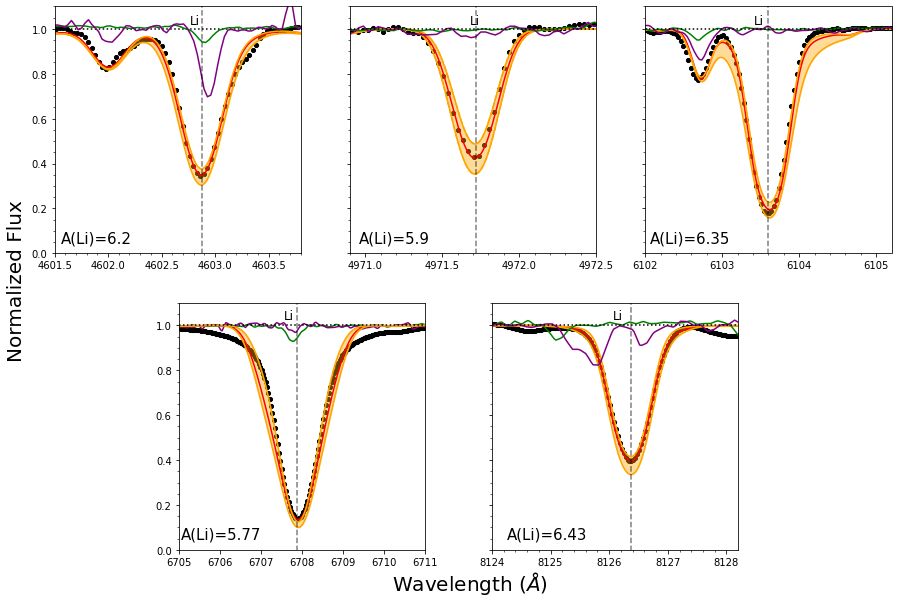}
\caption{\label{fig:Li_synth} Line profile observations, as well as 1D, LTE synthetic spectral fits for the 5 strong lithium lines detected in \mystar. The $\pm \sigma = 0.25$ abundance fits are also shown in the shaded orange. For comparison, we display the same Li lines for the two metal-poor stars at similar \feh\ as \mystar, HE\,0048$-$1109 (purple) with A(Li)= 2.34, and HE\,0324$-$0122 (green) with A(Li)= 0.78.}
\end{center}
\end{figure*}

\begin{figure*}[ht!]
\begin{center}
\hspace*{-1.6cm}
\includegraphics[scale=0.55]{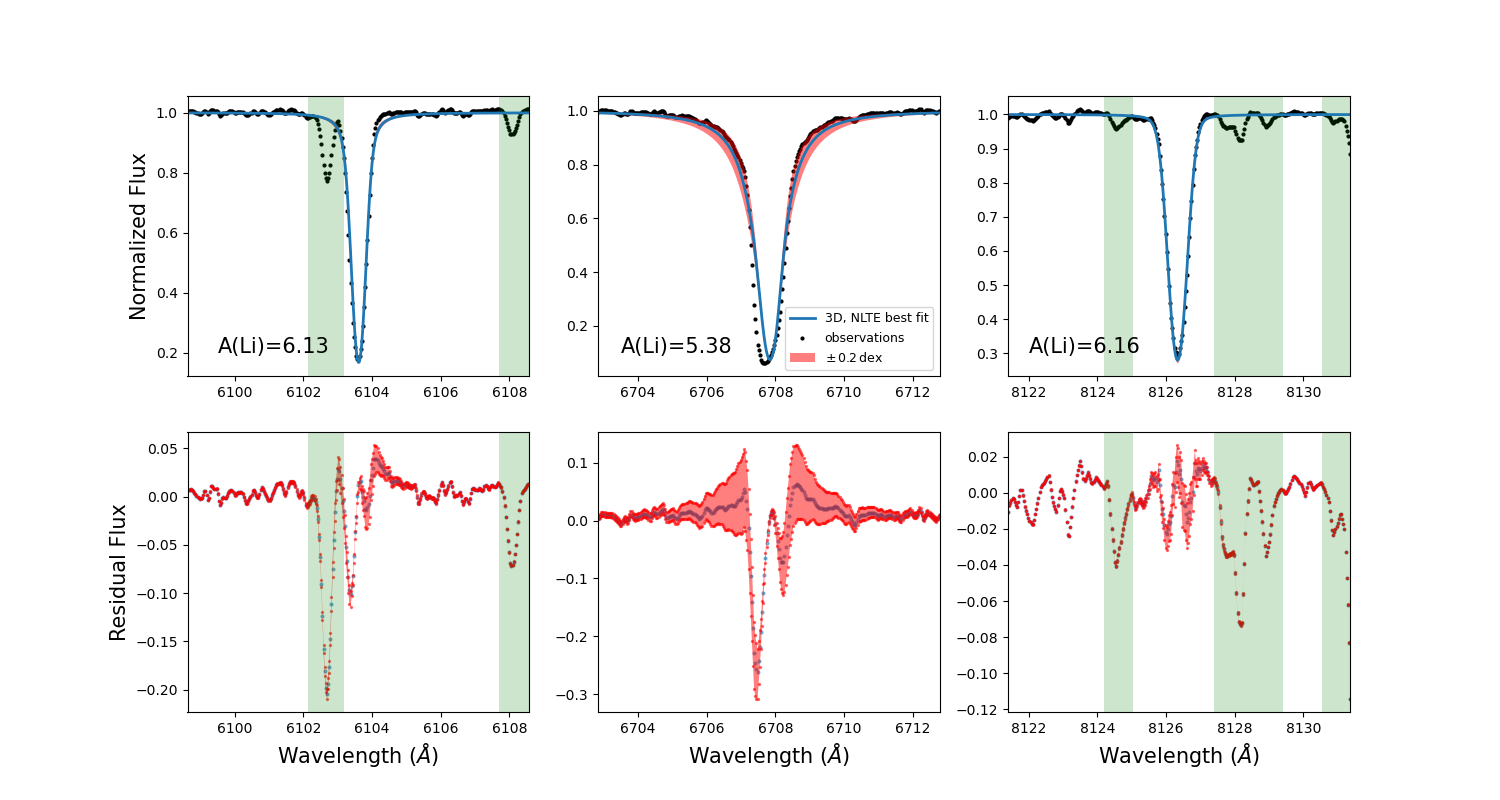}
\caption{\label{fig:Li_synth_3dnlte} Upper panel: Spectral line observations (black dots) and 3D, NLTE fits (blue lines) for the Li lines at 6103\,{\AA}, 6707\,{\AA} and 8126\,{\AA}, respectively. The red shaded area shows the spectral lines at $\pm 0.2$\,dex from the best fits. Lower panel: Residuals between the observed flux and best fit flux (blue points), as well as the $\pm 0.2$\,dex spectra from the best fit (red shaded area). The vertical green shaded areas in both panels show the masked lines excluded from the fits.}
\end{center}
\end{figure*}

\begin{figure*}[ht!]
\begin{center}
\hspace*{-1.5cm}
\includegraphics[scale=0.55]{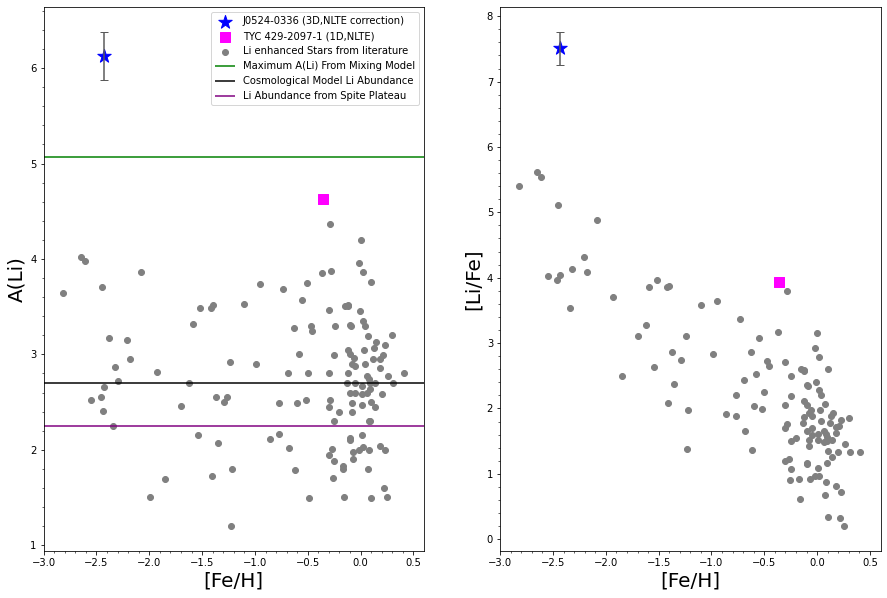}
\caption{\label{fig:Li_Plot} Absolute lithium abundance, A(Li), (left panel) and [Li/Fe] (right panel) for \mystar\ (derived using 1D, LTE and 3D, NLTE corrected values in red and blue stars, respectively) plots versus \feh, and other lithium-rich giant stars from the literature. The star with the next highest Li abundance after \mystar, TYC\,429-2097-1 \citep{Yan2018}, is also shown on the plots by the magenta square. The primordial cosmological Li abundance (black line), the Spite Plateau (magenta line) as well as the maximum Li enhancement from mixing models (green line) from \citet{Yan2018} are also shown.  References from: \citealp{LUCK,WALLER,HANNI,BROWN,GRATTON,MCWILLIAM,CARNY,JASN,SMITH,HILL,BALACH,REYN,DRAKE,CANTO,LEBRE,GONZALEZ,CARLBERG,ACALA,Andrievsky1999,KUMAR2015,RUCHTI,MONACO,KIRBY2,TWAROG,MARTELL2013,ADAMOW,LIU,MONACO3,ADAMOW2,CARLBURG2,DORAZI,JOFRE2,KIRBY3,CASEY,Yan2018}.}
\end{center}
\end{figure*}

\begin{figure*}[ht]
\begin{center}
\hspace*{-1cm}
\includegraphics[scale=0.8]{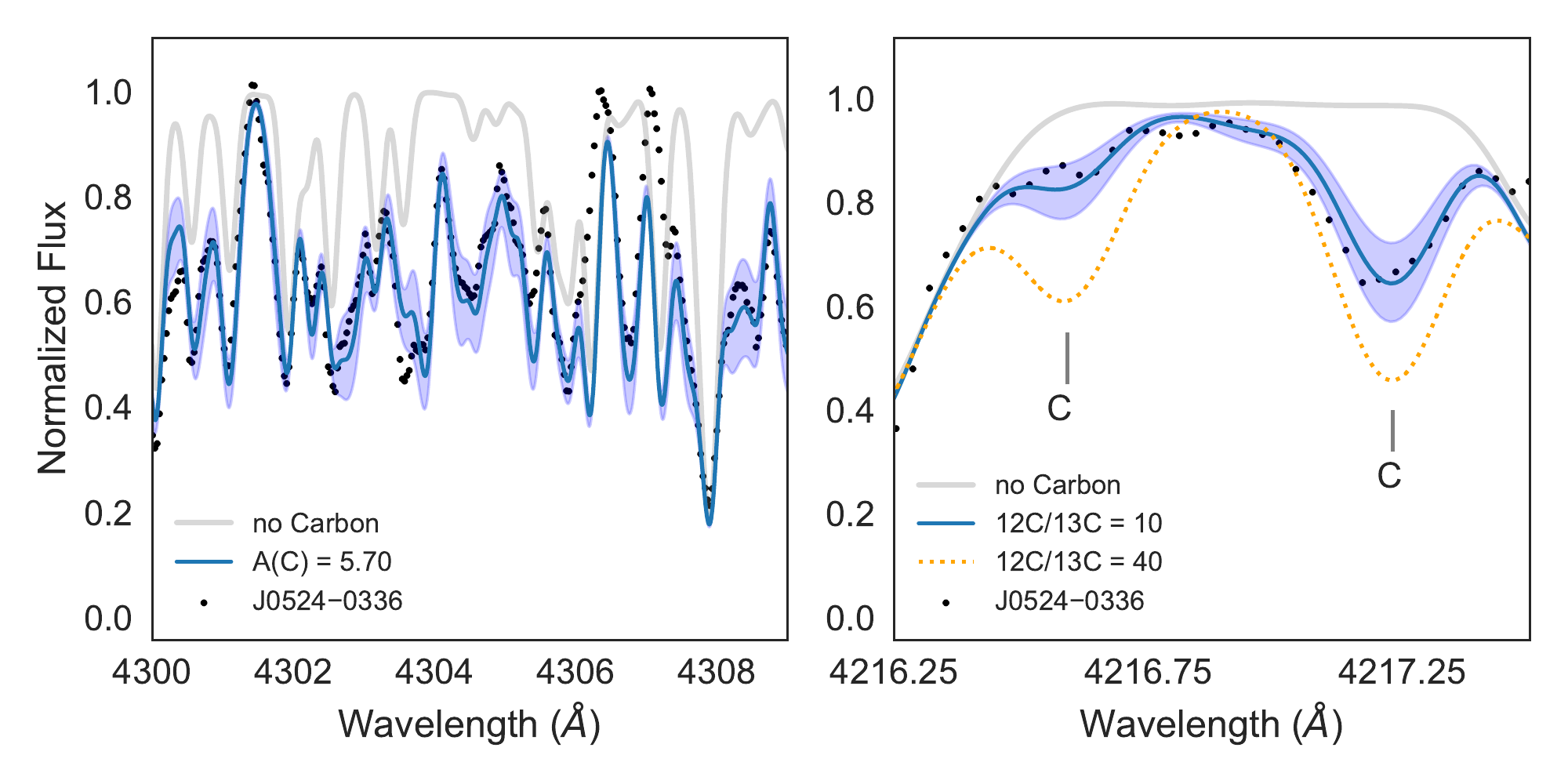}
\caption{\label{fig:C_plot}Left panel: spectral synthesis of the CH G band for \mystar. The filled black circles represent the observed high-resolution spectrum, the solid blue is the
best abundance fit, and the blue shaded area
encompasses a 0.2 dex difference in $A(C)$. The light-gray line shows the synthesized spectrum in the absence of carbon. Right panel: 
Determination of the carbon
isotopic ratio, $^{12}C/^{13}C$. The filled circles represent the observed spectrum, the solid blue line is the best fit for the $^{12}C/^{13}C = 10$, and the dashed orange line is for  $^{12}C/^{13}C=40$. The blue shaded area
encompasses a 0.2 dex difference in $A(C)$. The light-gray line shows the synthesized spectrum in the absence of carbon.}
\end{center}
\end{figure*}

\begin{figure*}[ht!]
\begin{center}
\hspace{-0.5cm}
\includegraphics[scale=0.65]{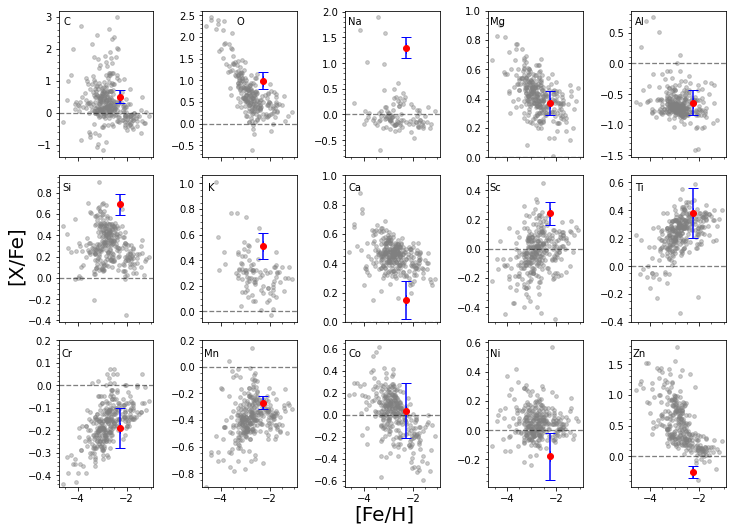}
\caption{\label{fig:eleplot} The [X/Fe] abundance ratios of C through Zn in \mystar\,, compared to Milky Way halo stars from \citet{yong2013} and \citet{roederer2014b}.} 
\end{center}
\end{figure*}
\section{Chemical Abundances}\label{sec:abund}


We derive abundance measurements, as well as upper limit estimates for the light,  $\alpha$, and Fe-peak elements for \mystar\ using the same radiative transfer models and spectroscopic tools described in Section\,\ref{sec:stell_param}. We computed the abundance ratios relative to H and Fe, adopting the solar photospheric abundances from \citet{asplund2009}. The abundances were derived using line-by-line EW and COG analysis, except for Li, C, O, Al, and the neutron-capture elements, where spectral profile synthesis were fit to each line. The linelist was adopted from \citet{roederer2018} (see their Table\,2 for atomic data references) for lines which could be detected in \mystar. Isotopic ratios were included for Li, C (see below for details), as well as for several neutron-capture elements from \citet{sneden2008}\footnote{https://github.com/vmplacco/linemake} and \citet{2021RNAAS...5...92P}. Hyper-fine structure (HFS) was considered for the Fe-peak elements, including Sc, V, Mn, and Co, when necessary. Upper limits were determined by matching the noise levels in the observed spectral lines with the corresponding synthetic spectral lines. Upon examining abundances from multiple lines for each element, outlying abundances {(outside of $1\sigma$)} were removed and an the average abundance and standard deviation was recorded for each element. For elements with only 2-5 lines measured, we estimated standard deviations by multiplying the range of values covered by our line abundances with the $k$-factor following \citet{keeping1962}. For
elements with one line only, we adopt an uncertainty between 0.1 and 0.3 dex, depending on the data and fit quality. The abundance averages and the corresponding standard deviations are reported in Table \ref{tab:abund}. The line-by-line abundances of each of these elements is recorded in Table \ref{tab:line_by_line}. 
We record systematic uncertainties in our
chemical abundances resulting from uncertainties in the model atmospheric parameters, by varying the stellar parameters by typical uncertainties in the positive direction (i.e., $\Delta$\teff = +150\,K, $\Delta$\logg = +0.30\,dex, $\Delta$[Fe/H] = +0.30\,dex, and $\Delta$\vt=0.2\,\kms), and recording the change in abundance. This is shown in Table \ref{tab:err}. All changes are within expected error bars.

\subsection{Lithium Abundance in \mystar}\label{sec:li_abund}
 
Lithium abundance was derived from synthesizing the Li lines at 4602\,{\AA}, 4971\,{\AA}, 6103\,{\AA}, 6707\,{\AA} and 8126\,{\AA} {(atomic data for these lines were adopted from \citealt{kramida21}). We adopted an isotopic ratio of $^{6}$Li/$^{7}$Li\,=\,0. While \citet{lind2013} measured a negligible $^{6}$Li/$^{7}$Li\,=\,0.005 in two out of four of their Li-rich metal-poor stars, \citet{Wang2022} was not able to detect any $^{6}$Li in any of the same stars using higher resolution ESPRESSO spectra ($R \approx 146,000$) and a 3D, NLTE analysis.}
As can be seen in Figure\,\ref{fig:Li_synth}, all the Li lines in this star are strikingly strong, with the line at 6707\,{\AA} being especially prominent. Li was also detected in the lines at 4273\,{\AA} and 4132\,{\AA}, however, as they were blended they were not used in the final abundance average. The five remaining lines yielded lithium abundances of A(4602\AA)=$6.20$, A(4973\AA)=$5.90$, A(6103\AA)=$6.35$, A(6707\AA)=$5.77$,  A(8126\AA)=$6.43$, respectively. The observations and corresponding synthesis of the lines are shown in Figure\,\ref{fig:Li_synth}. For comparison, we also show on this figure the same Li lines for two other Li-normal metal-poor stars with similar stellar parameters as \mystar.  The mere presence of seven detectable lithium lines is highly unusual in such a metal-poor star, as it is more common that only the line at 6707\,{\AA} can sometimes be detected, even in Li-rich giants. 
We note that the 6707\,{\AA} resonance line, however, yields a lower Li abundance than the rest of the lines by 0.5\,dex. This dispersion can be attributed to the asymmetric nature of this particular line and the strength of its wings. \citet{wang2021} investigated this (and other) Li line in details using a 3D, NLTE analysis. They indicate that as the A(Li) increases, the line formation pushes toward higher layers in the stellar photosphere with significantly lower temperatures that deeper layers which can drive the models away from hydrostatic equilibrium. 
By excluding the 6707\,{\AA} line from the Li average, \mystar\ yields an abundance of A(Li)$_\mathrm{1D,LTE}$=6.22 and [Li/Fe]$_\mathrm{1D,LTE}$=$+7.60$. 

{NLTE and 3D effects can be substantial for metal-poor giant stars, and particularly for strong Li lines, as is the case in Li-rich stars \citep{lind2013,wang2021}. The A(Li) of J0524-0336 is significantly higher than any existing 3D NLTE Li grids (e.g., \citealt{Harutyunyan2018}). Therefore, we calculated custom spectra at higher abundances using the same models as described in \citealt{wang2021} interpolated to the derived stellar parameters of \mystar. The 3D, NLTE best fit spectral lines to the observations are shown in Figure\,\ref{fig:Li_synth_3dnlte} for the Li lines 6103\,{\AA}, 6707\,{\AA}, and 8126\,{\AA}. Our models do not include computations for the 4602\,{\AA} and 4973\,{\AA} lines. We determine an abundance A(Li)$_{\mathrm{3D, NLTE}}$ of 6.13\,dex and 6.16\,dex for the 6103\,{\AA} and 8126\,{\AA} lines, respectively. On the other hand, the 6707\,{\AA} line yields a significantly lower abundance A(Li)$_{\mathrm{3D, NLTE}}$=5.38. As shown in Figure\,\ref{fig:Li_synth_3dnlte}, the broad core of this line suggests that it is so strong that it might have formed in the chromosphere, which cannot be fit with our current 3D, NLTE models as they are not accounted for in any of our model atmospheres and their implementation is beyond the current scope of the study. Given the excellent fits for the 6103\,{\AA} and 8126\,{\AA} lines and the agreement between their abundance values within $0.03$\,dex, we therefore adopt a final 3D, NLTE abundance for \mystar\ of A(Li)$_\mathrm{(3D, NLTE)}$= $6.15$\,dex from their average abundances, which renders [Li/Fe]$_\mathrm{(3D, NLTE)}$=+7.64 (adopting a solar lithium abundance A(Li)$_{\odot}$=1.05; \citealp{asplund2009}).} 

We compare the Li abundance in \mystar\ to other Li-rich star in the literature, as shown in Figure\,\ref{fig:Li_Plot}. To the best of our knowledge, both A(Li) and [Li/Fe] in \mystar\ are significantly higher than any other Li-rich star reported in the literature to date. It can be seen in Figure \ref{fig:Li_Plot} that the next highest lithium-enhanced giant, TYC\,429-2097-1 \citep{Yan2018}, has A(Li)(LTE)=$4.63$ and is at a much higher metallicity ([Fe/H]=$-0.36$) than our star. Thus, \mystar\ is the first star ever discovered to have an a lithium abundance of A(Li)$>5$ and [Li/Fe]$>+6$ at such a low metallicity. This significant enhancement relative to other Li-rich Milky Way stars suggests a novel or different method of lithium accretion or production within \mystar, \CC{or a similar process occurring on extremely short timescales}.

\input{Synthesis}

\subsubsection{Carbon and Oxygen}\label{sec:carbon}
We derive O abundance from the forbidden [O\,I] line at 6300\,{\AA}. While this line is commonly found to be weak, and often blended with a telluric feature in metal-poor stars, we find a very good fit and derive [O/Fe]=0.99. We derive the C abundance in \mystar\ by fitting the CH G-band at 4313\,{\AA} {following \citet{masseron2014}}. We estimate a carbon abundance of A(C)=5.7 and a ratio relative to metallicity of [C/Fe]=$-0.13$. As can be seen in Figure \ref{fig:eleplot}, this is not unusual compared to other low-metallicity halo stars from \citet{yong2013} and \citet{Roederer_2014}. We were unable to fit the weak NH band between 3355–3365\,{\AA}, and thus could not estimate an N abundance. 

The $^{12}$C/$^{13}$C 
isotopic ratio is a strong indicator of the extent of mixing processes in the red giant branch (RGB) stage of stellar evolution \citep{1998A&A...332..204C,2007A&A...467L..15C}. To derive this ratio, we fix the carbon abundance derived from the 4313\,{\AA} feature to A(C)=5.7, and derive $^{12}$C/$^{13}$C =$10\pm3$ from the 4217\,{\AA} line. This value suggests substantial processing of $^{12}$C into $^{13}$C  in \mystar. Carbon is usually depleted throughout the lives of giant stars. We thus determine and correct for the evolutionary depletion of C due to processing in \mystar. We determine a correction of $0.64$ dex using the online tool\footnote{http://vplacco.pythonanywhere.com/} described in \citet{placco2014}. This renders a final [C/Fe]=+0.51, which classifies \mystar\ slightly below being a Carbon Enhanced Metal-Poor (CEMP) star ([C/Fe]\,$>0.7$).

\subsubsection{Light Elements}
In addition to Li, C, and O, we also derive light ($Z<30$) element abundances of Na, Mg, Al, Si, K, Ca, Sc, Ti, Cr, Mn, Co, Ni, and Zn for \mystar\,, following our abundance analysis of light elements described in \cite{rpa3}. The derived abundances are listed in Table\,\ref{tab:abund}, and are compared to other MW field stars from \citet{yong2013} and \citet{Roederer_2014} in Figure\,\ref{fig:eleplot}.

As already mentioned above, NLTE corrections can be important for non-dominant species in the atmospheres of metal-poor giant stars. We thus investigate NLTE corrections for several elements, when available from the \texttt{INSPECT}\footnote{http://www.inspect-stars.com/} database \citep{lind2011}. We determine an average NLTE correction for
Na\,I of $-0.2$ dex from the seven Na\,I lines we detected in \mystar.
We derive a $-0.7$\,dex NLTE correction for K\,I line at 7698\,{\AA} from \citet{takeda2002}. Similarly, we interpolate NLTE corrections for our Mg\,I abundance from \citet{osorio2015}, and we find a negligible average correction of $<0.05$\,dex for \mystar. For Al\,I, we determine the abundance from the 3961\,{\AA} line, which is heavily affected by NLTE in cool metal-poor stars \citep{nordlander2017}. We estimate a NLTE correction of $\sim 0.5$\,dex based on their published grids and the stellar parameters of \mystar. Finally, we estimate a $\sim 0.1$\,dex NLTE correction for Ca\,I from \citet{mashonkina2016}. For the consistency of comparisons with literature abundances, derived in LTE, we only show our LTE abundances of \mystar\ in 
Figure\,\ref{fig:eleplot}, however, we list both LTE and NLTE values whenever relevant in Table\,\ref{tab:abund}. 

The abundances of all light elements ($Z<30$) in \mystar\ agree with the trends typically observed in metal-poor MW halo stars, except for Na, which seems to be enhanced related to the MW, with [Na/Fe](1D, LTE)$=+1.30$. Additionally, both Ni and Zn seem to be low as compared to the MW field stars. We note that no connection between the peculiarity of the abundances of these elements and the Li enhancement has, however, been established.   
Our element abundance derivations also agree well with those derived for \mystar\ in \citet{rpa3}.

\begin{figure*}[ht!]
\begin{center}
\includegraphics[scale=0.8]{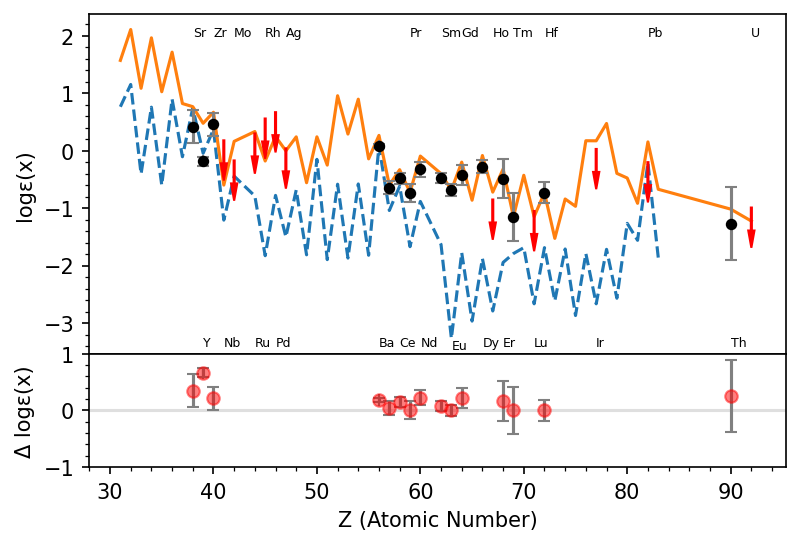}
\caption{{\label{fig:rproc} Derived neutron-capture abundances (black circles), as well as upper limits (red downward arrows) in \mystar\ overplotted against the scaled solar \rproc\ (orange solid line) and \sproc\ (blue dashed line) abundance patterns from \citet{sneden2008}. The lower panel shows the residual plot for the abundances we measured with respect to the solar \rproc\ pattern.}}
\end{center}
\end{figure*}

\subsection{Neutron-Capture Element Abundances}
We also measure abundances of neutron-capture elements in our star. \mystar\ was originally identified and classified as a highly enhanced r-II star ([Eu/Fe]$>+0.7$; \citealp{rpa4}) in \citet{rpa3}, from measurements of Sr, Ba, and Eu abundances.  We re-derived abundances of these elements, in addition of other neutron-capture elements from our high-resolution, high-S/N spectrum. The Eu abundance was derived from 12 \euii{} lines, which rendered A(Eu)$=-0.69 \pm 0.10$ and [Eu/Fe]=$+1.34\pm0.10$. This agrees with the value derived in \citet{rpa3}. We also derive [Ba/Fe]=$+0.47\pm 0.10$, with [Ba/Eu]= $-0.87 \pm 0.20$, suggesting a major contribution of neutron-capture elements in \mystar\ from an \rproc\ origin. It is worth noting that this \rproc\ enhancement is unlikely to be related to the Li enhancement in \mystar.

In addition to Eu and Ba, we measure abundances for Sr, Y, Zr, La, Ce, Pr, Nd, Sm, Gd, Dy, Er, Tm, Hf, and Th, as well as upper limits for Nb, Mo, Ru, Rh, Pd, Ag, Tb, Lu, Ir, Pb, and U. Figure\,\ref{fig:rproc} shows the neutron-capture abundance pattern for \mystar\ with the scaled solar \rproc\ pattern relative to Eu and the scaled \sproc\ pattern relative to Ba from \citet{doi:10.1146/annurev.astro.46.060407.145207}.  The abundance patterns of \mystar\ closely matches that of the scaled \rproc, except for Sr and Y from the first \rproc\ peak, which is consistent with abundances of these elements derived in other \rproc\ enhanced stars, indicating the possible onset of a secondary of weak r-process origin (e.g., \citealp{hansen2012}).

\input{Stellar_error}




\begin{figure*}[ht!]
\begin{center}
\includegraphics[scale=0.55]{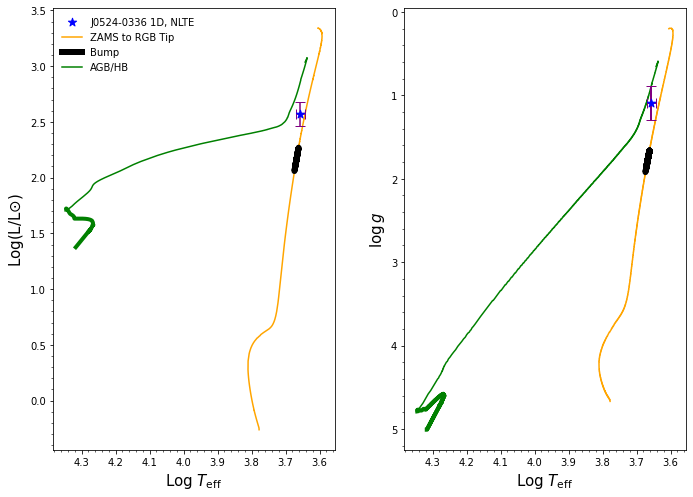}
\caption{\label{fig:evo} \mystar\ overplotted on \CC{the} stellar evolutionary track {in the HRD and the Kiel diagram (left and right panels, respectively) of a 0.8~M$_{\odot}$ model computed with the initial [Fe/H], [C/Fe], [O/Fe], and [Na/Fe] of \mystar. 
} 
Different stellar evolutionary phases are highlighted on each diagram.
\CC{The orange line goes from the zero age main sequence up to the RGB tip, and the green one from the zero age horizontal branch to the occurrence of the first thermal pulse on the AGB.} 
The RGB bump (black bold), and \CC{the location where central He-burning occurs on the horizontal branch (bold green)} 
are also shown on the plots. 
} 
\end{center}
\end{figure*}

\section{Stellar evolutionary Status of \mystar}\label{sec:evo}
To understand the ultra Li enhancement in \mystar\ (A(Li)$>5$), it is crucial to establish its evolutionary status. We compute tailored stellar evolution models using the latest version of
the stellar evolution code STAREVOL (see
\citealt[][]{2021A&A...646A..48D} for references and details on the
equation of state, opacities, and nuclear reactions).
The initial chemical composition accounts for the values of [Fe/H], [C/Fe], [O/Fe], and [Na/Fe] derived in this work (see Section\,\ref{sec:abund} and Table\,\ref{tab:abund} for values).
We use a gray atmosphere and define the stellar effective temperature
and radius at the optical depth $\tau = 2/3$.
The mass-loss rate follows   \citet{1975MSRSL...8..369R} empirical
relation (with $\eta_R = 0.5$) from the zero-age main sequence up to central
He exhaustion and switches to  \citet{1993ApJ...413..641V}'s prescription on
the asymptotic giant branch (AGB).
We adopt a mixing length parameter $\alpha_{\rm MLT} = 1.5$, and we assume the Schwarzschild criteria for convective stability.
We include the effects of the thermohaline instability as described in
\citet[][]{2007A&A...467L..15C} and \citet[][]{2012A&A...543A.108L}.
We also compute models with thermohaline mixing and solid-body rotation, to predict the evolution of the surface stellar rotation rate under conservative assumptions for such a low-mass star.

The position of \mystar\ on the HR diagram \CC{and on the Kiel diagram} is well fit by the evolutionary track of a model with initial mass of 0.8~M$_{\odot}$ (see Figure\,\ref{fig:evo}). {Given the uncertainties on the stellar parameters,  \mystar\ appears to be located either on RGB above the predicted location of the bump, or on the e-AGB before the occurrence of the thermal pulses on the AGB (TP-AGB). 
It has a much cooler temperature than the position of the
horizontal branch (HB), which is the metal-poor counterpart of the red clump (RC) for metal-rich giants. Although no asteroseismic constraint is available for \mystar, we can safely conclude that it is currently {\it{not}}  undergoing central He-burning. 

{We computed different global asteroseismic quantities ($\nu\mathrm{max}$, frequency at which the oscillation modes reach their strongest amplitudes; 
$\Delta\Pi$, asymptotic period spacing of g-modes (for $l=1$), total accoustic radius and acoustic radius at the base of the convective envelope) all along the evolution with the same prescriptions as in \citet{2012A&A...543A.108L}.  
At the two evolutionary points on the RGB and early-AGB where our model reaches the luminosity of J0524-0336 and has an effective temperature compatible with the derived values within the error bars, 
their values are very similar, and 2 to 3 orders of magnitude lower than when the model undergoes central He-burning at a very high effective temperature on the horizontal branch (see \citet{2012A&A...543A.108L} for a description of the variations of the global asteroseismic quantities along the evolution of low-mass stars from the pre-main sequence to the end of the TP-AGB).}

} 

\CC{Within the error bars on} the luminosity and effective temperature of \mystar, \CC{both on the RGB and the e-AGB, the model predicts a value of 9.6 for the
$^{12}$C/$^{13}$C ratio (to compare to the value of 42 obtained after the first dredge-up), due to the effect of the thermohaline mixing that is predicted to occur at the RGB bump.  This} is in very good agreement with the value we derived of $^{12}$C/$^{13}$C=10, \CC{confirming that \mystar\ has already undergone thermohaline mixing when it has previously crossed the RGB bump.} 
The typical rotation rate for stars at this evolution stage and metallicity is 1.2 \kms\ \citep{cortes2009}, which is much lower than that derived in this study (see Section\,\ref{sec:rot}). 

\section{Discussion}\label{sec:disc}

\CC{The evolutionary status of Li-rich giants has long been debated. \cite{Charbonnel2000} found an accumulation of such stars around the RGB bump and on the e-AGB, this later phase being in agreement with our conclusions for \mystar\ . Interestingly, the next most highly enhanced lithium star (A(Li)=4.51) is 
probably located at the RGB bump \citep{Yan2018}. Others studies, sometimes using additional constraints based on asteroseismology parameters, concluded that a large fraction of enhanced lithium giants 
are actually red clump (RC) stars presently burning He in their central convective core, the rest being either close to the RGB bump or randomly distributed along the advanced phases \citep{Kumar2011,2014ApJ...784L..16S,Casey_2019,Yan_2020,2020JApA...41...49K,Martell_2021,2021NatAs...5...86Y}.  Recent studies 
have shown that there could be different paths to Li-enrichment. For example, 
\citet{2022ApJ...933...58C}
demonstrated that lithium-rich RC giants are likely to be a younger and more massive population than
lithium-rich RGB stars. \citet{sayeed2023} performed an empirical study of Li-rich giants from the GALAH survey and showed that while Li-rich stars are prevalent on the red clump, other culprits such as binary spin-up and mass-transfer could also be likely mechanisms to enrich stars with Li at the RGB phase.  As of today, though, there is still no clear consensus, and different paths to Li-enrichment remain to be explored and confirmed observationally. 
}

Several mechanisms have been suggested in the literature to try to explain Li-richness in giant stars. These can be broadly divided into external and internal mechanisms. External mechanisms include the presence or interaction with a stellar or sub-stellar companion (such as planetary engulfment or binary interactions), while internal mechanisms include non-traditional or enhanced mixing processes. In what follows, we discuss some of these different mechanisms, linking them to the observable properties of \mystar, to try to explain its extreme lithium enhancement relative to other stars in the Milky Way.

\subsection{External Enrichment}

\subsubsection{Planetary Engulfment}
One mechanism that has been proposed for lithium enhancement in giant stars is planetary engulfment
\citep{Siess1999}. 
Planets below a certain mass threshold can dissolve in the convective envelope of a host star, which can (in some cases) lead to lithium (as well as other elemental) enhancement, and is postulated to be observed in the photospheric abundances of giant stars with $1.5 \leq$ A(Li) $ \leq 2.2$. \citet{Aguilera2016} have attempted to model lithium enrichment via \CC{the} engulfment \CC{of a planet or a brown dwarf} across a range of stellar metallicities and sub-stellar component masses, where they showed that planetary engulfment is unable to account for giant stars with A(Li) $\geq 2.2$. This has been further confirmed by a follow-up study {\citep{Aguilera_G_mez_2020}. It is unlikely that a star at such a low metallicity as \mystar\ would be able to form a planet {\citep{fischer2005}}, however, as an academic exercise, we nevertheless attempt to derive the substellar component parameters (including the mass and initial Li abundance) that would be theoretically required to account for the Li abundance in \mystar. We adopt the equations in \citet{Carlberg_2012} (see their Appendix), starting with a star of similar mass to \mystar\ (0.8\,M$_{\odot}$) to estimate the Li abundance the star would have after the substellar component engulfment event. We adopt a ratio of the mass of the planet to the mass of the stellar envelope of 0.19, with the mass of the envelope being 0.05 the mass of the star. This would result in a planet of size ~8 M$_J$, although we allowed the mass and thus the former ratio to vary from 1 to 32 M$_J$ (thus the ratio of the mass of the planet to the stellar envelope varies from 0.02 to 0.76). We show the results in Figure \ref{fig:planet}. The left panel shows the ``Final'' expected Li abundance of the star, starting with an ``initial'' Li abundance of A(Li)=2.2, after engulfing a planet with different intrinsic Li abundances, ranging from A$_p$(Li)=3.3 to 7.5, as a function of planet mass (in Jupiter masses, M$_J$).  The right panel of Figure\,\ref{fig:planet} shows the same thing, however, starting from a highly enriched ``intrinsic'' stellar Li abundance of A(Li)=4.0, for example, a star that has been previously enriched by an internal-mixing process.  
According to the the models outlined in \citet{Carlberg_2012}, regardless of the ``initial'' stellar Li abundances (A(Li)=2.2 and A(Li)=4.0), the ``final'' stellar photospheric Li abundances are expected to increase as a function of planet mass and intrinsic planetary Li content. For both cases, however, an unrealistically high intrinsic planetary Li content (A$_p$(Li)$>$7.0) is needed to reproduce the Li abundance in \mystar. Additionally, it would take a large planet  ($\ge$ 10\,M$_J$) to account for this level of Li enrichment in \mystar. Both scenarios seem unlikely, given our current understanding of intrinsic planetary lithium content.  We can thus dismiss planetary engulfment as a primary mechanism of Li enrichment in \mystar.

\begin{figure*}[ht!]
\begin{center}
\includegraphics[scale=0.55]{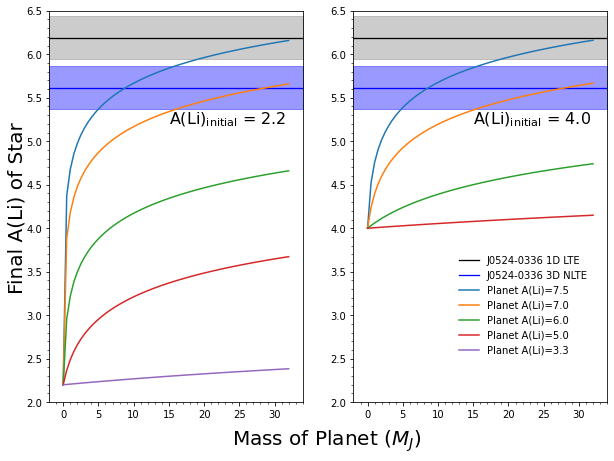}
\caption{\label{fig:planet} Final Li abundance after a theoretical planetary engulfment event, based on models from from \citet{Carlberg_2012}, starting with an initial stellar Li abundance of A(Li)=2.2 (left panel), and a super high initial abundance (possibly due to enhanced internal mixing) of A(Li)=4.0 (right panel), as a function of planet mass. It should be noted that the deuterium
burning limit is at 13 M\textsubscript{J} and thus "planets" larger than this might more realistically be considered brown dwarfs.} 
\end{center}
\end{figure*}

\subsubsection{Interaction with a Binary Stellar Companion}
The presence and interaction with a binary companion has been proposed as a scenario to explain lithium enhancement in stars. For example, \citet{Casey_2019} proposed that tidal interactions from a companion star could create internal mixing and thus drive lithium enhancement via the \citet{Cameron1971} mechanism. It has also been proposed by \citet{Zhang_2020} and \citet{mallik2022} that lithium enhancement could result from mergers between RGB stars and helium white dwarf companions, {where the surface Li abundance would get enhanced during the common envelope (CE) phase of the merger during the in-spiraling of the binary components, which would eventually lead to the formation of a Li-rich He-burning core RC star.}

We derived the radial velocity (RV) of \mystar\ (as explained in detailed in Section\,\ref{sec:obs})  from three different spectroscopic observational epochs of \mystar\ over several years, and compared it to the Gaia DR3 radial velocity. We found an excellent agreement within $<2$\,\kms, which excludes any radial velocity variation. {Additional insight on RV variation can be obtained by looking at the renormalized unit weight error (RUWE) from the $Gaia$ database, which corresponds to the reduced chi-squared of the best-fitting five-parameter single-body astrometric solution. \citet{Lindenberg2021} suggested that a RUWE value $>1.4$ could be used to signal and identify possible non-single stars. \mystar\ has a RUWE value of 1.02, which further excludes at the present any evidence for a binary companion.}

Nevertheless, H$_{\alpha}$ emission profile changes and IR-excesses have been suggested to be due to 
binary mergers that trigger mass loss events  \citep{Castellani1993,zhang2013} 
In this case, excess in near-IR colors will be detected if the material was ejected recently \citep{mallik2022}. This signifies the possibility, that while no present binary component has been detected for \mystar, it might have experienced a recent binary merger that could have invoked extra mixing, leading to enhanced rotation and mass-loss events, as evidenced by the detection of variable emission in the H$_{\alpha}$ profile, and \textbf{first order} IR-flux excess.




\subsection{Internal Mixing}


\CC{Lithium production in red giants requires a mechanism to transport $^3$He inwards from the stellar convective envelope to the deeper and hotter radiation layers where the pp-II chain produces fresh $^7$Be, and then to transport $^7$Be outwards so that its electron capture to $^7$Li occurs in cooler regions where $^7$Li can not burn (the so-called Cameron-Fowler process; \citealp{Cameron1971}). Although the general idea is well understood and several driving mixing processes have been proposed, 
the physical origin of the  transport mechanism is far from being clear.}

\CC{As initially shown by \citet{1992ApJ...392L..71S}, the amount of Li that can  potentially be produced in evolved stars strongly depends on the assumed speed and geometry 
of the driving mixing mechanism as well as on its episodicity, but it is independent of the previous $^7$Li depletion history of the stars.  \citet{Yan2018} revisited this seminal work,} 
using updated nuclear reaction rates and asymmetric parameters \CC{for the upwards and downwards mixing flows between the base of the convective envelope and the $^7$Be production layers.} 
Their \CC{fine-tuning} model 
was able to reproduce the Li abundance (A(Li)=4.51) of their star, TYC 429$-$2097-1, \CC{which is more massive and more metal-rich (M=1.43$\pm$0.55~M$_{\odot}$, [Fe/H]=-0.36$\pm$0.06) than \mystar, and which is lying at the RGB bump (this was established by the authors based on Gaia DR1 parameters, and we checked that it is still the case with Gaia DR2 and Gaia DR3}. However, their parametric model only allowed for a maximum lithium abundance of A(Li)= $5.07$, \CC{which is a factor of $\sim$~4 below that of \mystar. We speculate that this small difference could be due to the differences in mass and metallicity between the two stars, calling for the same transport process to be more efficient in more metal-poor giants. This was actually already anticipated based on the observed $^{12}$C/$^{13}$C behavior along the RGB, which reveals the role of molecular weight gradient in the development of transport processes such as the thermohaline mixing in red giants \citep[][]{1998A&A...332..204C,2007A&A...467L..15C,2012A&A...543A.108L,2019A&A...621A..24L}.}


\subsubsection{Lithium Flash}\label{sec:li_flash}

\CC{The scarcity of extremely Li-rich red giants like \mystar\ and TYC 429$-$2097-1 calls for these stars being caught during a 
very brief episode at the beginning of the Li enrichment phase, as originally proposed to occur at the RGB bump and/or during the early-AGB by \citet{Charbonnel2000}.}
\CC{The occurrence of this so-called Li-flash was first modelled around the RGB bump by \citet{Palacios2001}.}  
\CC{They suggested that rotation-induced mixing, which they were then modeling with a very simplified formalism, would drive the Cameron-Fowler process,  
leading to the formation} a very thin, short-lived lithium-burning shell \CC{in the outskirts of the hydrogen-burning shell. They showed that under certain assumptions for the mixing efficiency,  the amount of nuclear energy released in the lithium-burning shell is such that a thermal  instability can be ignited.  Convection then develops in the thermally unstable layers to carry out the energy, which allows the quasi-instantaneous transport of fresh $^7$Be towards the convective envelope. 
During this short-lived ($\sim 2 \times 10^4$ yrs) Li flash, the total stellar luminosity temporarily increases by a factor of $\sim 5$ in their model.} 
This causes a near doubling of the mass loss, which could potentially form a dust shell around the lithium rich star.
\CC{Once the thermal instability is quenched, the stellar luminosity decreases to its original bump value. The transport efficiency decreases simultaneously, and the stellar envelope is not fed any more with fresh $^7$Be. The Li abundance and the carbon isotope ratio start to decrease, under e.g., the influence of thermohaline mixing. Whether this mechanism (which efficiency is strongly debated, see e.g., \citealt[][and references therein]{2019ApJ...870L...5H}), or any other instability in rotating stars could be the original trigger of the Li-flash at the RGB bump and/or on the early-AGB, remains to be studied.} 


We see evidence of many aspects of this process in the spectrum of \mystar\ \CC{and of TYC 429$-$2097$-$1}: 
(i) the stars' evolutionary status, as defined by their stellar parameters, \CC{is compatible with the two stages} at which the Lithium Flash is expected to occur; \CC{we note in particular that the stellar luminosity of \mystar\ is a factor of $\sim 5$ brighter than the location of the RGB bump predicted by our model, and a factor of $\sim 2$ brighter than the beginning of the early-AGB; TYC 429$-$2097$-$1 is at the RGB bump;} 
(ii) the unusually high projected rotational velocity ($v_{\sin i} \sim 11$\kms\, in both \mystar\ and TYC 429$-$2097$-$1) suggests that the transport mechanism simultaneous extracts angular momentum from the stellar interior; (iii) the $^{12}C/^{13}C$ $\sim$ 10 ratio in \mystar\ \CC{(12 in TYC 429$-$2097$-$1) indicates a decrease below the post-dredge up value (possibly due to thermohaline mixing) as observed in the majority of bright low-mass red giants; it is also possible that \mystar\ has acquired its low carbon isotopic value at the RGB bump and is presently undergoing the Li flash on the early-AGB;} 
(iv) the detection of IR-excess colors, as well as the strong and variable emission in the H$_{\alpha}$ wing profile of \mystar\ indicates the presence of a dust shell which \CC{could result from enhanced mass loss during the Li flash; we note however that no IR excess was found for TYC 429$-$2097$-$1};  
(v) the uniquely ultra lithium abundance observed in \mystar\ which requires \CC{an extremely efficient transport of $^7$Be in its convective envelope; the Li abundance of TYC 429$-$2097$-$1, which is one order of magnitude below that of \mystar\,, is however still  one order of magnitude above the meteoritic value}.
It is thus possible that \CC{both \mystar\ and TYC
429$-$2097$-$1 are indeed 
 just undergoing the {\it lithium flash}.
}

\section{Summary and Conclusions}\label{sec:conc}

{We report on the discovery of \mystar, an ultra Li-rich star, with A(Li)(3D, NLTE)=${6.15}$ and [Li/Fe]=${+7.64}$, from two strong lithium absorption lines.} To our knowledge, this is the most Li-rich giant star discovered to date. Additionally, we derive abundances of 16 neutron capture elements in \mystar\ with [Eu/Fe]$=+1.34$, classifying it as a highly enhanced \rproc\ star (rII - [Eu/Fe]$>0.7$; \citealp{rpa4}).
We conduct a detailed stellar parameters analysis of our star using a high-resolution ($R\sim 35,000$), high S/N spectrum. \CC{Based on its stellar parameters and Gaia DR3 distance, we find that \mystar\ lies either on the RGB between the bump and the tip, or on the early-AGB. In any case, it is not currently undergoing core He-burning, which is predicted to occur at a much higher temperature for such a low-mass, low-metallicity star. The star should thus have a relatively low asymptotic period spacing of g-modes. } Additionally, we determine a fast projected rotational velocity, $v_{\sin i} \sim 11$\kms\,, as compared to typical values in red giant stars. Furthermore, both an IR-excess, as well as variable emission in the wings of the H$_{\alpha}$ profile were detected. 

We investigated both internal and external processes of possible lithium enhancement that might explain such an ultra-high lithium abundance in a red giant star. 
Fast rotational velocity could point toward an external source of enrichment, either due to a sub-stellar companion (e.g., planetary) engulfment or binary interaction. No variation was detected in the star's radial velocity over several observational epochs, from which no present binary companion can be inferred. 
Nevertheless, H$_{\alpha}$ emission and IR-excess has been postulated to be due to mass-loss events due to a binary merger with giant stars.  We show that a sub-stellar companion engulfment cannot produce the Li abundance observed in \mystar, and that a scenario in which this would be possible would require a high planetary mass ($M_p \ge 10$M$_J$) and an unrealistically high intrinsic planetary Li content (A$_p$(Li)$>7$). On the other hand, it could be possible that the ultra high Li in \mystar\ could be due to a previous merger with a binary star triggering and Li-producing extra mixing in the star. Additional studies on the exact amount of Li that could be produced during these interactions are, however, needed to confirm or refute such an event in \mystar.

We \CC{also investigate possible mixing mechanisms that could drive the so-called Cameron-Fowler process in the radiative layers between the H-burning shell and the convective envelope of \mystar, and account for its extreme Li abundance.} 
Recent \CC{parametric models, taking into account asymmetric upwards and downwards mixing flows} 
and updated nuclear reaction rates, 
\CC{could reach} A(Li)=5.07 \CC{for the case of a more massive and more metal-rich red giant sitting at the RGB bump. Similar computations should be done for the case of a low-metallicity, low-mass star such as \mystar, to investigate the mass and metallicity dependence of the driving mixing mechanism.}  We also considered the possibility of Li production in \mystar\ during the {\it lithium flash} as proposed by \citet{Palacios2001}, \CC{and which is expected to occur at the RGB bump and/or on the early-AGB \citep{Charbonnel2000}}. 
\mystar's evolutionary status, 
fast rotational velocity of $v_{\sin i} \sim 11$\,\kms, IR excess, H$_{\alpha}$ emission, and unusually high lithium content are predicted indicators of the lithium flash as described in \citet{Palacios2001}. It is therefore possible that \mystar\ was observed during this rare phenomenon, \CC{as well as TYC 429-2097-1, which is the next highest A(Li) red giant star, and which is sitting at the bump.} However, additional models and upper limits on the production of Li during the Li flash (preferably in 3D),  
are needed to confirm this scenario. 

\mystar\ sets a new benchmark for lithium-rich metal-poor stars, being the first giant discovered with an A(Li)$>6.0$ and [Li/Fe]$>+7$. 
It provides a great opportunity to investigate the origin and evolution of Li in the Galaxy, and could be the first of a new population of ultra-rich lithium-enhanced stars expected to be discovered in current and future high-resolution spectroscopic surveys such as the $R$-Process Alliance, as well as others. Asteroseismologic follow-up of  metal-poor Li-rich targets will be key to pin down the exact evolutionary status of our and similar stars. 
\CC{However, distinguishing whether \mystar\ is climbing the RGB or the early-AGB would be extremely challenging, since the global asteroseismic parameters should not differ significantly between these two evolution phases. Finally, we hope that the discovery of \mystar\ will open a new avenue to understand the instabilities that can develop and transport matter and angular momentum in red giant stars.}



\begin{acknowledgments}
We thank Laurent Eyer, Peter Hoeflich,  Pavel Dennisenkov and Jamie Tayar for helpful discussions.
J.K. thanks the University of Florida and the Center for Undergraduate Research and the University Scholars Program for funding his research. R.E. acknowledges support from NSF grant AST-2206263.
\CC{CC acknowledges support by the Swiss National Science Foundation (Project 200020-192039) and thanks N.Lagarde, A.Palacios, and N.Prantzos for useful discussions.}  I.U.R. acknowledges support from the NSF, grants AST~1815403/1815767 and 2205847, and the NASA Astrophysics Data Analysis Program, grant 80NSSC21K0627. A.F. acknowledges support from NSF grant AST-1716251. We acknowledge support from JINA-CEE (Joint Institute for Nuclear Astrophysics - Center for the Evolution of the Elements), funded by the NSF under Grant No. PHY-1430152. The work of V.M.P. is supported by NOIRLab, which is managed by the
Association of Universities for Research in Astronomy (AURA) under a cooperative agreement with the U.S. National Science Foundation. TTH acknowledges support from the Swedish Research Council (VR 2021-05556). This work was supported by computational resources provided by the Australian Government through the National Computational Infrastructure (NCI) under the National Computational Merit Allocation Scheme and the ANU Merit Allocation Scheme (project y89) and HPC-AI Talent Programme Scholarship (project hl99).
\end{acknowledgments}

%
%
This work made use of NASA's Astrophysics Data System Bibliographic Services, and the SIMBAD database, operated at CDS, Strasbourg, France \citep{simbad}. This work has made use of data from the European Space Agency (ESA) mission
{\it Gaia} (\url{https://www.cosmos.esa.int/gaia}), processed by the {\it Gaia}
Data Processing and Analysis Consortium (DPAC,
\url{https://www.cosmos.esa.int/web/gaia/dpac/consortium}). 

\facilities{Magellan-Clay (MIKE, \citealt{bernstein2003})}

\software{IRAF~\citep{iraf,Fitzpatrick2024}, matplotlib~\citep{matplotlib}, MOOG~\citep{sneden1973,sobeck2011}}, LOTUS \citep{lotus}
\clearpage

\bibliography{ref}
\newpage

\input{lbl4.tex}
\end{document}

%% file: Basics.tex
\begin{deluxetable}{c c}
\tablewidth{0pt}
\tabletypesize{\footnotesize}
\tablecaption{\label{tab:basic} Properties of \mystar
}
\tablehead{Label & Value}
\startdata
2MASS star ID & 2MASS~J05241392$-$0336543  \\
$Gaia$ DR3 ID & 3210839729979320064  \\
R.A. (J2000) & 05:24:13.900 \\
Decl. (J2000) & $-$03:37:00.300 \\
$G_{\mathrm{mag}}$ ($Gaia$ DR3) & 13.343 \\
$V_{\mathrm{mag}}$  ($Gaia$ DR3)\tablenotemark{a} &  13.904 \\
$V_{\mathrm{mag}}$  (LAMOST) &  13.894 \\
v$_{\mathrm{rad}}^{\mathrm{helio}}$(This work)(\kms) & $103.10 \pm 0.80$    \\
v$_{\mathrm{rad}}^{\mathrm{helio}}$($Gaia$ DR3)(\kms) & $102.79 \pm 1.35$  \\
\teff\ (K) & $4540 \pm \mathbf{150}$  \\
\logg\  & $1.09 \pm \mathbf{0.26}$  \\
\vt\ (\kms) & $2.37 \pm 0.08$  \\
\feh\  & $-2.54 \pm 0.17$ \\
$v_{\sin i}$ (\kms) & $11 \pm 2$\,\\
$\log(L/L_{\odot})$ & ${2.57 \pm 0.11}$\\
$R$\,($R_{\odot}$) & ${32 \pm 5}$\,\\
\enddata
\tablenotetext{a}{V magnitude determined from
the $Gaia$ DR3 $G$ magnitudes and $BP-RP$ color conversions.} 
\end{deluxetable}

%% file: StellarParameters.tex
\begin{deluxetable}{c c r r}
\tablewidth{0pt}
\tabletypesize{\footnotesize}
\tablecaption{\label{tab:stell_param} Stellar Atmospheric Parameters of \mystar }
\tablehead{\colhead{Parameter} & \colhead{LTE}  & \colhead{NLTE}  & \colhead{Adopted}}
\startdata
\teff\ (K) & $4300 \pm 150$      &   ${4540 \pm 150}$ & ${4540 \pm 150}$
\\
\logg  & $0.02 \pm 0.30$  &  $1.09 \pm {0.26}$ &  $1.09 \pm {0.26}$
\\ 
\vt\ (\kms) & $3.14 \pm 0.20$  & $2.37 \pm 0.08$ & $2.37 \pm 0.08$
\\
\feh\ & $-2.57 \pm 0.14$    & $-2.54 \pm 0.17$ & $-2.54 \pm 0.17$
\\   
\enddata
\end{deluxetable}

%% file: Synthesis.tex
\begin{deluxetable*}{l c c r r r c}
\tablewidth{0pt}
\tabletypesize{\footnotesize}
\tablecaption{\label{tab:abund} Element Abundances in \mystar}
\tablehead{
\colhead{El.} & \colhead{N} & \colhead{$\log \epsilon$ (X)$_{\odot}$} & \colhead{log $\epsilon$(X)} & \colhead{[X/H]} & \colhead{[X/Fe]} & \colhead{$\sigma$[X/H]}}
\startdata
Li(1D,LTE)  & 4   & 1.05 & 6.22  & +5.17  & +7.60   & 0.25         \\
Li(3D,NLTE)  & 2   & 1.05 & 6.15  & +5.10   & +7.64   & 0.20         \\
CH   & 1   &8.43 & 5.70    & $-$2.73   & $-0.13$   & 0.20   \\  
CH(corr.)\tablenotemark{a}  & 1   &8.43 & 6.34    & $-2.09$   & +0.51   & 0.20 \\
O I       & 1   & 8.69 & 7.41       & $-$1.28   & +0.99   & 0.20      \\
Na I(1D,LTE) & 1 & 6.24 & 5.26 & $-0.98$ & $+1.30$  & 0.20 \\
Na I(1D,NLTE) & 1 & 6.24 & 5.06 & $-1.18$ & $+1.10$  & 0.20 \\
Mg I(1D,LTE)      & 7   & 7.60 & 5.70       & $-$1.90   & +0.37   & 0.19          \\
Mg I(1D,NLTE)      & 7   & 7.60 & 5.75       & $-$1.85   & +0.42   & 0.19          \\
Al I(1D,LTE)      & 1   & 6.45 & 3.53        & $-$2.92   & $-$0.64   & 0.20   \\
Al I(1D,NLTE)      & 1   & 6.45 & 4.03        & $-$2.42   & $-$0.14   & 0.20   \\
Si I      & 2   & 7.51 & 5.93      & $-$1.58   & +0.69   & 0.17             \\
K I(1D,LTE)       & 2   & 5.03 & 3.27    & $-$1.76   & +0.51   & 0.06             \\
K I(1D,NLTE)       & 2   & 5.03 & 2.57    & $-$2.46   & $-$0.21   & 0.06             \\
Ca I(1D,LTE)      & 21  & 6.34 & 4.21       & $-$2.13   & +0.15   & 0.21            \\
Ca I(1D,NLTE)      & 21  & 6.34 & 4.31       & $-$2.03   & +0.25   & 0.21            \\
Sc II     & 11  & 3.15 & 1.11        & $-$2.04   & +0.24   & 0.28           \\
Ti II     & 43  & 4.95 & 3.05       & $-$1.90   & +0.38   & 0.35            \\
Cr I      & 16  & 5.64 & 3.18        & $-$2.46   & $-$0.19   & 0.17           \\
Mn I      & 4   & 5.43 & 2.88      & $-$2.55   & $-$0.27   & 0.37            \\
Fe I      & 178 & 7.50 & 5.24        & $-$2.26   & +0.01   & 0.36           \\
Fe II     & 17  & 7.50 & 5.11        & $-$2.39   & $-$0.12   & 0.15           \\
Co I      & 3   & 4.99 & 2.75     & $-$2.24   & +0.04   & 0.10           \\
Ni I      & 16  & 6.22 & 3.77     & $-$2.45   & $-$0.18   & 0.30           \\
Zn I      & 2   & 4.56 & 2.03    & $-$2.53   & $-$0.25  & 0.03   \\
Sr  & 2   & 2.87 & 0.42        & $-2.45$   & +0.10   & 0.29         \\
Y   & 15  & 2.21 & $-0.19$        & $-2.40$   & +0.14   & 0.08           \\
Zr  & 12  & 2.58 & 0.46        & $-2.12$   & +0.43   & 0.20           \\
Nb  & 1   & $1.46$ & $<0.18$   & $<-1.28$   & $<+1.27$ & 0.30     \\
Mo  & 1   & $1.88$ & $<-0.17$   & $<-2.05$   & $<+0.50$ & 0.30   \\
Ru  & 1   & $1.75$ & $<0.30$   & $<-1.45$   & $<+1.10$ & 0.30  \\
Rh  & 1   & $0.91$ & $<0.56$   & $<-0.35$   & $<+2.20$ & 0.20    \\
Pd  & 1   & $1.57$ & $<0.67$   & $<-0.90$   & $<+1.66$ & 0.30    \\
Ag  & 1   & $0.94$ & $<0.04$   & $<-0.90$   & $<+1.65$ & 0.30     \\
Ba  & 5   & 2.18 & 0.08        & $-2.10$   & +0.47   & 0.10           \\
La  & 19  & 1.10 & $-0.64$       & $-1.74$   & +0.81   & 0.12          \\
Ce  & 12  & 1.58 & $-0.48$        & $-2.06$   & +0.49   & 0.09           \\
Pr  & 9  & 0.72 & $-0.74$      & $-1.46$   & +1.09   & 0.08           \\
Nd  & 26  & 1.42 & $-0.32$       & $-1.74$   & +0.81   & 0.13          \\
Sm  & 10  & 0.96 & $-0.48$      & $-1.44$   & +1.13   & 0.09           \\
Eu  & 12  & 0.52 & $-0.69$       & $-1.21$   & +1.34   & 0.10         \\
Gd  & 7   & 1.07 & $-0.42$       & $-1.49$   & +1.07   & 0.09           \\
Dy  & 12  & 1.10 & $-0.28$      & -1.38   & +1.18   & 0.11           \\
Ho  & 1 & 0.48 & $<-$0.85 & $<-$1.33 & $<+$0.92 & 0.3 \\
Er  & 7   & 0.92 & $-0.49$      & -1.41   & +1.16   & 0.36           \\
Tm  & 2   & 0.10 & $-1.15$       & -1.25   & +1.30   & 0.14                   \\
Lu  & 1   & $0.10$ & $<-1.05$       & $<-1.15$   & $<+1.40$   & 0.10      \\
Hf  & 3   & 0.85 & $-0.73$       & $-1.58$   & +0.99   & 0.06           \\
Ir  & 1   & 1.38 & $<0.03$       & $<-1.35$   & $<+1.22$   & 0.30       \\
Pb  & 1   & 1.75 & $<-0.20$      & $<-1.95$   & $<+0.62$   & 0.30         \\
Th  & 2   & 0.02 & $-1.28$      & $-1.30$   & +1.27   & 0.21          \\
U   & 1   & $-0.54$ & $<-0.99$      & $<-0.45$   & $<+2.12$   & 0.20      \\
\enddata
\tablenotetext{a}{Following \citet{placco2014}.}
\end{deluxetable*}

%% file: Stellar_error.tex
\begin{deluxetable}{c c c c c c c}
\tablewidth{0pt}
\tabletypesize{\footnotesize}
\tablecaption{\label{tab:err} Abundance uncertainties due to fundamental atmospheric stellar parameter uncertainties.}
\tablehead{
\colhead{El.} & \colhead{$\Delta$ \teff\ $(\pm \sigma)$} & \colhead{$\Delta$ \logg\ $(\pm \sigma)$} & \colhead{$\Delta$ \vt\ $(\pm \sigma)$} & \colhead{$\Delta$\feh\ $(\pm \sigma)$} }
\startdata
Li    & $\pm 0.33$ & $\pm 0.10$ & $\pm 0.01$ & $\pm 0.10$       \\
\oi   & $\pm 0.09$ & $\pm 0.03$ & $\pm 0.01$ & $\pm 0.03$       \\
\mgi  & $\pm 0.15$ & $\pm 0.07$ & $\pm 0.03$ & $\pm 0.02$        \\
\ali  & $\pm 0.21$ & $\pm 0.12$ & $\pm 0.03$ & $\pm 0.07$         \\
\sii  & $\pm 0.11$ & $\pm 0.05$ & $\pm 0.02$ & $\pm 0.03$          \\
\ki   & $\pm 0.14$ & $\pm 0.03$ & $\pm 0.02$ & $\pm 0.01$          \\
\cai  & $\pm 0.12$ & $\pm 0.03$ & $\pm 0.01$ & $\pm 0.02$        \\
\scii & $\pm 0.03$ & $\pm 0.05$ & $\pm 0.03$ & $\pm 0.01$          \\
\tiii & $\pm 0.02$ & $\pm 0.03$ & $\pm 0.03$ & $\pm 0.01$         \\
\cri  & $\pm 0.21$ & $\pm 0.05$ & $\pm 0.03$ & $\pm 0.03$          \\
\mni  & $\pm 0.22$ & $\pm 0.06$ & $\pm 0.02$ & $\pm 0.04$           \\
\fei  & $\pm 0.19$ & $\pm 0.05$ & $\pm 0.03$ & $\pm 0.03$          \\
\feii & $\pm 0.04$ & $\pm 0.06$ & $\pm 0.02$ & $\pm 0.02$           \\
\coi  & $\pm 0.22$ & $\pm 0.09$ & $\pm 0.05$ & $\pm 0.09$           \\
\nii  & $\pm 0.15$ & $\pm 0.04$ & $\pm 0.01$ & $\pm 0.02$      \\
\zni  & $\pm 0.03$ & $\pm 0.03$ & $\pm 0.00$ & $\pm 0.01$ \\
\enddata
\end{deluxetable}

%% file: lbl4.tex
\startlongtable
\begin{deluxetable*}{ c c c c c c }
\tablewidth{0pt}
\tabletypesize{\footnotesize}
\tablecaption{\label{tab:line_by_line} Line by Line Element Abundances}
\tablehead{
\colhead{$\lambda$ ({\AA})} & \colhead{Species} & \colhead{$\chi$ (eV)} & \colhead{$\log{gf}$} & \colhead{EW (m{\AA})} & \colhead{$\log$\, $\epsilon$(X) (dex)}}
\startdata
6300.30 &       O I &      0.00 &    $-$9.82 &     30.15 &      7.41 \\
 4167.27 &      Mg I &      4.35 &    $-$0.71 &     65.42 &      5.51 \\
 4571.09 &      Mg I &      0.00 &    $-$5.68 &    135.03 &      6.09 \\
 4702.99 &      Mg I &      4.33 &    $-$0.38 &    105.49 &      5.60 \\
 5172.68 &      Mg I &      2.71 &    $-$0.45 &    310.27 &      5.70 \\
 5183.60 &      Mg I &      2.72 &    $-$0.23 &    363.18 &      5.70 \\
 5528.40 &      Mg I &      4.34 &    $-$0.49 &    118.84 &      5.78 \\
 5711.09 &      Mg I &      4.34 &    $-$1.72 &     21.35 &      5.51 \\
 3961.52 &      Al I &      0.01 &    $-$0.34 &    170.81 &      $<$ 3.53 \\
 4102.93 &      Si I &      1.91 &    $-$3.14 &    131.52 &      6.10 \\
 5708.39 &      Si I &      4.93 &    $-$1.47 &     16.96 &      5.76 \\
 7664.90 &       K I &      0.00 &      0.13 &    112.55 &      3.32 \\
 7698.96 &       K I &      0.00 &    $-$0.16 &     82.34 &      3.21 \\
 4425.44 &      Ca I &      1.88 &    $-$0.35 &     62.39 &      3.96 \\
 4454.78 &      Ca I &      1.90 &      0.26 &    131.61 &      4.66 \\
 5265.55 &      Ca I &      2.52 &    $-$0.26 &     63.30 &      4.52 \\
 5349.46 &      Ca I &      2.71 &    $-$0.31 &     20.74 &      4.03 \\
 5581.97 &      Ca I &      2.52 &    $-$0.55 &     26.55 &      4.17 \\
 5588.76 &      Ca I &      2.52 &      0.21 &     79.56 &      4.26 \\
 5590.12 &      Ca I &      2.52 &    $-$0.57 &     17.39 &      3.95 \\
 5594.48 &      Ca I &      2.52 &      0.09 &     66.63 &      4.17 \\
 5598.48 &      Ca I &      2.52 &    $-$0.08 &     55.24 &      4.18 \\
 5601.28 &      Ca I &      2.53 &    $-$0.52 &     23.28 &      4.06 \\
 5857.45 &      Ca I &      2.93 &      0.23 &     43.16 &      4.16 \\
 6102.72 &      Ca I &      1.88 &    $-$0.79 &     75.21 &      4.33 \\
 6122.22 &      Ca I &      1.89 &    $-$0.31 &    117.88 &      4.51 \\
 6162.17 &      Ca I &      1.90 &    $-$0.08 &    138.57 &      4.61 \\
 6169.05 &      Ca I &      2.52 &    $-$0.79 &     17.04 &      4.13 \\
 6169.55 &      Ca I &      2.53 &    $-$0.47 &     28.28 &      4.09 \\
 6439.07 &      Ca I &      2.52 &     0.47 &     98.82 &      4.20 \\
 6449.81 &      Ca I &      2.52 &    $-$0.50 &     34.06 &      4.19 \\
 6499.64 &      Ca I &      2.52 &    $-$0.81 &     16.79 &      4.12 \\
 6717.68 &      Ca I &      2.71 &    $-$0.52 &     26.98 &      4.29 \\
 4314.08 &     Sc II &      0.62 &    $-$0.10 &    160.44 &      1.61 \\
 4324.99 &     Sc II &      0.59 &    $-$0.44 &    143.71 &      1.55 \\
 4400.38 &     Sc II &      0.61 &    $-$0.54 &    125.07 &      1.24 \\
 4415.54 &     Sc II &      0.59 &    $-$0.67 &    127.64 &      1.39 \\
 5239.81 &     Sc II &      1.46 &    $-$0.77 &     48.47 &      0.99 \\
 5526.78 &     Sc II &      1.77 &     0.02 &     67.79 &      0.83 \\
 5641.00 &     Sc II &      1.50 &    $-$1.13 &     23.56 &      0.92 \\
 5657.90 &     Sc II &      1.51 &    $-$0.60 &     55.80 &      0.94 \\
 5684.21 &     Sc II &      1.51 &    $-$1.07 &     27.35 &      0.95 \\
 6604.57 &     Sc II &      1.36 &    $-$1.31 &     20.49 &      0.77 \\
3761.32 &     Ti II &      0.57 &      0.18 &    232.61 &      3.19 \\
 3913.46 &     Ti II &      1.12 &    $-$0.42 &    179.94 &      3.72 \\
 4028.33 &     Ti II &      1.89 &    $-$0.96 &     58.23 &      2.53 \\
 4053.82 &     Ti II &      1.89 &    $-$1.21 &     33.77 &      2.33 \\
 4161.52 &     Ti II &      1.08 &    $-$2.16 &     82.40 &      3.11 \\
 4163.63 &     Ti II &      2.59 &    $-$0.40 &     62.77 &      2.88 \\
 5669.05 &     Sc II &      1.50 &    $-$1.20 &     24.74 &      1.01 \\
 4184.30 &     Ti II &      1.08 &    $-$2.51 &     55.23 &      2.96 \\
 4330.72 &     Ti II &      1.18 &    $-$2.06 &     76.81 &      2.95 \\
 4337.91 &     Ti II &      1.08 &    $-$0.96 &    174.83 &      3.75 \\
 4394.05 &     Ti II &      1.22 &    $-$1.78 &     68.83 &      2.56 \\
 4395.03 &     Ti II &      1.08 &    $-$0.54 &    198.53 &      3.69 \\
 4395.83 &     Ti II &      1.24 &    $-$1.93 &     64.30 &      2.67 \\
 4399.76 &     Ti II &      1.24 &    $-$1.19 &    125.53 &      3.13 \\
 4417.71 &     Ti II &      1.17 &    $-$1.19 &    144.99 &      3.44 \\
 4418.33 &     Ti II &      1.24 &    $-$1.97 &     78.49 &      2.94 \\
 4441.73 &     Ti II &      1.18 &    $-$2.41 &     57.21 &      2.94 \\
 4443.80 &     Ti II &      1.08 &    $-$0.72 &    173.75 &      3.42 \\
 4444.55 &     Ti II &      1.12 &    $-$2.24 &     80.79 &      3.10 \\
 4450.48 &     Ti II &      1.08 &    $-$1.52 &    121.41 &      3.13 \\
 4464.44 &     Ti II &      1.16 &    $-$1.81 &     85.70 &      2.80 \\
 4470.85 &     Ti II &      1.17 &    $-$2.02 &     80.72 &      2.92 \\
 4488.34 &     Ti II &      3.12 &    $-$0.82 &     12.63 &      2.79 \\
 4468.51 &     Ti II &      1.13 &    $-$0.60 &    199.24 &      3.78 \\
 4493.52 &     Ti II &      1.08 &    $-$3.02 &     31.87 &      3.00 \\
 4501.27 &     Ti II &      1.12 &    $-$0.77 &    183.49 &      3.65 \\
 4529.48 &     Ti II &      1.57 &    $-$2.03 &     66.94 &      3.19 \\
 4563.77 &     Ti II &      1.22 &    $-$0.96 &    171.39 &      3.71 \\
 4589.91 &     Ti II &      1.24 &    $-$1.79 &    106.86 &      3.24 \\
 4636.32 &     Ti II &      1.16 &    $-$3.02 &     21.04 &      2.83 \\
 4657.20 &     Ti II &      1.24 &    $-$2.24 &     73.73 &      3.05 \\
 4708.66 &     Ti II &      1.24 &    $-$2.34 &     56.68 &      2.88 \\
 4779.97 &     Ti II &      2.05 &    $-$1.37 &     58.93 &      2.95 \\
 4798.53 &     Ti II &      1.08 &    $-$2.68 &     70.82 &      3.21 \\
 4805.08 &     Ti II &      2.06 &    $-$1.10 &     76.87 &      2.97 \\
 4865.61 &     Ti II &      1.12 &    $-$2.81 &     46.92 &      3.02 \\
 4911.17 &     Ti II &      3.12 &    $-$0.34 &     19.90 &      2.49 \\
 5005.15 &     Ti II &      1.57 &    $-$2.73 &     14.96 &      2.83 \\
 5129.15 &     Ti II &      1.89 &    $-$1.24 &     73.38 &      2.77 \\
 5185.90 &     Ti II &      1.89 &    $-$1.49 &     61.18 &      2.83 \\
 5188.68 &     Ti II &      1.58 &    $-$1.05 &    142.23 &      3.39 \\
 5336.78 &     Ti II &      1.58 &    $-$1.59 &     89.00 &      2.93 \\
 5381.02 &     Ti II &      1.57 &    $-$1.92 &     66.07 &      2.90 \\
 5418.76 &     Ti II &      1.58 &    $-$2.00 &     44.11 &      2.67 \\
 4289.72 &      Cr I &      0.00 &    $-$0.37 &    188.37 &      3.71 \\
 4545.95 &      Cr I &      0.94 &    $-$1.37 &     32.35 &      2.96 \\
 4580.05 &      Cr I &      0.94 &    $-$1.65 &     33.04 &      3.24 \\
 4600.75 &      Cr I &      1.00 &    $-$1.26 &     48.69 &      3.19 \\
 4616.13 &      Cr I &      0.98 &    $-$1.19 &     48.41 &      3.09 \\
 4626.18 &      Cr I &      0.97 &    $-$1.32 &     39.88 &      3.06 \\
 4646.15 &      Cr I &      1.03 &    $-$0.74 &     81.07 &      3.24 \\
 4651.28 &      Cr I &      0.98 &    $-$1.46 &     29.98 &      3.03 \\
 4652.15 &      Cr I &      1.00 &    $-$1.03 &     54.84 &      3.05 \\
 5247.56 &      Cr I &      0.96 &    $-$1.64 &     36.19 &      3.21 \\
 5296.69 &      Cr I &      0.98 &    $-$1.36 &     50.57 &      3.18 \\
 5298.28 &      Cr I &      0.98 &    $-$1.14 &     77.38 &      3.36 \\
 5300.74 &      Cr I &      0.98 &    $-$2.00 &     12.85 &      3.04 \\
 5345.80 &      Cr I &      1.00 &    $-$0.95 &     79.13 &      3.21 \\
 5348.31 &      Cr I &      1.00 &    $-$1.21 &     51.64 &      3.06 \\
 5409.77 &      Cr I &      1.03 &    $-$0.67 &     95.67 &      3.23 \\
 4034.48 &      Mn I &      0.00 &    $-$0.81 &    182.48 &      3.51 \\
 4754.04 &      Mn I &      2.28 &    $-$0.08 &     35.15 &      2.58 \\
 4783.43 &      Mn I &      2.30 &      0.04 &     46.29 &      2.66 \\
 4823.52 &      Mn I &      2.32 &      0.14 &     57.80 &      2.76 \\
 3852.57 &      Fe I &      2.18 &    $-$1.18 &     78.01 &      4.82 \\
 3949.95 &      Fe I &      2.18 &    $-$1.25 &     65.56 &      4.56 \\
 4001.66 &      Fe I &      2.18 &    $-$1.90 &     51.11 &      4.90 \\
 4032.62 &      Fe I &      1.49 &    $-$2.38 &     67.30 &      4.80 \\
 4058.21 &      Fe I &      3.21 &    $-$1.11 &     27.83 &      4.92 \\
 4062.44 &      Fe I &      2.85 &    $-$0.86 &     49.74 &      4.65 \\
 4067.97 &      Fe I &      3.21 &    $-$0.47 &     47.51 &      4.66 \\
 4073.76 &      Fe I &      3.27 &    $-$0.90 &     36.24 &      4.95 \\
 4076.62 &      Fe I &      3.21 &    $-$0.37 &     66.66 &      4.92 \\
 4098.17 &      Fe I &      3.24 &    $-$0.88 &     41.33 &      4.98 \\
 4109.80 &      Fe I &      2.85 &    $-$0.94 &     60.44 &      4.91 \\
 4114.44 &      Fe I &      2.83 &    $-$1.30 &     26.73 &      4.60 \\
 4121.80 &      Fe I &      2.83 &    $-$1.45 &     29.13 &      4.80 \\
 4134.67 &      Fe I &      2.83 &    $-$0.64 &     71.54 &      4.80 \\
 4136.99 &      Fe I &      3.42 &    $-$0.45 &     32.64 &      4.59 \\
 4139.92 &      Fe I &      0.99 &    $-$3.62 &     55.77 &      5.16 \\
 4147.66 &      Fe I &      1.48 &    $-$2.07 &    112.74 &      5.39 \\
 4154.49 &      Fe I &      2.83 &    $-$0.68 &     83.85 &      5.09 \\
 4154.80 &      Fe I &      3.37 &    $-$0.40 &     49.67 &      4.79 \\
 4156.79 &      Fe I &      2.83 &    $-$0.80 &     73.25 &      4.98 \\
 4157.78 &      Fe I &      3.42 &    $-$0.40 &     47.66 &      4.82 \\
 4158.79 &      Fe I &      3.43 &    $-$0.67 &     21.66 &      4.57 \\
 4174.91 &      Fe I &      0.91 &    $-$2.93 &    132.56 &      5.93 \\
 4181.75 &      Fe I &      2.83 &    $-$0.37 &     86.72 &      4.82 \\
 4184.89 &      Fe I &      2.83 &    $-$0.86 &     44.61 &      4.50 \\
 4187.03 &      Fe I &      2.45 &    $-$0.51 &    119.64 &      5.15 \\
 4187.79 &      Fe I &      2.42 &    $-$0.51 &    135.57 &      5.40 \\
 4195.32 &      Fe I &      3.33 &    $-$0.49 &     49.55 &      4.82 \\
 4216.18 &      Fe I &      0.00 &    $-$3.35 &    193.17 &      6.27 \\
 4227.42 &      Fe I &      3.33 &      0.26 &     99.30 &      5.07 \\
 4233.60 &      Fe I &      2.48 &    $-$0.57 &     98.99 &      4.80 \\
 4238.81 &      Fe I &      3.40 &    $-$0.23 &     49.48 &      4.63 \\
 4260.47 &      Fe I &      2.40 &      0.07 &    153.14 &      5.02 \\
 4337.04 &      Fe I &      1.56 &    $-$1.69 &    130.81 &      5.36 \\
 4352.73 &      Fe I &      2.22 &    $-$1.29 &    120.03 &      5.61 \\
 4415.12 &      Fe I &      1.61 &    $-$0.62 &    199.74 &      5.33 \\
 4422.56 &      Fe I &      2.85 &    $-$1.11 &     73.35 &      5.21 \\
 4430.61 &      Fe I &      2.22 &    $-$1.65 &     85.86 &      5.18 \\
 4442.33 &      Fe I &      2.20 &    $-$1.22 &    120.79 &      5.39 \\
 4443.19 &      Fe I &      2.86 &    $-$1.04 &     64.01 &      4.97 \\
 4447.71 &      Fe I &      2.22 &    $-$1.33 &     92.49 &      4.98 \\
 4459.11 &      Fe I &      2.18 &    $-$1.27 &    126.76 &      5.51 \\
 4476.01 &      Fe I &      2.85 &    $-$0.82 &     90.55 &      5.23 \\
 4489.73 &      Fe I &      0.12 &    $-$3.89 &    153.35 &      6.02 \\
 4494.56 &      Fe I &      2.20 &    $-$1.14 &    128.04 &      5.40 \\
 4528.61 &      Fe I &      2.18 &    $-$0.82 &    168.74 &      5.63 \\
 4531.14 &      Fe I &      1.48 &    $-$2.10 &    135.86 &      5.63 \\
 4647.43 &      Fe I &      2.95 &    $-$1.35 &     50.34 &      5.12 \\
 4678.84 &      Fe I &      3.60 &    $-$0.83 &     34.09 &      5.11 \\
 4710.28 &      Fe I &      3.02 &    $-$1.61 &     33.94 &      5.17 \\
 4733.59 &      Fe I &      1.49 &    $-$2.98 &     85.41 &      5.48 \\
 4736.77 &      Fe I &      3.21 &    $-$0.75 &     77.89 &      5.27 \\
 4859.74 &      Fe I &      2.88 &    $-$0.76 &     93.52 &      5.12 \\
 4871.31 &      Fe I &      2.87 &    $-$0.36 &    117.83 &      5.15 \\
 4872.13 &      Fe I &      2.88 &    $-$0.56 &    100.27 &      5.05 \\
 4890.75 &      Fe I &      2.88 &    $-$0.39 &    127.83 &      5.37 \\
 4891.49 &      Fe I &      2.85 &    $-$0.11 &    131.07 &      5.11 \\
 4903.31 &      Fe I &      2.88 &    $-$0.92 &     90.68 &      5.22 \\
 4918.99 &      Fe I &      2.85 &    $-$0.34 &    123.67 &      5.19 \\
 4924.77 &      Fe I &      2.28 &    $-$2.11 &     64.25 &      5.21 \\
 4938.81 &      Fe I &      2.88 &    $-$1.07 &     73.14 &      5.06 \\
 4939.68 &      Fe I &      0.86 &    $-$3.25 &    136.46 &      5.81 \\
 4946.38 &      Fe I &      3.37 &    $-$1.17 &     38.09 &      5.20 \\
 4966.08 &      Fe I &      3.33 &    $-$0.87 &     56.68 &      5.15 \\
 4973.10 &      Fe I &      3.96 &    $-$0.95 &      9.44 &      4.94 \\
 4994.13 &      Fe I &      0.92 &    $-$2.96 &    140.66 &      5.66 \\
 5001.87 &      Fe I &      3.88 &      0.05 &     44.16 &      4.70 \\
 5006.11 &      Fe I &      2.83 &    $-$0.61 &    104.69 &      5.07 \\
 5012.06 &      Fe I &      0.86 &    $-$2.64 &    174.48 &      5.83 \\
 5014.94 &      Fe I &      3.94 &    $-$0.30 &     28.89 &      4.84 \\
 5022.23 &      Fe I &      3.98 &    $-$0.53 &     25.66 &      5.05 \\
 5028.12 &      Fe I &      3.56 &    $-$1.12 &     16.61 &      4.90 \\
 5041.07 &      Fe I &      0.96 &    $-$3.09 &    151.13 &      6.02 \\
 5041.75 &      Fe I &      1.49 &    $-$2.20 &    148.85 &      5.80 \\
 5044.21 &      Fe I &      2.85 &    $-$2.01 &     19.39 &      5.01 \\
 5049.82 &      Fe I &      2.28 &    $-$1.35 &    122.73 &      5.44 \\
 5051.63 &      Fe I &      0.92 &    $-$2.76 &    162.12 &      5.80 \\
 5060.08 &      Fe I &      0.00 &    $-$5.46 &     52.39 &      5.45 \\
 5068.76 &      Fe I &      2.94 &    $-$1.04 &     74.15 &      5.08 \\
 5074.74 &      Fe I &      4.22 &    $-$0.20 &     30.14 &      5.11 \\
 5079.22 &      Fe I &      2.20 &    $-$2.10 &     84.41 &      5.38 \\
 5079.74 &      Fe I &      0.99 &    $-$3.24 &    141.35 &      6.01 \\
 5083.33 &      Fe I &      0.96 &    $-$2.84 &    155.50 &      5.81 \\
 5098.69 &      Fe I &      2.18 &    $-$2.03 &     95.76 &      5.49 \\
 5123.72 &      Fe I &      1.01 &    $-$3.05 &    152.50 &      6.03 \\
 5125.11 &      Fe I &      4.22 &    $-$0.14 &     28.61 &      5.01 \\
 5127.36 &      Fe I &      0.92 &    $-$3.24 &    147.06 &      6.00 \\
 5131.46 &      Fe I &      2.22 &    $-$2.51 &     50.68 &      5.30 \\
 5133.68 &      Fe I &      4.18 &      0.14 &     52.82 &      5.10 \\
 5141.73 &      Fe I &      2.42 &    $-$2.23 &     46.84 &      5.21 \\
 5142.92 &      Fe I &      0.96 &    $-$3.08 &    152.70 &      5.99 \\
 5150.83 &      Fe I &      0.99 &    $-$3.03 &    130.39 &      5.57 \\
 5151.91 &      Fe I &      1.01 &    $-$3.32 &    141.23 &      6.08 \\
 5162.27 &      Fe I &      4.18 &      0.02 &     34.92 &      4.92 \\
 5166.28 &      Fe I &      0.00 &    $-$4.12 &    176.61 &      6.11 \\
 5191.45 &      Fe I &      3.04 &    $-$0.55 &    115.99 &      5.42 \\
 5192.34 &      Fe I &      3.00 &    $-$0.42 &    114.78 &      5.22 \\
 5194.94 &      Fe I &      1.56 &    $-$2.02 &    151.43 &      5.68 \\
 5198.71 &      Fe I &      2.22 &    $-$2.09 &     67.46 &      5.11 \\
 5202.33 &      Fe I &      2.18 &    $-$1.87 &    105.62 &      5.47 \\
 5216.27 &      Fe I &      1.61 &    $-$2.08 &    134.65 &      5.49 \\
 5217.39 &      Fe I &      3.21 &    $-$1.16 &     44.93 &      5.07 \\
 5225.52 &      Fe I &      0.11 &    $-$4.75 &    123.10 &      5.96 \\
 5232.94 &      Fe I &      2.94 &    $-$0.05 &    136.45 &      5.15 \\
 5242.49 &      Fe I &      3.63 &    $-$0.96 &     22.10 &      4.96 \\
 5247.05 &      Fe I &      0.09 &    $-$4.94 &    111.61 &      5.92 \\
 5250.21 &      Fe I &      0.12 &    $-$4.93 &    103.67 &      5.82 \\
 5250.64 &      Fe I &      2.20 &    $-$2.18 &     85.58 &      5.45 \\
 5254.95 &      Fe I &      0.11 &    $-$4.76 &    126.61 &      6.02 \\
 5263.30 &      Fe I &      3.27 &    $-$0.87 &     45.31 &      4.87 \\
 5266.55 &      Fe I &      3.00 &    $-$0.38 &    107.59 &      5.04 \\
 5281.79 &      Fe I &      3.04 &    $-$0.83 &     76.03 &      5.00 \\
 5283.62 &      Fe I &      3.24 &    $-$0.52 &     89.49 &      5.16 \\
 5302.30 &      Fe I &      3.28 &    $-$0.72 &     70.81 &      5.10 \\
 5307.36 &      Fe I &      1.61 &    $-$2.91 &     80.34 &      5.34 \\
 5322.04 &      Fe I &      2.28 &    $-$2.80 &     21.19 &      5.09 \\
 5324.17 &      Fe I &      3.21 &    $-$0.10 &    112.44 &      5.09 \\
 5328.03 &      Fe I &      0.92 &    $-$1.46 &    275.72 &      5.75 \\
 5328.53 &      Fe I &      1.56 &    $-$1.85 &    167.08 &      5.72 \\
 5332.90 &      Fe I &      1.55 &    $-$2.77 &     96.70 &      5.39 \\
 5339.93 &      Fe I &      3.27 &    $-$0.72 &     72.85 &      5.12 \\
 5364.87 &      Fe I &      4.45 &      0.22 &     34.10 &      5.00 \\
 5365.40 &      Fe I &      3.56 &    $-$1.02 &     14.99 &      4.72 \\
 5367.46 &      Fe I &      4.42 &      0.44 &     38.56 &      4.84 \\
 5369.96 &      Fe I &      4.37 &      0.53 &     50.18 &      4.87 \\
 5383.36 &      Fe I &      4.31 &     0.64 &     62.39 &      4.87 \\
 5393.16 &      Fe I &      3.24 &    $-$0.91 &     73.59 &      5.27 \\
 5410.91 &      Fe I &      4.47 &      0.39 &     37.65 &      4.92 \\
 5415.19 &      Fe I &      4.39 &      0.64 &     52.59 &      4.82 \\
 5424.06 &      Fe I &      4.32 &      0.52 &     64.71 &      5.04 \\
 5497.51 &      Fe I &      1.01 &    $-$2.82 &    163.23 &      5.84 \\
 5501.46 &      Fe I &      0.96 &    $-$3.04 &    160.42 &      5.95 \\
 5506.77 &      Fe I &      0.99 &    $-$2.78 &    172.42 &      5.92 \\
 5569.61 &      Fe I &      3.42 &    $-$0.54 &     70.06 &      5.05 \\
 5572.84 &      Fe I &      3.40 &    $-$0.27 &     88.97 &      5.06 \\
 5576.08 &      Fe I &      3.43 &    $-$1.00 &     44.33 &      5.13 \\
 5586.75 &      Fe I &      3.37 &    $-$0.14 &     98.95 &      5.04 \\
 5615.64 &      Fe I &      3.33 &      0.05 &    129.93 &      5.32 \\
 5624.54 &      Fe I &      3.42 &    $-$0.75 &     54.76 &      5.04 \\
 5658.81 &      Fe I &      3.40 &    $-$0.79 &     62.36 &      5.15 \\
 5662.51 &      Fe I &      4.18 &    $-$0.57 &     12.95 &      4.92 \\
 5686.52 &      Fe I &      4.55 &    $-$0.45 &      6.12 &      4.89 \\
 5701.54 &      Fe I &      2.56 &    $-$2.14 &     39.48 &      5.10 \\
 5753.12 &      Fe I &      4.26 &    $-$0.68 &     10.54 &      5.03 \\
 5816.37 &      Fe I &      4.55 &    $-$0.60 &      4.53 &      4.89 \\
 6065.48 &      Fe I &      2.61 &    $-$1.41 &     89.14 &      5.13 \\
 6136.61 &      Fe I &      2.45 &    $-$1.41 &    130.41 &      5.59 \\
 6137.69 &      Fe I &      2.59 &    $-$1.34 &    102.84 &      5.25 \\
 6151.61 &      Fe I &      2.18 &    $-$3.37 &     17.79 &      5.37 \\
 6191.558 &      Fe I &      2.43 &    $-$1.41 &    124.73 &      5.47 \\
 6200.31 &      Fe I &      2.61 &    $-$2.43 &     31.54 &      5.28 \\
 6213.42 &      Fe I &      2.22 &    $-$2.48 &     58.47 &      5.25 \\
 6219.28 &      Fe I &      2.20 &    $-$2.44 &     68.86 &      5.34 \\
 6230.72 &      Fe I &      2.56 &    $-$1.27 &    119.95 &      5.41 \\
 6240.64 &      Fe I &      2.22 &    $-$3.17 &     14.62 &      5.12 \\
 6246.31 &      Fe I &      3.60 &    $-$0.87 &     37.17 &      5.05 \\
 6252.55 &      Fe I &      2.40 &    $-$1.68 &    101.70 &      5.31 \\
 6254.25 &      Fe I &      2.28 &    $-$2.44 &     62.00 &      5.33 \\
 6265.13 &      Fe I &      2.18 &    $-$2.54 &     60.53 &      5.28 \\
 6301.49 &      Fe I &      3.65 &    $-$0.71 &     31.46 &      4.85 \\
 6322.68 &      Fe I &      2.59 &    $-$2.46 &     34.02 &      5.32 \\
 6335.33 &      Fe I &      2.20 &    $-$2.18 &     81.92 &      5.24 \\
 6336.83 &      Fe I &      3.69 &    $-$1.05 &     27.97 &      5.16 \\
 6344.14 &      Fe I &      2.43 &    $-$2.87 &     20.72 &      5.26 \\
 6355.02 &      Fe I &      2.84 &    $-$2.29 &     19.01 &      5.13 \\
 6393.60 &      Fe I &      2.43 &    $-$1.57 &    108.29 &      5.33 \\
 6411.64 &      Fe I &      3.65 &    $-$0.59 &     46.91 &      4.97 \\
 6421.35 &      Fe I &      2.28 &    $-$2.01 &     93.84 &      5.34 \\
 6430.84 &      Fe I &      2.18 &    $-$1.94 &    109.48 &      5.38 \\
 6494.98 &      Fe I &      2.40 &    $-$1.23 &    133.28 &      5.34 \\
 6498.94 &      Fe I &      0.96 &    $-$4.69 &     36.23 &      5.52 \\
 6592.91 &      Fe I &      2.73 &    $-$1.47 &     81.04 &      5.17 \\
 6593.86 &      Fe I &      2.44 &    $-$2.36 &     47.25 &      5.22 \\
 6609.10 &      Fe I &      2.56 &    $-$2.66 &     20.37 &      5.17 \\
 6663.44 &      Fe I &      2.42 &    $-$2.47 &     46.62 &      5.30 \\
 6677.98 &      Fe I &      2.69 &    $-$1.41 &     97.69 &      5.30 \\
 6750.15 &      Fe I &      2.42 &    $-$2.58 &     39.34 &      5.28 \\
 6978.85 &      Fe I &      2.48 &    $-$2.45 &     42.64 &      5.26 \\
 4416.81 &     Fe II &      2.78 &    $-$2.60 &     57.94 &      5.07 \\
 4489.18 &     Fe II &      2.83 &    $-$2.97 &     29.09 &      4.96 \\
 4491.41 &     Fe II &      2.86 &    $-$2.71 &     32.93 &      4.82 \\
 4520.22 &     Fe II &      2.81 &    $-$2.60 &     53.63 &      5.01 \\
 4576.34 &     Fe II &      2.84 &    $-$2.95 &     29.28 &      4.95 \\
 4620.52 &     Fe II &      2.83 &    $-$3.21 &     17.14 &      4.89 \\
 4923.93 &     Fe II &      2.89 &    $-$1.32 &    140.72 &      5.28 \\
 5018.45 &     Fe II &      2.89 &    $-$1.22 &    155.56 &      5.38 \\
 5197.58 &     Fe II &      3.23 &    $-$2.22 &     63.07 &      5.17 \\
 5234.63 &     Fe II &      3.22 &    $-$2.18 &     64.19 &      5.14 \\
 5276.00 &     Fe II &      3.20 &    $-$2.01 &     75.20 &      5.10 \\
 5284.08 &     Fe II &      2.89 &    $-$3.19 &     28.97 &      5.15 \\
 5325.55 &     Fe II &      3.22 &    $-$3.16 &     14.37 &      5.15 \\
 5534.83 &     Fe II &      3.25 &    $-$2.93 &     28.66 &      5.30 \\
 6247.54 &     Fe II &      3.89 &    $-$2.51 &     14.19 &      5.24 \\
 6432.68 &     Fe II &      2.89 &    $-$3.71 &     14.07 &      5.21 \\
 6456.38 &     Fe II &      3.90 &    $-$2.08 &     21.03 &      5.01 \\
 3842.04 &      Co I &      0.92 &    $-$0.77 &     76.89 &      2.64 \\
 3845.46 &      Co I &      0.92 &      0.01 &    122.56 &      2.88 \\
 4121.31 &      Co I &      0.92 &    $-$0.32 &    110.45 &      2.73 \\
 3783.52 &      Ni I &      0.42 &    $-$1.42 &    159.67 &      4.38 \\
 4648.65 &      Ni I &      3.42 &    $-$0.16 &     14.28 &      3.46 \\
 4714.42 &      Ni I &      3.38 &      0.23 &     39.62 &      3.59 \\
 4855.41 &      Ni I &      3.54 &      0.00 &     20.24 &      3.59 \\
 4904.41 &      Ni I &      3.54 &    $-$0.17 &     18.79 &      3.72 \\
 4980.16 &      Ni I &      3.61 &    $-$0.11 &     20.79 &      3.79 \\
 5017.59 &      Ni I &      3.54 &    $-$0.08 &     15.00 &      3.50 \\
 5035.37 &      Ni I &      3.63 &      0.29 &     29.05 &      3.59 \\
 5080.52 &      Ni I &      3.65 &      0.13 &     25.58 &      3.70 \\
 5081.11 &      Ni I &      3.85 &      0.30 &     18.58 &      3.60 \\
 5084.08 &      Ni I &      3.68 &      0.03 &     11.06 &      3.41 \\
 5115.40 &      Ni I &      3.83 &    $-$0.11 &     16.49 &      3.92 \\
 5476.90 &      Ni I &      1.83 &    $-$0.89 &    146.85 &      4.49 \\
 5578.73 &      Ni I &      1.68 &    $-$2.64 &     20.87 &      3.87 \\
 6108.12 &      Ni I &      1.68 &    $-$2.45 &     22.36 &      3.67 \\
 6767.77 &      Ni I &      1.83 &    $-$2.17 &     47.05 &      3.96 \\ 
 4722.15 &      Zn I &      4.03 &    $-$0.39 &     13.62 &      2.00 \\
 4810.52 &      Zn I &      4.08 &    $-$0.13 &     22.25 &      2.06 \\
 4077.71 & Sr\,II & 0.00 & 0.15 & \nodata & 0.42 \\ 
 4215.52 & Sr\,II & 0.00 & $-$0.17 & \nodata & 0.42 \\ 
 3600.73 & Y\,II & 0.18 & 0.34 & \nodata & $-$0.14 \\ 
 3601.92 & Y\,II & 0.10 & $-$0.15 & \nodata & $-$0.19 \\ 
 3710.29 & Y\,II & 0.18 & 0.51 & \nodata & $-$0.24 \\ 
 3950.35 & Y\,II & 0.10 & $-$0.73 & \nodata & $-$0.19 \\ 
 4398.01 & Y\,II & 0.13 & $-$0.75 & \nodata & $-$0.24 \\ 
 4682.32 & Y\,II & 0.41 & $-$1.73 & \nodata & $-$0.14 \\ 
 4854.86 & Y\,II & 0.99 & $-$0.27 & \nodata & $-$0.38 \\ 
 4883.68 & Y\,II & 1.08 & 0.19 & \nodata & $-$0.19 \\ 
 4900.12 & Y\,II & 1.03 & 0.03 & \nodata & $-$0.02 \\ 
 5087.42 & Y\,II & 1.08 & $-$0.16 & \nodata & $-$0.24 \\ 
 5119.11 & Y\,II & 0.99 & $-$1.33 & \nodata & $-$0.24 \\ 
 5200.41 & Y\,II & 0.99 & $-$0.47 & \nodata & $-$0.16 \\ 
 5205.72 & Y\,II & 1.03 & $-$0.28 & \nodata & $-$0.19 \\ 
 5289.82 & Y\,II & 1.03 & $-$1.68 & \nodata & $-$0.14 \\ 
 5473.38 & Y\,II & 1.74 & $-$0.78 & \nodata & $-$0.13 \\ 
 3998.96 & Zr\,II & 0.56 & $-$0.52 & \nodata & 0.23 \\ 
 4050.32 & Zr\,II & 0.71 & $-$1.06 & \nodata & 0.53 \\ 
 4077.05 & Zr\,II & 0.96 & $-$1.69 & \nodata & 0.73 \\
 4149.20 & Zr\,II & 0.80 & $-$0.04 & \nodata & 0.28 \\ 
 4156.23 & Zr\,II & 0.71 & $-$0.78 & \nodata & 0.33 \\ 
 4161.20 & Zr\,II & 0.71 & $-$0.59 & \nodata & 0.28 \\ 
 4208.98 & Zr\,II & 0.71 & $-$0.51 & \nodata & 0.53 \\ 
 4317.31 & Zr\,II & 0.71 & $-$1.45 & \nodata & 0.23 \\ 
 4442.99 & Zr\,II & 1.49 & $-$0.42 & \nodata & 0.50 \\ 
 4496.96 & Zr\,II & 0.71 & $-$0.89 & \nodata & 0.38 \\ 
 4613.95 & Zr\,II & 0.97 & $-$1.54 & \nodata & 0.73 \\ 
 5112.27 & Zr\,II & 1.66 & $-$0.85 & \nodata & 0.73 \\ 
 3740.72 & Nb\,II & 1.62 & $-$0.31 & \nodata & $<$ 0.18 \\ 
 3864.10 & Mo\,I & 0.00 & $-$0.01 & \nodata & $<-$0.17 \\ 
 3798.90 & Ru\,I & 0.15 & $-$0.04 & \nodata & $<$ 0.30 \\ 
 3958.86 & Rh\,I & 0.97 & 0.01 & \nodata & $<$ 0.56 \\ 
 3404.58 & Pd\,I & 0.81 & 0.32 & \nodata & $<$ 0.67 \\ 
 3382.89 & Ag\,I & 0.00 & $-$0.33 & \nodata & $<$ 0.04 \\ 
 3891.78 & Ba\,II & 2.51 & 0.29 & \nodata & $-$0.01 \\ 
 4130.65 & Ba\,II & 2.72 & 0.52 & \nodata & $-$0.02 \\ 
 4166.00 & Ba\,II & 2.72 & $-$0.43 & \nodata & 0.18 \\ 
 4524.93 & Ba\,II & 2.51 & $-$0.39 & \nodata & 0.18 \\ 
 5853.68 & Ba\,II & 0.60 & $-$0.91 & \nodata & 0.08 \\ 
 3949.10 & La\,II & 0.40 & 0.49 & \nodata & $-$0.75 \\ 
 3995.74 & La\,II & 0.17 & $-$0.06 & \nodata & $-$0.70 \\ 
 4086.71 & La\,II & 0.00 & $-$0.07 & \nodata & $-$0.75 \\ 
 4322.50 & La\,II & 0.17 & $-$0.93 & \nodata & $-$0.85 \\ 
 4333.75 & La\,II & 0.17 & $-$0.06 & \nodata & $-$0.45 \\ 
 4662.50 & La\,II & 0.00 & $-$1.24 & \nodata & $-$0.65 \\ 
 4804.04 & La\,II & 0.24 & $-$1.49 & \nodata & $-$0.65 \\ 
 4809.00 & La\,II & 0.24 & $-$1.4 & \nodata & $-$0.60 \\ 
 4920.98 & La\,II & 0.13 & $-$0.58 & \nodata & $-$0.70 \\ 
 4921.78 & La\,II & 0.24 & $-$0.45 & \nodata & $-$0.65 \\ 
 4986.82 & La\,II & 0.17 & $-$1.3 & \nodata & $-$0.75 \\ 
 5122.99 & La\,II & 0.32 & $-$0.91 & \nodata & $-$0.45 \\ 
 5163.61 & La\,II & 0.24 & $-$1.81 & \nodata & $-$0.65 \\ 
 5290.82 & La\,II & 0.00 & $-$1.65 & \nodata & $-$0.65 \\ 
 5303.53 & La\,II & 0.32 & $-$1.35 & \nodata & $-$0.75 \\ 
 5482.27 & La\,II & 0.00 & $-$2.23 & \nodata & $-$0.75 \\ 
 5797.57 & La\,II & 0.24 & $-$1.36 & \nodata & $-$0.45 \\ 
 6262.29 & La\,II & 0.40 & $-$1.22 & \nodata & $-$0.55 \\ 
 6390.48 & La\,II & 0.32 & $-$1.41 & \nodata & $-$0.45 \\ 
 3999.24 & Ce\,II & 0.30 & 0.06 & \nodata & $-$0.47 \\ 
 4073.47 & Ce\,II & 0.48 & 0.21 & \nodata & $-$0.57 \\ 
 4075.70 & Ce\,II & 0.70 & 0.23 & \nodata & $-$0.27 \\ 
 4083.22 & Ce\,II & 0.70 & 0.27 & \nodata & $-$0.37 \\ 
 4127.36 & Ce\,II & 0.68 & 0.31 & \nodata & $-$0.57 \\ 
 4137.64 & Ce\,II & 0.52 & 0.4 & \nodata & $-$0.47 \\
 4142.40 & Ce\,II & 0.70 & 0.22 & \nodata & $-$0.34 \\ 
 4418.78 & Ce\,II & 0.86 & 0.27 & \nodata & $-$0.57 \\ 
 4628.16 & Ce\,II & 0.52 & 0.14 & \nodata & $-$0.37 \\ 
 4882.46 & Ce\,II & 0.52 & 0.19 & \nodata & $-$0.47 \\ 
 5187.46 & Ce\,II & 1.53 & 0.17 & \nodata & $-$0.57 \\ 
 5274.23 & Ce\,II & 1.21 & 0.13 & \nodata & $-$0.47 \\ 
 3994.80 & Pr\,II & 1.04 & $-$0.03 & \nodata & $-$0.63 \\ 
 4044.80 & Pr\,II & 0.06 & $-$0.29 & \nodata & $-$0.73 \\ 
 4062.80 & Pr\,II & 0.00 & 0.33 & \nodata & $-$0.63 \\ 
 4189.49 & Pr\,II & 0.42 & 0.43 & \nodata & $-$0.83 \\ 
 4222.95 & Pr\,II & 0.06 & 0.23 & \nodata & $-$0.83 \\ 
 4496.33 & Pr\,II & 0.22 & $-$0.76 & \nodata & $-$0.73 \\ 
 5135.15 & Pr\,II & 0.95 & 0.01 & \nodata & $-$0.73 \\ 
 5173.91 & Pr\,II & 0.97 & 0.36 & \nodata & $-$0.83 \\ 
 5352.40 & Pr\,II & 0.48 & $-$0.74 & \nodata & $-$0.73 \\ 
 3838.98 & Nd\,II & 0.00 & $-$0.24 & \nodata & $-$0.23 \\ 
 3991.74 & Nd\,II & 0.00 & $-$0.26 & \nodata & $-$0.23 \\ 
 4004.00 & Nd\,II & 0.06 & $-$0.57 & \nodata & $-$0.28 \\ 
 4018.82 & Nd\,II & 0.06 & $-$0.85 & \nodata & $-$0.48 \\ 
 4023.00 & Nd\,II & 0.56 & 0.04 & \nodata & $-$0.53 \\ 
 4059.95 & Nd\,II & 0.20 & $-$0.52 & \nodata & $-$0.43 \\ 
 4069.26 & Nd\,II & 0.06 & $-$0.57 & \nodata & $-$0.43 \\ 
 4075.27 & Nd\,II & 0.06 & $-$0.76 & \nodata & $-$0.33 \\ 
 4109.45 & Nd\,II & 0.32 & 0.35 & \nodata & $-$0.33 \\ 
 4177.31 & Nd\,II & 0.06 & $-$0.1 & \nodata & $-$0.43 \\ 
 4385.66 & Nd\,II & 0.20 & $-$0.3 & \nodata & $-$0.28 \\ 
 4462.98 & Nd\,II & 0.56 & 0.04 & \nodata & $-$0.23 \\ 
 4542.60 & Nd\,II & 0.74 & $-$0.28 & \nodata & $-$0.33 \\ 
 4706.54 & Nd\,II & 0.00 & $-$0.71 & \nodata & $-$0.13 \\ 
 4797.15 & Nd\,II & 0.56 & $-$0.69 & \nodata & $-$0.48 \\ 
 5063.72 & Nd\,II & 0.98 & $-$0.62 & \nodata & $-$0.23 \\ 
 5130.59 & Nd\,II & 1.30 & 0.45 & \nodata & $-$0.23 \\ 
 5212.36 & Nd\,II & 0.20 & $-$0.96 & \nodata & $-$0.13 \\ 
 5255.51 & Nd\,II & 0.20 & $-$0.67 & \nodata & $-$0.13 \\ 
 5311.45 & Nd\,II & 0.98 & $-$0.42 & \nodata & $-$0.53 \\ 
 5319.81 & Nd\,II & 0.55 & $-$0.14 & \nodata & $-$0.13 \\ 
 5356.97 & Nd\,II & 1.26 & $-$0.28 & \nodata & $-$0.43 \\ 
 5371.93 & Nd\,II & 1.41 & 0.00 & \nodata & $-$0.43 \\ 
 5431.52 & Nd\,II & 1.12 & $-$0.47 & \nodata & $-$0.33 \\ 
 5485.70 & Nd\,II & 1.26 & $-$0.12 & \nodata & $-$0.25 \\ 
 5740.86 & Nd\,II & 1.16 & $-$0.53 & \nodata & $-$0.40 \\ 
 4318.94 & Sm\,II & 0.28 & $-$0.25 & \nodata & $-$0.39 \\ 
 4424.34 & Sm\,II & 0.48 & 0.14 & \nodata & $-$0.49 \\ 
 4434.32 & Sm\,II & 0.38 & $-$0.07 & \nodata & $-$0.59 \\ 
 4467.34 & Sm\,II & 0.66 & 0.15 & \nodata & $-$0.49 \\ 
 4519.63 & Sm\,II & 0.54 & $-$0.35 & \nodata & $-$0.29 \\ 
 4537.94 & Sm\,II & 0.48 & $-$0.48 & \nodata & $-$0.49 \\ 
 4566.20 & Sm\,II & 0.33 & $-$0.59 & \nodata & $-$0.49 \\ 
 4642.23 & Sm\,II & 0.38 & $-$0.46 & \nodata & $-$0.49 \\ 
 4669.64 & Sm\,II & 0.28 & $-$0.53 & \nodata & $-$0.59 \\ 
 5103.08 & Sm\,II & 1.16 & $-$0.35 & \nodata & $-$0.49 \\ 
 3971.97 & Eu\,II & 0.21 & 0.27 & \nodata & $-$0.63 \\ 
 4129.72 & Eu\,II & 0.00 & 0.22 & \nodata & $-$0.53 \\ 
 4205.04 & Eu\,II & 0.00 & 0.21 & \nodata & $-$0.63 \\ 
 4435.58 & Eu\,II & 0.21 & $-$0.11 & \nodata & $-$0.63 \\ 
 5966.06 & Eu\,II & 1.25 & $-$1.04 & \nodata & $-$0.63 \\ 
 6049.51 & Eu\,II & 1.28 & $-$0.8 & \nodata & $-$0.63 \\ 
 6173.03 & Eu\,II & 1.32 & $-$0.86 & \nodata & $-$0.63 \\ 
 6437.64 & Eu\,II & 1.32 & $-$0.32 & \nodata & $-$0.83 \\ 
 6645.06 & Eu\,II & 1.38 & 0.12 & \nodata & $-$0.83 \\ 
 7077.09 & Eu\,II & 1.25 & $-$0.72 & \nodata & $-$0.83 \\ 
 7217.56 & Eu\,II & 1.23 & $-$0.35 & \nodata & $-$0.73 \\ 
 7370.22 & Eu\,II & 1.32 & $-$0.29 & \nodata & $-$0.71 \\ 
 3481.8 & Gd\,II & 0.49 & 0.12 & \nodata & $-$0.38 \\ 
 3796.38 & Gd\,II & 0.03 & 0.02 & \nodata & $-$0.55 \\ 
 4049.85 & Gd\,II & 0.99 & 0.48 & \nodata & $-$0.50 \\ 
 4063.38 & Gd\,II & 0.99 & 0.33 & \nodata & $-$0.40 \\ 
 4251.73 & Gd\,II & 0.38 & $-$0.22 & \nodata & $-$0.48 \\ 
 4316.05 & Gd\,II & 0.66 & $-$0.45 & \nodata & $-$0.38 \\ 
 4506.34 & Gd\,II & 0.50 & $-$1.03 & \nodata & $-$0.28 \\ 
 3694.81 & Dy\,II & 0.10 & $-$0.11 & \nodata & $-$0.35 \\ 
 3757.37 & Dy\,II & 0.10 & $-$0.17 & \nodata & $-$0.25 \\ 
 3944.68 & Dy\,II & 0.00 & 0.11 & \nodata & $-$0.45 \\ 
 4041.98 & Dy\,II & 0.93 & $-$0.9 & \nodata & $-$0.15 \\ 
 4050.57 & Dy\,II & 0.59 & $-$0.47 & \nodata & $-$0.45 \\
 4077.97 & Dy\,II & 0.10 & $-$0.04 & \nodata & $-$0.15 \\ 
 4103.31 & Dy\,II & 0.10 & $-$0.38 & \nodata & $-$0.25 \\
 4124.63 & Dy\,II & 0.10 & $-$0.66 & \nodata & $-$0.15 \\ 
 4129.42 & Dy\,II & 0.92 & $-$0.61 & \nodata & $-$0.25 \\ 
 4409.38 & Dy\,II & 0.54 & $-$1.24 & \nodata & $-$0.30 \\ 
 4449.70 & Dy\,II & 0.00 & $-$1.03 & \nodata & $-$0.40 \\ 
 4620.04 & Dy\,II & 0.10 & $-$1.93 & \nodata & $-$0.20 \\ 
4045.45 & Ho\,II & 0.00 & $-$0.05 & \nodata & $<-$0.85 \\ 
 3580.52 & Er\,II & 0.06 & $-$0.62 & \nodata & $-$0.18 \\ 
 3729.52 & Er\,II & 0.00 & $-$0.59 & \nodata & $-$0.53 \\ 
 3781.01 & Er\,II & 0.67 & $-$0.66 & \nodata & $-$0.23 \\ 
 3786.84 & Er\,II & 0.00 & $-$0.52 & \nodata & $-$0.13 \\ 
 3797.06 & Er\,II & 0.06 & $-$1.03 & \nodata & $-$0.43 \\ 
 3830.48 & Er\,II & 0.00 & $-$0.22 & \nodata & $-$0.83 \\ 
 3896.23 & Er\,II & 0.06 & $-$0.12 & \nodata & $-$1.08 \\
 3795.76 & Tm\,II & 0.03 & $-$0.23 & \nodata & $-$1.25 \\ 
 3848.02 & Tm\,II & 0.00 & $-$0.14 & \nodata & $-$1.05 \\ 
 6221.86 & Lu\,II & 1.54 & $-$0.76 & \nodata & $<-$1.05 \\ 
 3719.28 & Hf\,II & 0.61 & $-$0.81 & \nodata & $-$0.70 \\ 
 3918.09 & Hf\,II & 0.45 & $-$1.14 & \nodata & $-$0.80 \\ 
 4093.15 & Hf\,II & 0.45 & $-$1.15 & \nodata & $-$0.70 \\ 
 3800.12 & Ir\,I & 0.00 & $-$1.43 & \nodata & $<$ 0.03 \\ 
 4057.81 & Pb\,I & 1.32 & $-$0.22 & \nodata & $<-$0.20 \\ 
 4019.13 & Th\,II & 0.00 & $-$0.23 & \nodata & $-$1.43 \\ 
 4094.75 & Th\,II & 0.00 & $-$0.88 & \nodata & $-$1.13 \\ 
 4241.66 & U\,II & 0.57 & $-$0.10 & \nodata & $<-$0.99 \\
 \enddata
\end{deluxetable*}

%% file: Main.bbl
\begin{thebibliography}{}
\expandafter\ifx\csname natexlab\endcsname\relax\def\natexlab#1{#1}\fi
\providecommand{\url}[1]{\href{#1}{#1}}
\providecommand{\dodoi}[1]{doi:~\href{http://doi.org/#1}{\nolinkurl{#1}}}
\providecommand{\doeprint}[1]{\href{http://ascl.net/#1}{\nolinkurl{http://ascl.net/#1}}}
\providecommand{\doarXiv}[1]{\href{https://arxiv.org/abs/#1}{\nolinkurl{https://arxiv.org/abs/#1}}}

\bibitem[{BRO(1989)}]{BROWN}
 1989, \apjs, 71, 293, \dodoi{10.1086/191375}

\bibitem[{{Adam{\'o}w} {et~al.}(2015){Adam{\'o}w}, {Niedzielski}, {Villaver},
  {Wolszczan}, {Kowalik}, {Nowak}, {Adamczyk}, \&
  {Deka-Szymankiewicz}}]{ADAMOW2}
{Adam{\'o}w}, M., {Niedzielski}, A., {Villaver}, E., {et~al.} 2015, \aap, 581,
  A94, \dodoi{10.1051/0004-6361/201526582}

\bibitem[{{Adam{\'o}w} {et~al.}(2014){Adam{\'o}w}, {Niedzielski}, {Villaver},
  {Wolszczan}, \& {Nowak}}]{ADAMOW}
{Adam{\'o}w}, M., {Niedzielski}, A., {Villaver}, E., {Wolszczan}, A., \&
  {Nowak}, G. 2014, \aap, 569, A55, \dodoi{10.1051/0004-6361/201423400}

\bibitem[{{Aguilera-G{\'o}mez} {et~al.}(2016){Aguilera-G{\'o}mez},
  {Chanam{\'e}}, {Pinsonneault}, \& {Carlberg}}]{2016ApJ...829..127A}
{Aguilera-G{\'o}mez}, C., {Chanam{\'e}}, J., {Pinsonneault}, M.~H., \&
  {Carlberg}, J.~K. 2016, \apj, 829, 127, \dodoi{10.3847/0004-637X/829/2/127}

\bibitem[{Aguilera-G{\'{o}}mez {et~al.}(2016)Aguilera-G{\'{o}}mez,
  Chanam{\'{e}}, Pinsonneault, \& Carlberg}]{Aguilera2016}
Aguilera-G{\'{o}}mez, C., Chanam{\'{e}}, J., Pinsonneault, M.~H., \& Carlberg,
  J.~K. 2016, The Astrophysical Journal, 829, 127,
  \dodoi{10.3847/0004-637x/829/2/127}

\bibitem[{{Aguilera-G{\'o}mez} {et~al.}(2022){Aguilera-G{\'o}mez}, {Monaco},
  {Mucciarelli}, {Salaris}, {Villanova}, \& {Pancino}}]{2022A&A...657A..33A}
{Aguilera-G{\'o}mez}, C., {Monaco}, L., {Mucciarelli}, A., {et~al.} 2022, \aap,
  657, A33, \dodoi{10.1051/0004-6361/202141750}

\bibitem[{Aguilera-Gómez {et~al.}(2020)Aguilera-Gómez, Chanamé, \&
  Pinsonneault}]{Aguilera_G_mez_2020}
Aguilera-Gómez, C., Chanamé, J., \& Pinsonneault, M.~H. 2020, The
  Astrophysical Journal, 897, L20, \dodoi{10.3847/2041-8213/ab9d26}

\bibitem[{{Alcal{\'a}} {et~al.}(2011){Alcal{\'a}}, {Biazzo}, {Covino},
  {Frasca}, \& {Bedin}}]{ACALA}
{Alcal{\'a}}, J.~M., {Biazzo}, K., {Covino}, E., {Frasca}, A., \& {Bedin},
  L.~R. 2011, \aap, 531, L12, \dodoi{10.1051/0004-6361/201117174}

\bibitem[{{Alexander}(1967)}]{1967Obs....87..238A}
{Alexander}, J.~B. 1967, The Observatory, 87, 238

\bibitem[{{Amarsi} {et~al.}(2016){Amarsi}, {Lind}, {Asplund}, {Barklem}, \&
  {Collet}}]{amarsi2016}
{Amarsi}, A.~M., {Lind}, K., {Asplund}, M., {Barklem}, P.~S., \& {Collet}, R.
  2016, \mnras, 463, 1518, \dodoi{10.1093/mnras/stw2077}

\bibitem[{{Andrievsky} {et~al.}(1999){Andrievsky}, {Gorlova}, {Klochkova},
  {Kovtyukh}, \& {Panchuk}}]{Andrievsky1999}
{Andrievsky}, S.~M., {Gorlova}, N.~I., {Klochkova}, V.~G., {Kovtyukh}, V.~V.,
  \& {Panchuk}, V.~E. 1999, Astronomische Nachrichten, 320, 35

\bibitem[{{Angelou} {et~al.}(2015){Angelou}, {D'Orazi}, {Constantino},
  {Church}, {Stancliffe}, \& {Lattanzio}}]{angelou2015}
{Angelou}, G.~C., {D'Orazi}, V., {Constantino}, T.~N., {et~al.} 2015, \mnras,
  450, 2423, \dodoi{10.1093/mnras/stv770}

\bibitem[{{Anthony-Twarog} {et~al.}(2013){Anthony-Twarog}, {Deliyannis},
  {Rich}, \& {Twarog}}]{TWAROG}
{Anthony-Twarog}, B.~J., {Deliyannis}, C.~P., {Rich}, E., \& {Twarog}, B.~A.
  2013, \apjl, 767, L19, \dodoi{10.1088/2041-8205/767/1/L19}

\bibitem[{{Asplund} {et~al.}(2009){Asplund}, {Grevesse}, {Sauval}, \&
  {Scott}}]{asplund2009}
{Asplund}, M., {Grevesse}, N., {Sauval}, A.~J., \& {Scott}, P. 2009, \araa, 47,
  481, \dodoi{10.1146/annurev.astro.46.060407.145222}

\bibitem[{{Bailer-Jones} {et~al.}(2021){Bailer-Jones}, {Rybizki}, {Fouesneau},
  {Demleitner}, \& {Andrae}}]{bailer-jones2021}
{Bailer-Jones}, C.~A.~L., {Rybizki}, J., {Fouesneau}, M., {Demleitner}, M., \&
  {Andrae}, R. 2021, \aj, 161, 147, \dodoi{10.3847/1538-3881/abd806}

\bibitem[{{Balachandran} {et~al.}(2000){Balachandran}, {Fekel}, {Henry}, \&
  {Uitenbroek}}]{BALACH}
{Balachandran}, S.~C., {Fekel}, F.~C., {Henry}, G.~W., \& {Uitenbroek}, H.
  2000, \apj, 542, 978, \dodoi{10.1086/317055}

\bibitem[{{Barklem}(2018)}]{barklem2018}
{Barklem}, P.~S. 2018, \aap, 612, A90, \dodoi{10.1051/0004-6361/201732365}

\bibitem[{{Bernstein} {et~al.}(2003){Bernstein}, {Shectman}, {Gunnels},
  {Mochnacki}, \& {Athey}}]{bernstein2003}
{Bernstein}, R., {Shectman}, S.~A., {Gunnels}, S.~M., {Mochnacki}, S., \&
  {Athey}, A.~E. 2003, in \procspie, Vol. 4841, Instrument Design and
  Performance for Optical/Infrared Ground-based Telescopes, ed. M.~{Iye} \&
  A.~F.~M. {Moorwood}, 1694--1704, \dodoi{10.1117/12.461502}

\bibitem[{{Bharat Kumar} {et~al.}(2015){Bharat Kumar}, {Reddy},
  {Muthumariappan}, \& {Zhao}}]{2015A&A...577A..10B}
{Bharat Kumar}, Y., {Reddy}, B.~E., {Muthumariappan}, C., \& {Zhao}, G. 2015,
  \aap, 577, A10, \dodoi{10.1051/0004-6361/201425076}

\bibitem[{{Binks} {et~al.}(2022){Binks}, {Jeffries}, {Sacco}, {Jackson}, {Cao},
  {Bayo}, {Bergemann}, {Bonito}, {Gilmore}, {Gonneau}, {Jimin{\'e}z-Esteban},
  {Morbidelli}, {Randich}, {Roccatagliata}, {Smiljanic}, \&
  {Zaggia}}]{2022MNRAS.513.5727B}
{Binks}, A.~S., {Jeffries}, R.~D., {Sacco}, G.~G., {et~al.} 2022, \mnras, 513,
  5727, \dodoi{10.1093/mnras/stac1245}

\bibitem[{{Brown} {et~al.}(1989){Brown}, {Sneden}, {Lambert}, \&
  {Dutchover}}]{BROWN1989}
{Brown}, J.~A., {Sneden}, C., {Lambert}, D.~L., \& {Dutchover}, Edward, J.
  1989, \apjs, 71, 293, \dodoi{10.1086/191375}

\bibitem[{{Bruntt} {et~al.}(2010){Bruntt}, {Bedding}, {Quirion}, {Lo Curto},
  {Carrier}, {Smalley}, {Dall}, {Arentoft}, {Bazot}, \& {Butler}}]{bruntt}
{Bruntt}, H., {Bedding}, T.~R., {Quirion}, P.~O., {et~al.} 2010, \mnras, 405,
  1907, \dodoi{10.1111/j.1365-2966.2010.16575.x}

\bibitem[{{Cai} {et~al.}(2023){Cai}, {Kong}, {Shi}, {Gao}, {Bu}, \&
  {Yi}}]{cai2023}
{Cai}, B., {Kong}, X., {Shi}, J., {et~al.} 2023, \aj, 165, 52,
  \dodoi{10.3847/1538-3881/aca098}

\bibitem[{{Cameron} \& {Fowler}(1971)}]{Cameron1971}
{Cameron}, A.~G.~W., \& {Fowler}, W.~A. 1971, \apj, 164, 111,
  \dodoi{10.1086/150821}

\bibitem[{{Canto Martins} {et~al.}(2006){Canto Martins}, {L{\`e}bre}, {de
  Laverny}, {Melo}, {Do Nascimento}, {Richard}, \& {de Medeiros}}]{CANTO}
{Canto Martins}, B.~L., {L{\`e}bre}, A., {de Laverny}, P., {et~al.} 2006, \aap,
  451, 993, \dodoi{10.1051/0004-6361:20053334}

\bibitem[{{Canto Martins} {et~al.}(2011){Canto Martins}, {L{\`e}bre},
  {Palacios}, {de Laverny}, {Richard}, {Melo}, {Do Nascimento}, \& {de
  Medeiros}}]{2011A&A...527A..94C}
{Canto Martins}, B.~L., {L{\`e}bre}, A., {Palacios}, A., {et~al.} 2011, \aap,
  527, A94, \dodoi{10.1051/0004-6361/201015015}

\bibitem[{Carlberg {et~al.}(2012)Carlberg, Cunha, Smith, \&
  Majewski}]{Carlberg_2012}
Carlberg, J.~K., Cunha, K., Smith, V.~V., \& Majewski, S.~R. 2012, The
  Astrophysical Journal, 757, 109, \dodoi{10.1088/0004-637x/757/2/109}

\bibitem[{{Carlberg} {et~al.}(2010){Carlberg}, {Smith}, {Cunha}, {Majewski}, \&
  {Rood}}]{CARLBERG}
{Carlberg}, J.~K., {Smith}, V.~V., {Cunha}, K., {Majewski}, S.~R., \& {Rood},
  R.~T. 2010, \apjl, 723, L103, \dodoi{10.1088/2041-8205/723/1/L103}

\bibitem[{{Carlberg} {et~al.}(2015){Carlberg}, {Smith}, {Cunha}, {Majewski},
  {M{\'e}sz{\'a}ros}, {Shetrone}, {Allende Prieto}, {Bizyaev}, {Stassun},
  {Fleming}, {Zasowski}, {Hearty}, {Nidever}, {Schneider}, {Holtzman}, \&
  {Frinchaboy}}]{CARLBURG2}
{Carlberg}, J.~K., {Smith}, V.~V., {Cunha}, K., {et~al.} 2015, \apj, 802, 7,
  \dodoi{10.1088/0004-637X/802/1/7}

\bibitem[{{Carlsson}(1986)}]{Carlsson1986}
{Carlsson}, M. 1986, Uppsala Astronomical Observatory Reports, 33

\bibitem[{{Carlsson}(1992)}]{carlsson1992}
{Carlsson}, M. 1992, in Astronomical Society of the Pacific Conference Series,
  Vol.~26, Cool Stars, Stellar Systems, and the Sun, ed. M.~S. {Giampapa} \&
  J.~A. {Bookbinder}, 499

\bibitem[{{Carney} {et~al.}(1998){Carney}, {Fry}, \& {Gonzalez}}]{CARNY}
{Carney}, B.~W., {Fry}, A.~M., \& {Gonzalez}, G. 1998, \aj, 116, 2984,
  \dodoi{10.1086/300630}

\bibitem[{{Casey}(2014)}]{casey2014}
{Casey}, A.~R. 2014, PhD thesis, Australian National University,
  \dodoi{10.5281/zenodo.49493}

\bibitem[{{Casey} {et~al.}(2016){Casey}, {Ruchti}, {Masseron}, {Randich},
  {Gilmore}, {Lind}, {Kennedy}, {Koposov}, {Hourihane}, {Franciosini}, {Lewis},
  {Magrini}, {Morbidelli}, {Sacco}, {Worley}, {Feltzing}, {Jeffries},
  {Vallenari}, {Bensby}, {Bragaglia}, {Flaccomio}, {Francois}, {Korn},
  {Lanzafame}, {Pancino}, {Recio-Blanco}, {Smiljanic}, {Carraro}, {Costado},
  {Damiani}, {Donati}, {Frasca}, {Jofr{\'e}}, {Lardo}, {de Laverny}, {Monaco},
  {Prisinzano}, {Sbordone}, {Sousa}, {Tautvai{\v{s}}ien{\.{e}}}, {Zaggia},
  {Zwitter}, {Delgado Mena}, {Chorniy}, {Martell}, {Silva Aguirre}, {Miglio},
  {Chiappini}, {Montalban}, {Morel}, \& {Valentini}}]{CASEY}
{Casey}, A.~R., {Ruchti}, G., {Masseron}, T., {et~al.} 2016, \mnras, 461, 3336,
  \dodoi{10.1093/mnras/stw1512}

\bibitem[{{Casey} {et~al.}(2019){Casey}, {Ho}, {Ness}, {Hogg}, {Rix},
  {Angelou}, {Hekker}, {Tout}, {Lattanzio}, {Karakas}, {Woods}, {Price-Whelan},
  \& {Schlaufman}}]{casey2019}
{Casey}, A.~R., {Ho}, A. Y.~Q., {Ness}, M., {et~al.} 2019, \apj, 880, 125,
  \dodoi{10.3847/1538-4357/ab27bf}

\bibitem[{Casey {et~al.}(2019)Casey, Ho, Ness, Hogg, Rix, Angelou, Hekker,
  Tout, Lattanzio, Karakas, \& et~al.}]{Casey_2019}
Casey, A.~R., Ho, A. Y.~Q., Ness, M., {et~al.} 2019, The Astrophysical Journal,
  880, 125, \dodoi{10.3847/1538-4357/ab27bf}

\bibitem[{{Castellani} \& {Castellani}(1993)}]{Castellani1993}
{Castellani}, M., \& {Castellani}, V. 1993, \apj, 407, 649,
  \dodoi{10.1086/172547}

\bibitem[{{Castelli} \& {Kurucz}(2004)}]{castelli2004}
{Castelli}, F., \& {Kurucz}, R.~L. 2004

\bibitem[{{Chanam{\'e}} {et~al.}(2022){Chanam{\'e}}, {Pinsonneault},
  {Aguilera-G{\'o}mez}, \& {Zinn}}]{2022ApJ...933...58C}
{Chanam{\'e}}, J., {Pinsonneault}, M.~H., {Aguilera-G{\'o}mez}, C., \& {Zinn},
  J.~C. 2022, \apj, 933, 58, \dodoi{10.3847/1538-4357/ac70c8}

\bibitem[{{Charbonneau} \& {Michaud}(1990)}]{1990ApJ...352..681C}
{Charbonneau}, P., \& {Michaud}, G. 1990, \apj, 352, 681,
  \dodoi{10.1086/168570}

\bibitem[{{Charbonnel} \& {Balachandran}(2000)}]{Charbonnel2000}
{Charbonnel}, C., \& {Balachandran}, S.~C. 2000, \aap, 359, 563.
\newblock \doarXiv{astro-ph/0005280}

\bibitem[{{Charbonnel} {et~al.}(1998){Charbonnel}, {Brown}, \&
  {Wallerstein}}]{1998A&A...332..204C}
{Charbonnel}, C., {Brown}, J.~A., \& {Wallerstein}, G. 1998, \aap, 332, 204.
\newblock \doarXiv{astro-ph/9712207}

\bibitem[{{Charbonnel} {et~al.}(2000){Charbonnel}, {Deliyannis}, \&
  {Pinsonneault}}]{2000astro.ph..6280C}
{Charbonnel}, C., {Deliyannis}, C.~P., \& {Pinsonneault}, M.~H. 2000, arXiv
  e-prints, astro.
\newblock \doarXiv{astro-ph/0006280}

\bibitem[{{Charbonnel} \& {Lagarde}(2010)}]{2010A&A...522A..10C}
{Charbonnel}, C., \& {Lagarde}, N. 2010, \aap, 522, A10,
  \dodoi{10.1051/0004-6361/201014432}

\bibitem[{{Charbonnel} \& {Talon}(2005)}]{2005Sci...309.2189C}
{Charbonnel}, C., \& {Talon}, S. 2005, Science, 309, 2189,
  \dodoi{10.1126/science.1116849}

\bibitem[{{Charbonnel} \& {Zahn}(2007)}]{2007A&A...467L..15C}
{Charbonnel}, C., \& {Zahn}, J.~P. 2007, \aap, 467, L15,
  \dodoi{10.1051/0004-6361:20077274}

\bibitem[{Charbonnel {et~al.}(2020)Charbonnel, Lagarde, Jasniewicz, North,
  Shetrone, Krugler~Hollek, Smith, Smiljanic, Palacios, \&
  Ottoni}]{Charbonnel_2020}
Charbonnel, C., Lagarde, N., Jasniewicz, G., {et~al.} 2020, Astronomy \&
  Astrophysics, 633, A34, \dodoi{10.1051/0004-6361/201936360}

\bibitem[{{Cort{\'e}s} {et~al.}(2009){Cort{\'e}s}, {Silva}, {Recio-Blanco},
  {Catelan}, {Do Nascimento}, \& {De Medeiros}}]{cortes2009}
{Cort{\'e}s}, C., {Silva}, J.~R.~P., {Recio-Blanco}, A., {et~al.} 2009, \apj,
  704, 750, \dodoi{10.1088/0004-637X/704/1/750}

\bibitem[{de~la Reza {et~al.}(1996)de~la Reza, Drake, \&
  da~Silva}]{de_la_Reza__1996}
de~la Reza, R., Drake, N.~A., \& da~Silva, L. 1996, The Astrophysical Journal,
  456, \dodoi{10.1086/309874}

\bibitem[{{de la Reza} {et~al.}(1997){de la Reza}, {Drake}, {da Silva},
  {Torres}, \& {Martin}}]{1997ApJ...482L..77D}
{de la Reza}, R., {Drake}, N.~A., {da Silva}, L., {Torres}, C.~A.~O., \&
  {Martin}, E.~L. 1997, \apjl, 482, L77, \dodoi{10.1086/310685}

\bibitem[{{De Silva} {et~al.}(2015){De Silva}, {Freeman}, {Bland-Hawthorn},
  {Martell}, {de Boer}, {Asplund}, {Keller}, {Sharma}, {Zucker}, {Zwitter},
  {Anguiano}, {Bacigalupo}, {Bayliss}, {Beavis}, {Bergemann}, {Campbell},
  {Cannon}, {Carollo}, {Casagrande}, {Casey}, {Da Costa}, {D'Orazi}, {Dotter},
  {Duong}, {Heger}, {Ireland}, {Kafle}, {Kos}, {Lattanzio}, {Lewis}, {Lin},
  {Lind}, {Munari}, {Nataf}, {O'Toole}, {Parker}, {Reid}, {Schlesinger},
  {Sheinis}, {Simpson}, {Stello}, {Ting}, {Traven}, {Watson}, {Wittenmyer},
  {Yong}, \& {{\v{Z}}erjal}}]{Desilva2015}
{De Silva}, G.~M., {Freeman}, K.~C., {Bland-Hawthorn}, J., {et~al.} 2015,
  \mnras, 449, 2604, \dodoi{10.1093/mnras/stv327}

\bibitem[{{Deal} \& {Martins}(2021)}]{Deal_Martins_2021}
{Deal}, M., \& {Martins}, C.~J.~A.~P. 2021, \aap, 653, A48,
  \dodoi{10.1051/0004-6361/202140725}

\bibitem[{{Deepak} \& {Lambert}(2021)}]{deepak2021}
{Deepak}, \& {Lambert}, D.~L. 2021, \mnras, 505, 642,
  \dodoi{10.1093/mnras/stab1195}

\bibitem[{{Deepak} {et~al.}(2020){Deepak}, {Lambert}, \&
  {Reddy}}]{2020MNRAS.494.1348D}
{Deepak}, {Lambert}, D.~L., \& {Reddy}, B.~E. 2020, \mnras, 494, 1348,
  \dodoi{10.1093/mnras/staa729}

\bibitem[{{Deliyannis} {et~al.}(2000){Deliyannis}, {Pinsonneault}, \&
  {Charbonnel}}]{2000IAUS..198...61D}
{Deliyannis}, C.~P., {Pinsonneault}, M.~H., \& {Charbonnel}, C. 2000, in The
  Light Elements and their Evolution, ed. L.~{da Silva}, R.~{de Medeiros}, \&
  M.~{Spite}, Vol. 198, 61

\bibitem[{{Deng} {et~al.}(2012){Deng}, {Newberg}, {Liu}, {Carlin}, {Beers},
  {Chen}, {Chen}, {Christlieb}, {Grillmair}, {Guhathakurta}, {Han}, {Hou},
  {Lee}, {L{\'e}pine}, {Li}, {Liu}, {Pan}, {Sellwood}, {Wang}, {Wang}, {Yang},
  {Yanny}, {Zhang}, {Zhang}, {Zheng}, \& {Zhu}}]{LAMOST}
{Deng}, L.-C., {Newberg}, H.~J., {Liu}, C., {et~al.} 2012, Research in
  Astronomy and Astrophysics, 12, 735, \dodoi{10.1088/1674-4527/12/7/003}

\bibitem[{Denissenkov \& Weiss(2000)}]{Denissenkov}
Denissenkov, P., \& Weiss, A. 2000, Astronomy and Astrophysics, 358

\bibitem[{{Denissenkov} \& {Herwig}(2004)}]{Denissenkov2004}
{Denissenkov}, P.~A., \& {Herwig}, F. 2004, \apj, 612, 1081,
  \dodoi{10.1086/422575}

\bibitem[{{D'Orazi} {et~al.}(2015){D'Orazi}, {Gratton}, {Angelou}, {Bragaglia},
  {Carretta}, {Lattanzio}, {Lucatello}, {Momany}, \& {Sollima}}]{DORAZI}
{D'Orazi}, V., {Gratton}, R.~G., {Angelou}, G.~C., {et~al.} 2015, \apjl, 801,
  L32, \dodoi{10.1088/2041-8205/801/2/L32}

\bibitem[{{Drake} {et~al.}(2002){Drake}, {de la Reza}, {da Silva}, \&
  {Lambert}}]{DRAKE}
{Drake}, N.~A., {de la Reza}, R., {da Silva}, L., \& {Lambert}, D.~L. 2002,
  \aj, 123, 2703, \dodoi{10.1086/339968}

\bibitem[{{Dumont} {et~al.}(2021{\natexlab{a}}){Dumont}, {Charbonnel},
  {Palacios}, \& {Borisov}}]{2021A&A...654A..46D}
{Dumont}, T., {Charbonnel}, C., {Palacios}, A., \& {Borisov}, S.
  2021{\natexlab{a}}, \aap, 654, A46, \dodoi{10.1051/0004-6361/202141094}

\bibitem[{{Dumont} {et~al.}(2021{\natexlab{b}}){Dumont}, {Palacios},
  {Charbonnel}, {Richard}, {Amard}, {Augustson}, \&
  {Mathis}}]{2021A&A...646A..48D}
{Dumont}, T., {Palacios}, A., {Charbonnel}, C., {et~al.} 2021{\natexlab{b}},
  \aap, 646, A48, \dodoi{10.1051/0004-6361/202039515}

\bibitem[{{Ezzeddine} {et~al.}(2017){Ezzeddine}, {Frebel}, \&
  {Plez}}]{ezzeddine2017}
{Ezzeddine}, R., {Frebel}, A., \& {Plez}, B. 2017, \apj, 847, 142,
  \dodoi{10.3847/1538-4357/aa8875}

\bibitem[{{Ezzeddine} {et~al.}(2016){Ezzeddine}, {Plez}, {Merle}, {Gebran}, \&
  {Th{\'e}venin}}]{Ezzeddine2016b}
{Ezzeddine}, R., {Plez}, B., {Merle}, T., {Gebran}, M., \& {Th{\'e}venin}, F.
  2016, ArXiv e-prints.
\newblock \doarXiv{1612.09302}

\bibitem[{Ezzeddine {et~al.}(2020)Ezzeddine, Rasmussen, Frebel, Chiti,
  Hinojisa, Placco, Ji, Beers, Hansen, Roederer, \& et~al.}]{rpa3}
Ezzeddine, R., Rasmussen, K., Frebel, A., {et~al.} 2020, The Astrophysical
  Journal, 898, 150, \dodoi{10.3847/1538-4357/ab9d1a}

\bibitem[{{Fekel} \& {Watson}(1998)}]{1998AJ....116.2466F}
{Fekel}, F.~C., \& {Watson}, L.~C. 1998, \aj, 116, 2466, \dodoi{10.1086/300614}

\bibitem[{{Fischer} \& {Valenti}(2005)}]{fischer2005}
{Fischer}, D.~A., \& {Valenti}, J. 2005, \apj, 622, 1102,
  \dodoi{10.1086/428383}

\bibitem[{{Fitzpatrick} {et~al.}(2024){Fitzpatrick}, {Placco}, {Bolton},
  {Merino}, {Ridgway}, \& {Stanghellini}}]{Fitzpatrick2024}
{Fitzpatrick}, M., {Placco}, V., {Bolton}, A., {et~al.} 2024, arXiv e-prints,
  arXiv:2401.01982, \dodoi{10.48550/arXiv.2401.01982}

\bibitem[{{Forestini} \& {Charbonnel}(1997)}]{1997A&AS..123..241F}
{Forestini}, M., \& {Charbonnel}, C. 1997, \aaps, 123, 241,
  \dodoi{10.1051/aas:1997348}

\bibitem[{{Gao} {et~al.}(2019{\natexlab{a}}){Gao}, {Shi}, {Yan}, {Yan},
  {Xiang}, {Zhou}, {Li}, \& {Zhao}}]{2019ApJS..245...33G}
{Gao}, Q., {Shi}, J.-R., {Yan}, H.-L., {et~al.} 2019{\natexlab{a}}, \apjs, 245,
  33, \dodoi{10.3847/1538-4365/ab505c}

\bibitem[{{Gao} {et~al.}(2019{\natexlab{b}}){Gao}, {Shi}, {Yan}, {Yan},
  {Xiang}, {Zhou}, {Li}, \& {Zhao}}]{gao2019}
---. 2019{\natexlab{b}}, \apjs, 245, 33, \dodoi{10.3847/1538-4365/ab505c}

\bibitem[{{Gao} {et~al.}(2020){Gao}, {Lind}, {Amarsi}, {Buder},
  {Bland-Hawthorn}, {Campbell}, {Asplund}, {Casey}, {de Silva}, {Freeman},
  {Hayden}, {Lewis}, {Martell}, {Simpson}, {Sharma}, {Zucker}, {Zwitter},
  {Horner}, {Munari}, {Nordlander}, {Stello}, {Ting}, {Traven}, {Wittenmyer},
  \& {GALAH Collaboration}}]{Gao2020}
{Gao}, X., {Lind}, K., {Amarsi}, A.~M., {et~al.} 2020, \mnras, 497, L30,
  \dodoi{10.1093/mnrasl/slaa109}

\bibitem[{{Gonzalez} {et~al.}(2009){Gonzalez}, {Zoccali}, {Monaco}, {Hill},
  {Cassisi}, {Minniti}, {Renzini}, {Barbuy}, {Ortolani}, \& {Gomez}}]{GONZALEZ}
{Gonzalez}, O.~A., {Zoccali}, M., {Monaco}, L., {et~al.} 2009, \aap, 508, 289,
  \dodoi{10.1051/0004-6361/200912469}

\bibitem[{{Gratton} \& {D'Antona}(1989)}]{GRATTON}
{Gratton}, R.~G., \& {D'Antona}, F. 1989, \aap, 215, 66

\bibitem[{{Gratton} {et~al.}(2000){Gratton}, {Sneden}, {Carretta}, \&
  {Bragaglia}}]{Gratton2000}
{Gratton}, R.~G., {Sneden}, C., {Carretta}, E., \& {Bragaglia}, A. 2000, \aap,
  354, 169

\bibitem[{{Gruyters} {et~al.}(2016){Gruyters}, {Lind}, {Richard}, {Grundahl},
  {Asplund}, {Casagrande}, {Charbonnel}, {Milone}, {Primas}, \&
  {Korn}}]{Gruyters2016}
{Gruyters}, P., {Lind}, K., {Richard}, O., {et~al.} 2016, \aap, 589, A61,
  \dodoi{10.1051/0004-6361/201527948}

\bibitem[{{Gustafsson} {et~al.}(1975){Gustafsson}, {Bell}, {Eriksson}, \&
  {Nordlund}}]{gustafsson1975}
{Gustafsson}, B., {Bell}, R.~A., {Eriksson}, K., \& {Nordlund}, A. 1975, \aap,
  42, 407

\bibitem[{{Gustafsson} {et~al.}(2008){Gustafsson}, {Edvardsson}, {Eriksson},
  {J{\o}rgensen}, {Nordlund}, \& {Plez}}]{gustafsson2008}
{Gustafsson}, B., {Edvardsson}, B., {Eriksson}, K., {et~al.} 2008, \aap, 486,
  951, \dodoi{10.1051/0004-6361:200809724}

\bibitem[{{Hanni}(1984)}]{HANNI}
{Hanni}, L. 1984, Soviet Astronomy Letters, 10, 51

\bibitem[{{Hansen} {et~al.}(2012){Hansen}, {Primas}, {Hartman}, {Kratz},
  {Wanajo}, {Leibundgut}, {Farouqi}, {Hallmann}, {Christlieb}, \&
  {Nilsson}}]{hansen2012}
{Hansen}, C.~J., {Primas}, F., {Hartman}, H., {et~al.} 2012, \aap, 545, A31,
  \dodoi{10.1051/0004-6361/201118643}

\bibitem[{{Hansen} {et~al.}(2018){Hansen}, {Holmbeck}, {Beers}, {Placco},
  {Roederer}, {Frebel}, {Sakari}, {Simon}, \& {Thompson}}]{rpa1}
{Hansen}, T.~T., {Holmbeck}, E.~M., {Beers}, T.~C., {et~al.} 2018, \apj, 858,
  92, \dodoi{10.3847/1538-4357/aabacc}

\bibitem[{{Harrington} \& {Garaud}(2019)}]{2019ApJ...870L...5H}
{Harrington}, P.~Z., \& {Garaud}, P. 2019, \apjl, 870, L5,
  \dodoi{10.3847/2041-8213/aaf812}

\bibitem[{{Harutyunyan} {et~al.}(2018){Harutyunyan}, {Steffen}, {Mott},
  {Caffau}, {Israelian}, {Gonz{\'a}lez Hern{\'a}ndez}, \&
  {Strassmeier}}]{Harutyunyan2018}
{Harutyunyan}, G., {Steffen}, M., {Mott}, A., {et~al.} 2018, \aap, 618, A16,
  \dodoi{10.1051/0004-6361/201832852}

\bibitem[{Hekker \& Meléndez(2007)}]{Hekker_2007}
Hekker, S., \& Meléndez, J. 2007, Astronomy \& Astrophysics, 475, 1003–1009,
  \dodoi{10.1051/0004-6361:20078233}

\bibitem[{{Hill} \& {Pasquini}(1999)}]{HILL}
{Hill}, V., \& {Pasquini}, L. 1999, \aap, 348, L21.
\newblock \doarXiv{astro-ph/9907106}

\bibitem[{{Holmbeck} {et~al.}(2020){Holmbeck}, {Hansen}, {Beers}, {Placco},
  {Whitten}, {Rasmussen}, {Roederer}, {Ezzeddine}, {Sakari}, {Frebel}, {Drout},
  {Simon}, {Thompson}, {Bland-Hawthorn}, {Gibson}, {Grebel}, {Kordopatis},
  {Kunder}, {Mel{\'e}ndez}, {Navarro}, {Reid}, {Seabroke}, {Steinmetz},
  {Watson}, \& {Wyse}}]{rpa4}
{Holmbeck}, E.~M., {Hansen}, T.~T., {Beers}, T.~C., {et~al.} 2020, \apjs, 249,
  30, \dodoi{10.3847/1538-4365/ab9c19}

\bibitem[{Hunter(2007)}]{matplotlib}
Hunter, J.~D. 2007, Computing in Science \& Engineering, 9, 90,
  \dodoi{10.1109/MCSE.2007.55}

\bibitem[{{Iben}(1967)}]{Icko1967}
{Iben}, Icko, J. 1967, \apj, 147, 624, \dodoi{10.1086/149040}

\bibitem[{{Jasniewicz} {et~al.}(1999{\natexlab{a}}){Jasniewicz},
  {Parthasarathy}, {de Laverny}, \& {Th{\'e}venin}}]{Jasniewicz1999}
{Jasniewicz}, G., {Parthasarathy}, M., {de Laverny}, P., \& {Th{\'e}venin}, F.
  1999{\natexlab{a}}, \aap, 342, 831

\bibitem[{{Jasniewicz} {et~al.}(1999{\natexlab{b}}){Jasniewicz},
  {Parthasarathy}, {de Laverny}, \& {Th{\'e}venin}}]{JASN}
---. 1999{\natexlab{b}}, \aap, 342, 831

\bibitem[{{Ji} {et~al.}(2016){Ji}, {Frebel}, {Simon}, \& {Chiti}}]{ji2016b}
{Ji}, A.~P., {Frebel}, A., {Simon}, J.~D., \& {Chiti}, A. 2016, The
  Astrophysical Journal, 830, 93, \dodoi{10.3847/0004-637X/830/2/93}

\bibitem[{{Jofr{\'e}} {et~al.}(2015){Jofr{\'e}}, {Petrucci}, {Garc{\'\i}a}, \&
  {G{\'o}mez}}]{JOFRE2}
{Jofr{\'e}}, E., {Petrucci}, R., {Garc{\'\i}a}, L., \& {G{\'o}mez}, M. 2015,
  \aap, 584, L3, \dodoi{10.1051/0004-6361/201527337}

\bibitem[{{Keeping}(1962)}]{keeping1962}
{Keeping}, E.~S. 1962, Princeton, NJ: Van Nostrand

\bibitem[{{Kelson}(2003)}]{kelson2003}
{Kelson}, D.~D. 2003, \pasp, 115, 688, \dodoi{10.1086/375502}

\bibitem[{{Kirby} {et~al.}(2012){Kirby}, {Fu}, {Guhathakurta}, \&
  {Deng}}]{KIRBY2}
{Kirby}, E.~N., {Fu}, X., {Guhathakurta}, P., \& {Deng}, L. 2012, \apjl, 752,
  L16, \dodoi{10.1088/2041-8205/752/1/L16}

\bibitem[{{Kirby} {et~al.}(2016){Kirby}, {Guhathakurta}, {Zhang}, {Hong},
  {Guo}, {Guo}, {Cohen}, \& {Cunha}}]{KIRBY3}
{Kirby}, E.~N., {Guhathakurta}, P., {Zhang}, A.~J., {et~al.} 2016, \apj, 819,
  135, \dodoi{10.3847/0004-637X/819/2/135}

\bibitem[{{Korn} {et~al.}(2006){Korn}, {Grundahl}, {Richard}, {Barklem},
  {Mashonkina}, {Collet}, {Piskunov}, \& {Gustafsson}}]{Korn2006}
{Korn}, A.~J., {Grundahl}, F., {Richard}, O., {et~al.} 2006, \nat, 442, 657,
  \dodoi{10.1038/nature05011}

\bibitem[{{Kramida} {et~al.}(2021){Kramida}, {Ralchenko}, {Reader}, \& {NIST
  ASD Team}}]{kramida21}
{Kramida}, A., {Ralchenko}, Y., {Reader}, J., \& {NIST ASD Team}. 2021, NIST
  Atomic Spectra Database (ver. 5.9), [Online]. Available:
  {\tt{https://physics.nist.gov/asd}},  National Institute of Standards and
  Technology, Gaithersburg, MD.

\bibitem[{{Kumar} \& {Reddy}(2020)}]{2020JApA...41...49K}
{Kumar}, Y.~B., \& {Reddy}, B.~E. 2020, Journal of Astrophysics and Astronomy,
  41, 49, \dodoi{10.1007/s12036-020-09660-9}

\bibitem[{{Kumar} {et~al.}(2011){Kumar}, {Reddy}, \& {Lambert}}]{KUMAR2015}
{Kumar}, Y.~B., {Reddy}, B.~E., \& {Lambert}, D.~L. 2011, \apjl, 730, L12,
  \dodoi{10.1088/2041-8205/730/1/L12}

\bibitem[{Kumar {et~al.}(2011)Kumar, Reddy, \& Lambert}]{Kumar2011}
Kumar, Y.~B., Reddy, B.~E., \& Lambert, D.~L. 2011, The Astrophysical Journal,
  730, L12, \dodoi{10.1088/2041-8205/730/1/l12}

\bibitem[{{Lagarde} {et~al.}(2012){Lagarde}, {Decressin}, {Charbonnel},
  {Eggenberger}, {Ekstr{\"o}m}, \& {Palacios}}]{2012A&A...543A.108L}
{Lagarde}, N., {Decressin}, T., {Charbonnel}, C., {et~al.} 2012, \aap, 543,
  A108, \dodoi{10.1051/0004-6361/201118331}

\bibitem[{{Lagarde} {et~al.}(2019){Lagarde}, {Reyl{\'e}}, {Robin},
  {Tautvai{\v{s}}ien{\.{e}}}, {Drazdauskas}, {Mikolaitis},
  {Minkevi{\v{c}}i{\={u}}t{\.{e}}}, {Stonkut{\.{e}}}, {Chorniy}, {Bagdonas},
  {Miglio}, {Nasello}, {Gilmore}, {Randich}, {Bensby}, {Bragaglia},
  {Flaccomio}, {Francois}, {Korn}, {Pancino}, {Smiljanic}, {Bayo}, {Carraro},
  {Costado}, {Jim{\'e}nez-Esteban}, {Jofr{\'e}}, {Martell}, {Masseron},
  {Monaco}, {Morbidelli}, {Sbordone}, {Sousa}, \&
  {Zaggia}}]{2019A&A...621A..24L}
{Lagarde}, N., {Reyl{\'e}}, C., {Robin}, A.~C., {et~al.} 2019, \aap, 621, A24,
  \dodoi{10.1051/0004-6361/201732433}

\bibitem[{{L{\`e}bre} {et~al.}(1999){L{\`e}bre}, {de Laverny}, {de Medeiros},
  {Charbonnel}, \& {da Silva}}]{1999A&A...345..936L}
{L{\`e}bre}, A., {de Laverny}, P., {de Medeiros}, J.~R., {Charbonnel}, C., \&
  {da Silva}, L. 1999, \aap, 345, 936

\bibitem[{{L{\`e}bre} {et~al.}(2009){L{\`e}bre}, {Palacios}, {Do Nascimento},
  {Konstantinova-Antova}, {Kolev}, {Auri{\`e}re}, {de Laverny}, \& {de
  Medeiros}}]{LEBRE}
{L{\`e}bre}, A., {Palacios}, A., {Do Nascimento}, J.~D., J., {et~al.} 2009,
  \aap, 504, 1011, \dodoi{10.1051/0004-6361/200912038}

\bibitem[{{Li} \& {Ezzeddine}(2022)}]{lotus}
{Li}, Y., \& {Ezzeddine}, R. 2022, {LOTUS: 1D Non-LTE stellar parameter
  determination via Equivalent Width method}, Astrophysics Source Code Library,
  record ascl:2207.017.
\newblock \doeprint{2207.017}

\bibitem[{{Li} \& {Ezzeddine}(2023)}]{lotus2023}
---. 2023, \aj, 165, 145, \dodoi{10.3847/1538-3881/acb7f0}

\bibitem[{{Lind} {et~al.}(2011){Lind}, {Asplund}, {Barklem}, \&
  {Belyaev}}]{lind2011}
{Lind}, K., {Asplund}, M., {Barklem}, P.~S., \& {Belyaev}, A.~K. 2011, \aap,
  528, A103, \dodoi{10.1051/0004-6361/201016095}

\bibitem[{{Lind} {et~al.}(2012){Lind}, {Bergemann}, \& {Asplund}}]{lind2012}
{Lind}, K., {Bergemann}, M., \& {Asplund}, M. 2012, \mnras, 427, 50,
  \dodoi{10.1111/j.1365-2966.2012.21686.x}

\bibitem[{{Lind} {et~al.}(2013){Lind}, {Melendez}, {Asplund}, {Collet}, \&
  {Magic}}]{lind2013}
{Lind}, K., {Melendez}, J., {Asplund}, M., {Collet}, R., \& {Magic}, Z. 2013,
  \aap, 554, A96, \dodoi{10.1051/0004-6361/201321406}

\bibitem[{{Lind} {et~al.}(2009){Lind}, {Primas}, {Charbonnel}, {Grundahl}, \&
  {Asplund}}]{Lind2009}
{Lind}, K., {Primas}, F., {Charbonnel}, C., {Grundahl}, F., \& {Asplund}, M.
  2009, \aap, 503, 545, \dodoi{10.1051/0004-6361/200912524}

\bibitem[{{Lindegren} {et~al.}(2021){Lindegren}, {Bastian}, {Biermann},
  {Bombrun}, {de Torres}, {Gerlach}, {Geyer}, {Hern{\'a}ndez}, {Hilger},
  {Hobbs}, {Klioner}, {Lammers}, {McMillan}, {Ramos-Lerate},
  {Steidelm{\"u}ller}, {Stephenson}, \& {van Leeuwen}}]{Lindenberg2021}
{Lindegren}, L., {Bastian}, U., {Biermann}, M., {et~al.} 2021, \aap, 649, A4,
  \dodoi{10.1051/0004-6361/202039653}

\bibitem[{{Liu} {et~al.}(2014{\natexlab{a}}){Liu}, {Tan}, {Wang}, {Zhao},
  {Sato}, {Takeda}, \& {Li}}]{2014ApJ...785...94L}
{Liu}, Y.~J., {Tan}, K.~F., {Wang}, L., {et~al.} 2014{\natexlab{a}}, \apj, 785,
  94, \dodoi{10.1088/0004-637X/785/2/94}

\bibitem[{{Liu} {et~al.}(2014{\natexlab{b}}){Liu}, {Tan}, {Wang}, {Zhao},
  {Sato}, {Takeda}, \& {Li}}]{LIU}
---. 2014{\natexlab{b}}, \apj, 785, 94, \dodoi{10.1088/0004-637X/785/2/94}

\bibitem[{{Luck}(1982)}]{LUCK}
{Luck}, R.~E. 1982, \pasp, 94, 811, \dodoi{10.1086/131068}

\bibitem[{{Lyubimkov}(2016)}]{2016Ap.....59..411L}
{Lyubimkov}, L.~S. 2016, Astrophysics, 59, 411,
  \dodoi{10.1007/s10511-016-9446-5}

\bibitem[{{Lyubimkov} {et~al.}(2012){Lyubimkov}, {Lambert}, {Kaminsky},
  {Pavlenko}, {Poklad}, \& {Rachkovskaya}}]{2012MNRAS.427...11L}
{Lyubimkov}, L.~S., {Lambert}, D.~L., {Kaminsky}, B.~M., {et~al.} 2012, \mnras,
  427, 11, \dodoi{10.1111/j.1365-2966.2012.21617.x}

\bibitem[{{Magrini} {et~al.}(2021{\natexlab{a}}){Magrini}, {Lagarde},
  {Charbonnel}, {Franciosini}, {Randich}, {Smiljanic}, {Casali}, {Viscasillas
  V{\'a}zquez}, {Spina}, {Biazzo}, {Pasquini}, {Bragaglia}, {Van der Swaelmen},
  {Tautvai{\v{s}}ien{\.{e}}}, {Inno}, {Sanna}, {Prisinzano}, {Degl'Innocenti},
  {Prada Moroni}, {Roccatagliata}, {Tognelli}, {Monaco}, {de Laverny},
  {Delgado-Mena}, {Baratella}, {D'Orazi}, {Vallenari}, {Gonneau}, {Worley},
  {Jim{\'e}nez-Esteban}, {Jofre}, {Bensby}, {Fran{\c{c}}ois}, {Guiglion},
  {Bayo}, {Jeffries}, {Binks}, {Gilmore}, {Damiani}, {Korn}, {Pancino},
  {Sacco}, {Hourihane}, {Morbidelli}, \& {Zaggia}}]{2021A&A...651A..84M}
{Magrini}, L., {Lagarde}, N., {Charbonnel}, C., {et~al.} 2021{\natexlab{a}},
  \aap, 651, A84, \dodoi{10.1051/0004-6361/202140935}

\bibitem[{{Magrini} {et~al.}(2021{\natexlab{b}}){Magrini}, {Smiljanic},
  {Franciosini}, {Pasquini}, {Randich}, {Casali}, {Viscasillas V{\'a}zquez},
  {Bragaglia}, {Spina}, {Biazzo}, {Tautvai{\v{s}}ien{\.{e}}}, {Masseron}, {Van
  der Swaelmen}, {Pancino}, {Jim{\'e}nez-Esteban}, {Guiglion}, {Martell},
  {Bensby}, {D'Orazi}, {Baratella}, {Korn}, {Jofre}, {Gilmore}, {Worley},
  {Hourihane}, {Gonneau}, {Sacco}, \& {Morbidelli}}]{2021A&A...655A..23M}
{Magrini}, L., {Smiljanic}, R., {Franciosini}, E., {et~al.} 2021{\natexlab{b}},
  \aap, 655, A23, \dodoi{10.1051/0004-6361/202141275}

\bibitem[{{Mallick} {et~al.}(2022){Mallick}, {Reddy}, \&
  {Muthumariappan}}]{mallik2022}
{Mallick}, A., {Reddy}, B.~E., \& {Muthumariappan}, C. 2022, \mnras, 511, 3741,
  \dodoi{10.1093/mnras/stac224}

\bibitem[{Martell {et~al.}(2020)Martell, Simpson, Balasubramaniam, Buder,
  Sharma, Hon, Stello, Ting, Asplund, Bland-Hawthorn, Silva, Freeman, Hayden,
  Kos, Lewis, Lind, Zucker, Zwitter, Campbell, Cotar, Horner, Montet, \&
  Wittenmyer}]{martell2020galah}
Martell, S., Simpson, J., Balasubramaniam, A., {et~al.} 2020, The GALAH survey:
  Lithium-rich giant stars require multiple formation channels.
\newblock \doarXiv{2006.02106}

\bibitem[{{Martell} \& {Shetrone}(2013)}]{MARTELL2013}
{Martell}, S.~L., \& {Shetrone}, M.~D. 2013, \mnras, 430, 611,
  \dodoi{10.1093/mnras/sts661}

\bibitem[{Martell {et~al.}(2021)Martell, Simpson, Balasubramaniam, Buder,
  Sharma, Hon, Stello, Ting, Asplund, Bland-Hawthorn, \& et~al.}]{Martell_2021}
Martell, S.~L., Simpson, J.~D., Balasubramaniam, A.~G., {et~al.} 2021, Monthly
  Notices of the Royal Astronomical Society, \dodoi{10.1093/mnras/stab1356}

\bibitem[{{Mashonkina} {et~al.}(2016){Mashonkina}, {Sitnova}, \&
  {Pakhomov}}]{mashonkina2016}
{Mashonkina}, L.~I., {Sitnova}, T.~N., \& {Pakhomov}, Y.~V. 2016, Astronomy
  Letters, 42, 606, \dodoi{10.1134/S1063773716080028}

\bibitem[{{Masseron} {et~al.}(2014){Masseron}, {Plez}, {Van Eck}, {Colin},
  {Daoutidis}, {Godefroid}, {Coheur}, {Bernath}, {Jorissen}, \&
  {Christlieb}}]{masseron2014}
{Masseron}, T., {Plez}, B., {Van Eck}, S., {et~al.} 2014, \aap, 571, A47,
  \dodoi{10.1051/0004-6361/201423956}

\bibitem[{{McWilliam} \& {Rich}(1994)}]{MCWILLIAM}
{McWilliam}, A., \& {Rich}, R.~M. 1994, \apjs, 91, 749, \dodoi{10.1086/191954}

\bibitem[{{Monaco} {et~al.}(2011{\natexlab{a}}){Monaco}, {Villanova}, {Moni
  Bidin}, {Carraro}, {Geisler}, {Bonifacio}, {Gonzalez}, {Zoccali}, \&
  {Jilkova}}]{Monaco2011}
{Monaco}, L., {Villanova}, S., {Moni Bidin}, C., {et~al.} 2011{\natexlab{a}},
  \aap, 529, A90, \dodoi{10.1051/0004-6361/201016285}

\bibitem[{{Monaco} {et~al.}(2011{\natexlab{b}}){Monaco}, {Villanova}, {Moni
  Bidin}, {Carraro}, {Geisler}, {Bonifacio}, {Gonzalez}, {Zoccali}, \&
  {Jilkova}}]{MONACO}
---. 2011{\natexlab{b}}, \aap, 529, A90, \dodoi{10.1051/0004-6361/201016285}

\bibitem[{{Monaco} {et~al.}(2014){Monaco}, {Boffin}, {Bonifacio}, {Villanova},
  {Carraro}, {Caffau}, {Steffen}, {Ahumada}, {Beletsky}, \&
  {Beccari}}]{MONACO3}
{Monaco}, L., {Boffin}, H.~M.~J., {Bonifacio}, P., {et~al.} 2014, \aap, 564,
  L6, \dodoi{10.1051/0004-6361/201323348}

\bibitem[{{Mori} {et~al.}(2021){Mori}, {Kusakabe}, {Balantekin}, {Kajino}, \&
  {Famiano}}]{mori2021}
{Mori}, K., {Kusakabe}, M., {Balantekin}, A.~B., {Kajino}, T., \& {Famiano},
  M.~A. 2021, \mnras, 503, 2746, \dodoi{10.1093/mnras/stab595}

\bibitem[{{Mucciarelli} {et~al.}(2022){Mucciarelli}, {Monaco}, {Bonifacio},
  {Salaris}, {Deal}, {Spite}, {Richard}, \& {Lallement}}]{2022A&A...661A.153M}
{Mucciarelli}, A., {Monaco}, L., {Bonifacio}, P., {et~al.} 2022, \aap, 661,
  A153, \dodoi{10.1051/0004-6361/202142889}

\bibitem[{{Nordlander} {et~al.}(2012){Nordlander}, {Korn}, {Richard}, \&
  {Lind}}]{Nordlander2012}
{Nordlander}, T., {Korn}, A.~J., {Richard}, O., \& {Lind}, K. 2012, \apj, 753,
  48, \dodoi{10.1088/0004-637X/753/1/48}

\bibitem[{{Nordlander} \& {Lind}(2017)}]{nordlander2017}
{Nordlander}, T., \& {Lind}, K. 2017, \aap, 607, A75,
  \dodoi{10.1051/0004-6361/201730427}

\bibitem[{{Osorio} {et~al.}(2015){Osorio}, {Barklem}, {Lind}, {Belyaev},
  {Spielfiedel}, {Guitou}, \& {Feautrier}}]{osorio2015}
{Osorio}, Y., {Barklem}, P.~S., {Lind}, K., {et~al.} 2015, \aap, 579, A53,
  \dodoi{10.1051/0004-6361/201525846}

\bibitem[{{Palacios} {et~al.}(2001){Palacios}, {Charbonnel}, \&
  {Forestini}}]{Palacios2001}
{Palacios}, A., {Charbonnel}, C., \& {Forestini}, M. 2001, \aap, 375, L9,
  \dodoi{10.1051/0004-6361:20010903}

\bibitem[{{Placco} {et~al.}(2014){Placco}, {Frebel}, {Beers}, \&
  {Stancliffe}}]{placco2014}
{Placco}, V.~M., {Frebel}, A., {Beers}, T.~C., \& {Stancliffe}, R.~J. 2014,
  \apj, 797, 21, \dodoi{10.1088/0004-637X/797/1/21}

\bibitem[{{Placco} {et~al.}(2021){Placco}, {Sneden}, {Roederer}, {Lawler}, {Den
  Hartog}, {Hejazi}, {Maas}, \& {Bernath}}]{2021RNAAS...5...92P}
{Placco}, V.~M., {Sneden}, C., {Roederer}, I.~U., {et~al.} 2021, Research Notes
  of the American Astronomical Society, 5, 92, \dodoi{10.3847/2515-5172/abf651}

\bibitem[{{Rebull} {et~al.}(2015){Rebull}, {Carlberg}, {Gibbs}, {Deeb},
  {Larsen}, {Black}, {Altepeter}, {Bucksbee}, {Cashen}, {Clarke}, {Datta},
  {Hodgson}, \& {Lince}}]{rebull2015}
{Rebull}, L.~M., {Carlberg}, J.~K., {Gibbs}, J.~C., {et~al.} 2015, \aj, 150,
  123, \dodoi{10.1088/0004-6256/150/4/123}

\bibitem[{{Reimers}(1975)}]{1975MSRSL...8..369R}
{Reimers}, D. 1975, Memoires of the Societe Royale des Sciences de Liege, 8,
  369

\bibitem[{{Reyniers} \& {Van Winckel}(2001)}]{REYN}
{Reyniers}, M., \& {Van Winckel}, H. 2001, \aap, 365, 465,
  \dodoi{10.1051/0004-6361:20000146}

\bibitem[{{Richard} {et~al.}(2005){Richard}, {Michaud}, \&
  {Richer}}]{Richard2005}
{Richard}, O., {Michaud}, G., \& {Richer}, J. 2005, \apj, 619, 538,
  \dodoi{10.1086/426470}

\bibitem[{{Roederer} {et~al.}(2018){Roederer}, {Hattori}, \&
  {Valluri}}]{roederer2018}
{Roederer}, I.~U., {Hattori}, K., \& {Valluri}, M. 2018, \aj, 156, 179,
  \dodoi{10.3847/1538-3881/aadd9c}

\bibitem[{{Roederer} {et~al.}(2014){Roederer}, {Preston}, {Thompson},
  {Shectman}, {Sneden}, {Burley}, \& {Kelson}}]{roederer2014b}
{Roederer}, I.~U., {Preston}, G.~W., {Thompson}, I.~B., {et~al.} 2014, \aj,
  147, 136, \dodoi{10.1088/0004-6256/147/6/136}

\bibitem[{Roederer {et~al.}(2014)Roederer, Preston, Thompson, Shectman, Sneden,
  Burley, \& Kelson}]{Roederer_2014}
Roederer, I.~U., Preston, G.~W., Thompson, I.~B., {et~al.} 2014, The
  Astronomical Journal, 147, 136, \dodoi{10.1088/0004-6256/147/6/136}

\bibitem[{Roederer {et~al.}(2018)Roederer, Sakari, Placco, Beers, Ezzeddine,
  Frebel, \& Hansen}]{Roederer_2018}
Roederer, I.~U., Sakari, C.~M., Placco, V.~M., {et~al.} 2018, The Astrophysical
  Journal, 865, 129, \dodoi{10.3847/1538-4357/aadd92}

\bibitem[{{Ruchti} {et~al.}(2011){Ruchti}, {Fulbright}, {Wyse}, {Gilmore},
  {Grebel}, {Bienaym{\'e}}, {Bland-Hawthorn}, {Freeman}, {Gibson}, {Munari},
  {Navarro}, {Parker}, {Reid}, {Seabroke}, {Siebert}, {Siviero}, {Steinmetz},
  {Watson}, {Williams}, \& {Zwitter}}]{RUCHTI}
{Ruchti}, G.~R., {Fulbright}, J.~P., {Wyse}, R. F.~G., {et~al.} 2011, \apj,
  743, 107, \dodoi{10.1088/0004-637X/743/2/107}

\bibitem[{{Sackmann} \& {Boothroyd}(1992)}]{1992ApJ...392L..71S}
{Sackmann}, I.~J., \& {Boothroyd}, A.~I. 1992, \apjl, 392, L71,
  \dodoi{10.1086/186428}

\bibitem[{{Sackmann} \& {Boothroyd}(1999)}]{1999ApJ...510..217S}
---. 1999, \apj, 510, 217, \dodoi{10.1086/306545}

\bibitem[{{Sakari} {et~al.}(2018){Sakari}, {Placco}, {Farrell}, {Roederer},
  {Wallerstein}, {Beers}, {Ezzeddine}, {Frebel}, {Hansen}, {Holmbeck},
  {Sneden}, {Cowan}, {Venn}, {Davis}, {Matijevi{\v{c}}}, {Wyse},
  {Bland-Hawthorn}, {Chiappini}, {Freeman}, {Gibson}, {Grebel}, {Helmi},
  {Kordopatis}, {Kunder}, {Navarro}, {Reid}, {Seabroke}, {Steinmetz}, \&
  {Watson}}]{rpa2}
{Sakari}, C.~M., {Placco}, V.~M., {Farrell}, E.~M., {et~al.} 2018, \apj, 868,
  110, \dodoi{10.3847/1538-4357/aae9df}

\bibitem[{{Sayeed} {et~al.}(2023){Sayeed}, {Ness}, {Montet}, {Cantiello},
  {Casey}, {Buder}, {Bedell}, {Breivik}, {Metzger}, {Martell}, \&
  {McGee-Gold}}]{sayeed2023}
{Sayeed}, M., {Ness}, M.~K., {Montet}, B.~T., {et~al.} 2023, arXiv e-prints,
  arXiv:2306.03323, \dodoi{10.48550/arXiv.2306.03323}

\bibitem[{{Schlafly} \& {Finkbeiner}(2011)}]{Fink}
{Schlafly}, E.~F., \& {Finkbeiner}, D.~P. 2011, \apj, 737, 103,
  \dodoi{10.1088/0004-637X/737/2/103}

\bibitem[{{Shappee} {et~al.}(2014){Shappee}, {Prieto}, {Stanek}, {Kochanek},
  {Holoien}, {Jencson}, {Basu}, {Beacom}, {Szczygiel}, {Pojmanski},
  {Brimacombe}, {Dubberley}, {Elphick}, {Foale}, {Hawkins}, {Mullins},
  {Rosing}, {Ross}, \& {Walker}}]{shappee2014}
{Shappee}, B., {Prieto}, J., {Stanek}, K.~Z., {et~al.} 2014, in American
  Astronomical Society Meeting Abstracts, Vol. 223, American Astronomical
  Society Meeting Abstracts \#223, 236.03

\bibitem[{{Siess} \& {Livio}(1999)}]{Siess1999}
{Siess}, L., \& {Livio}, M. 1999, \mnras, 308, 1133,
  \dodoi{10.1046/j.1365-8711.1999.02784.x}

\bibitem[{{Silva Aguirre} {et~al.}(2014){Silva Aguirre}, {Ruchti}, {Hekker},
  {Cassisi}, {Christensen-Dalsgaard}, {Datta}, {Jendreieck}, {Jessen-Hansen},
  {Mazumdar}, {Mosser}, {Stello}, {Beck}, \& {de Ridder}}]{2014ApJ...784L..16S}
{Silva Aguirre}, V., {Ruchti}, G.~R., {Hekker}, S., {et~al.} 2014, \apjl, 784,
  L16, \dodoi{10.1088/2041-8205/784/1/L16}

\bibitem[{{Singh} {et~al.}(2019){Singh}, {Reddy}, \&
  {Kumar}}]{2019MNRAS.482.3822S}
{Singh}, R., {Reddy}, B.~E., \& {Kumar}, Y.~B. 2019, \mnras, 482, 3822,
  \dodoi{10.1093/mnras/sty2939}

\bibitem[{{Smiljanic} {et~al.}(2018){Smiljanic}, {Franciosini}, {Bragaglia},
  {Tautvai{\v{s}}ien{\.{e}}}, {Fu}, {Pancino}, {Adibekyan}, {Sousa}, {Randich},
  {Montalb{\'a}n}, {Pasquini}, {Magrini}, {Drazdauskas}, {Garc{\'\i}a},
  {Mathur}, {Mosser}, {R{\'e}gulo}, {de Assis Peralta}, {Hekker}, {Feuillet},
  {Valentini}, {Morel}, {Martell}, {Gilmore}, {Feltzing}, {Vallenari},
  {Bensby}, {Korn}, {Lanzafame}, {Recio-Blanco}, {Bayo}, {Carraro}, {Costado},
  {Frasca}, {Jofr{\'e}}, {Lardo}, {de Laverny}, {Lind}, {Masseron}, {Monaco},
  {Morbidelli}, {Prisinzano}, {Sbordone}, \& {Zaggia}}]{2018A&A...617A...4S}
{Smiljanic}, R., {Franciosini}, E., {Bragaglia}, A., {et~al.} 2018, \aap, 617,
  A4, \dodoi{10.1051/0004-6361/201833027}

\bibitem[{{Smith} {et~al.}(1999){Smith}, {Shetrone}, \& {Keane}}]{SMITH}
{Smith}, V.~V., {Shetrone}, M.~D., \& {Keane}, M.~J. 1999, \apjl, 516, L73,
  \dodoi{10.1086/312011}

\bibitem[{{Sneden} {et~al.}(2008){Sneden}, {Cowan}, \& {Gallino}}]{sneden2008}
{Sneden}, C., {Cowan}, J.~J., \& {Gallino}, R. 2008, \araa, 46, 241,
  \dodoi{10.1146/annurev.astro.46.060407.145207}

\bibitem[{Sneden {et~al.}(2008)Sneden, Cowan, \&
  Gallino}]{doi:10.1146/annurev.astro.46.060407.145207}
Sneden, C., Cowan, J.~J., \& Gallino, R. 2008, Annual Review of Astronomy and
  Astrophysics, 46, 241, \dodoi{10.1146/annurev.astro.46.060407.145207}

\bibitem[{{Sneden}(1973)}]{sneden1973}
{Sneden}, C.~A. 1973, PhD thesis, University of Texas at Austin

\bibitem[{{Sobeck} {et~al.}(2011){Sobeck}, {Kraft}, {Sneden}, {Preston},
  {Cowan}, {Smith}, {Thompson}, {Shectman}, \& {Burley}}]{sobeck2011}
{Sobeck}, J.~S., {Kraft}, R.~P., {Sneden}, C., {et~al.} 2011, \aj, 141, 175,
  \dodoi{10.1088/0004-6256/141/6/175}

\bibitem[{{Soderblom} {et~al.}(1993){Soderblom}, {Fedele}, {Jones}, {Stauffer},
  \& {Prosser}}]{1993AJ....106.1080S}
{Soderblom}, D.~R., {Fedele}, S.~B., {Jones}, B.~F., {Stauffer}, J.~R., \&
  {Prosser}, C.~F. 1993, \aj, 106, 1080, \dodoi{10.1086/116705}

\bibitem[{{Strassmeier} {et~al.}(2015){Strassmeier}, {Carroll}, {Weber}, \&
  {Granzer}}]{2015A&A...574A..31S}
{Strassmeier}, K.~G., {Carroll}, T.~A., {Weber}, M., \& {Granzer}, T. 2015,
  \aap, 574, A31, \dodoi{10.1051/0004-6361/201424130}

\bibitem[{{Susmitha} {et~al.}(2024){Susmitha}, {Mallick}, \&
  {Reddy}}]{susmitha2024}
{Susmitha}, A., {Mallick}, A., \& {Reddy}, B.~E. 2024, \apj, 966, 109,
  \dodoi{10.3847/1538-4357/ad35b9}

\bibitem[{{Takeda} {et~al.}(2002){Takeda}, {Zhao}, {Chen}, {Qiu}, \&
  {Takada-Hidai}}]{takeda2002}
{Takeda}, Y., {Zhao}, G., {Chen}, Y.-Q., {Qiu}, H.-M., \& {Takada-Hidai}, M.
  2002, \pasj, 54, 275, \dodoi{10.1093/pasj/54.2.275}

\bibitem[{{Talon} \& {Charbonnel}(2010)}]{2010IAUS..268..365T}
{Talon}, S., \& {Charbonnel}, C. 2010, in Light Elements in the Universe, ed.
  C.~{Charbonnel}, M.~{Tosi}, F.~{Primas}, \& C.~{Chiappini}, Vol. 268,
  365--374, \dodoi{10.1017/S1743921310004485}

\bibitem[{{Tayar} {et~al.}(2015){Tayar}, {Ceillier},
  {Garc{\'\i}a-Hern{\'a}ndez}, {Troup}, {Mathur}, {Garc{\'\i}a}, {Zamora},
  {Johnson}, {Pinsonneault}, {M{\'e}sz{\'a}ros}, {Allende Prieto}, {Chaplin},
  {Elsworth}, {Hekker}, {Nidever}, {Salabert}, {Schneider}, {Serenelli},
  {Shetrone}, \& {Stello}}]{tayar2015A}
{Tayar}, J., {Ceillier}, T., {Garc{\'\i}a-Hern{\'a}ndez}, D.~A., {et~al.} 2015,
  \apj, 807, 82, \dodoi{10.1088/0004-637X/807/1/82}

\bibitem[{{Tody}(1986)}]{tody}
{Tody}, D. 1986, in Society of Photo-Optical Instrumentation Engineers (SPIE)
  Conference Series, Vol. 627, Instrumentation in astronomy VI, ed. D.~L.
  {Crawford}, 733, \dodoi{10.1117/12.968154}

\bibitem[{{Tody}(1993)}]{iraf}
{Tody}, D. 1993, Astronomical Society of the Pacific Conference Series,
  Vol.~52, {IRAF in the Nineties}, ed. R.~J. {Hanisch}, R.~J.~V. {Brissenden},
  \& J.~{Barnes}, 173

\bibitem[{{Tognelli} {et~al.}(2021){Tognelli}, {Degl'Innocenti}, {Prada
  Moroni}, {Lamia}, {Pizzone}, {Tumino}, {Spitaleri}, \&
  {Chiavassa}}]{2021FrASS...8...22T}
{Tognelli}, E., {Degl'Innocenti}, S., {Prada Moroni}, P.~G., {et~al.} 2021,
  Frontiers in Astronomy and Space Sciences, 8, 22,
  \dodoi{10.3389/fspas.2021.604872}

\bibitem[{{Vassiliadis} \& {Wood}(1993)}]{1993ApJ...413..641V}
{Vassiliadis}, E., \& {Wood}, P.~R. 1993, \apj, 413, 641,
  \dodoi{10.1086/173033}

\bibitem[{{Wallerstein} \& {Sneden}(1982{\natexlab{a}})}]{Wallerstein1982}
{Wallerstein}, G., \& {Sneden}, C. 1982{\natexlab{a}}, \apj, 255, 577,
  \dodoi{10.1086/159859}

\bibitem[{{Wallerstein} \& {Sneden}(1982{\natexlab{b}})}]{WALLER}
---. 1982{\natexlab{b}}, \apj, 255, 577, \dodoi{10.1086/159859}

\bibitem[{{Wang} {et~al.}(2021){Wang}, {Nordlander}, {Asplund}, {Amarsi},
  {Lind}, \& {Zhou}}]{wang2021}
{Wang}, E.~X., {Nordlander}, T., {Asplund}, M., {et~al.} 2021, \mnras, 500,
  2159, \dodoi{10.1093/mnras/staa3381}

\bibitem[{{Wang} {et~al.}(2022){Wang}, {Nordlander}, {Asplund}, {Lind}, {Zhou},
  \& {Reggiani}}]{Wang2022}
---. 2022, \mnras, 509, 1521, \dodoi{10.1093/mnras/stab2924}

\bibitem[{{Wenger} {et~al.}(2000){Wenger}, {Ochsenbein}, {Egret}, {Dubois},
  {Bonnarel}, {Borde}, {Genova}, {Jasniewicz}, {Lalo{\"e}}, {Lesteven}, \&
  {Monier}}]{simbad}
{Wenger}, M., {Ochsenbein}, F., {Egret}, D., {et~al.} 2000, \aaps, 143, 9,
  \dodoi{10.1051/aas:2000332}

\bibitem[{{Wright} {et~al.}(2010){Wright}, {Eisenhardt}, {Mainzer}, {Ressler},
  {Cutri}, {Jarrett}, {Kirkpatrick}, {Padgett}, {McMillan}, {Skrutskie},
  {Stanford}, {Cohen}, {Walker}, {Mather}, {Leisawitz}, {Gautier}, {McLean},
  {Benford}, {Lonsdale}, {Blain}, {Mendez}, {Irace}, {Duval}, {Liu}, {Royer},
  {Heinrichsen}, {Howard}, {Shannon}, {Kendall}, {Walsh}, {Larsen}, {Cardon},
  {Schick}, {Schwalm}, {Abid}, {Fabinsky}, {Naes}, \& {Tsai}}]{WISE}
{Wright}, E.~L., {Eisenhardt}, P. R.~M., {Mainzer}, A.~K., {et~al.} 2010, \aj,
  140, 1868, \dodoi{10.1088/0004-6256/140/6/1868}

\bibitem[{{Yan} {et~al.}(2018){Yan}, {Shi}, {Zhou}, {Chen}, {Li}, {Zhang},
  {Bi}, {Wu}, {Li}, {Guo}, {Liu}, {Gao}, {Zhang}, {Zhou}, {Li}, \&
  {Zhao}}]{Yan2018}
{Yan}, H.-L., {Shi}, J.-R., {Zhou}, Y.-T., {et~al.} 2018, Nature Astronomy, 2,
  790, \dodoi{10.1038/s41550-018-0544-7}

\bibitem[{Yan {et~al.}(2020)Yan, Zhou, Zhang, Li, Gao, Shi, Zhao, Aoki,
  Matsuno, Li, \& et~al.}]{Yan_2020}
Yan, H.-L., Zhou, Y.-T., Zhang, X., {et~al.} 2020, Nature Astronomy, 5,
  86–93, \dodoi{10.1038/s41550-020-01217-8}

\bibitem[{{Yan} {et~al.}(2021{\natexlab{a}}){Yan}, {Zhou}, {Zhang}, {Li},
  {Gao}, {Shi}, {Zhao}, {Aoki}, {Matsuno}, {Li}, {Xu}, {Li}, {Wu}, {Jin},
  {Mosser}, {Bi}, {Fu}, {Pan}, {Suda}, {Liu}, {Zhao}, \& {Liang}}]{yan2021}
{Yan}, H.-L., {Zhou}, Y.-T., {Zhang}, X., {et~al.} 2021{\natexlab{a}}, Nature
  Astronomy, 5, 86, \dodoi{10.1038/s41550-020-01217-8}

\bibitem[{{Yan} {et~al.}(2021{\natexlab{b}}){Yan}, {Zhou}, {Zhang}, {Li},
  {Gao}, {Shi}, {Zhao}, {Aoki}, {Matsuno}, {Li}, {Xu}, {Li}, {Wu}, {Jin},
  {Mosser}, {Bi}, {Fu}, {Pan}, {Suda}, {Liu}, {Zhao}, \&
  {Liang}}]{2021NatAs...5...86Y}
---. 2021{\natexlab{b}}, Nature Astronomy, 5, 86,
  \dodoi{10.1038/s41550-020-01217-8}

\bibitem[{{Yong} {et~al.}(2013){Yong}, {Norris}, {Bessell}, {Christlieb},
  {Asplund}, {Beers}, {Barklem}, {Frebel}, \& {Ryan}}]{yong2013}
{Yong}, D., {Norris}, J.~E., {Bessell}, M.~S., {et~al.} 2013, \apj, 762, 26,
  \dodoi{10.1088/0004-637X/762/1/26}

\bibitem[{{Zhang} \& {Jeffery}(2013)}]{zhang2013}
{Zhang}, X., \& {Jeffery}, C.~S. 2013, \mnras, 430, 2113,
  \dodoi{10.1093/mnras/stt035}

\bibitem[{Zhang {et~al.}(2020)Zhang, Jeffery, Li, \& Bi}]{Zhang_2020}
Zhang, X., Jeffery, C.~S., Li, Y., \& Bi, S. 2020, The Astrophysical Journal,
  889, 33, \dodoi{10.3847/1538-4357/ab5e89}

\bibitem[{{Zhou} {et~al.}(2018){Zhou}, {Shi}, {Yan}, {Gao}, {Zhang}, {Zhao},
  {Pan}, \& {Kumar}}]{2018A&A...615A..74Z}
{Zhou}, Y.~T., {Shi}, J.~R., {Yan}, H.~L., {et~al.} 2018, \aap, 615, A74,
  \dodoi{10.1051/0004-6361/201730389}

\end{thebibliography}
